\documentclass[a4paper,11pt]{article}
\usepackage[utf8]{inputenc}
 \usepackage{graphicx}
\usepackage{cancel}
\usepackage{amsmath, amssymb, amsfonts}
\usepackage{bbold}
\usepackage{latexsym}
\usepackage{subfigure}
\usepackage[comma,numbers,sort&compress]{natbib}
\usepackage[font=small, format=plain, labelfont=bf,up, textfont=up]{caption}
\usepackage[top=2.9cm, bottom=3cm, left=2.1cm, right=2cm]{geometry}
\usepackage{setspace}
    \usepackage{sectsty}
    \sectionfont{\LARGE}
    \subsectionfont{\Large}
    \subsubsectionfont{\large}
\usepackage{color}
\makeatletter
\def\maketag@@@#1{\hbox{\m@th\normalfont\normalsize#1}}
\makeatother
\usepackage[colorlinks=true,citecolor=blue,urlcolor=blue,linktocpage=true,linkcolor=blue]{hyperref}
\hypersetup{pageanchor=false}

\newcommand{\ca}{c_{\alpha}}
\newcommand{\sa}{s_{\alpha}}
\newcommand{\cb}{c_{\beta}}
\newcommand{\sinb}{s_{\beta}}
\newcommand{\sw}{s_{W}}
\newcommand{\cw}{c_{W}}

\newcommand{\mo}{m_{\tilde{\chi}_1^0}}
\newcommand{\mf}{m_{\tilde{\chi}_4^0}}
\newcommand{\co}{\tilde{\chi}_1^0}
\newcommand{\cf}{\tilde{\chi}_4^0}
\newcommand{\ctw}{\tilde{\chi}_2^0}
\newcommand{\cth}{\tilde{\chi}_3^0}

\newcommand{\tb}{t_{\beta}}

\newcommand{\sib}{s_{\beta}}

\newcommand{\ci}{\tilde{\chi}_i^{0}}
\newcommand{\cj}{\tilde{\chi}_j^{0}}

\newcommand{\pl}{\omega_L}
\newcommand{\pr}{\omega_R}

\newcommand{\bpm}{\begin{pmatrix}}
\newcommand{\epm}{\end{pmatrix}}
\newcommand{\mhp}{M_{H^{\pm}}}
\newcommand{\qsq}{q^{2}}
\newcommand{\psq}{p^{2}}

\newcommand{\gh}{\hat{\Gamma}_}

\newcommand{\Zbf}{\hat{\textbf{Z}}}
\newcommand{\Mm}{\mathcal{M}}
\newcommand{\Pp}{\mathcal{P}}
\newcommand{\Dd}{\mathcal{D}}
\newcommand{\epsp}{\epsilon_+}
\newcommand{\epsm}{\epsilon_-}
\newcommand{\CP}{\mathcal{CP}}
\newcommand{\Zb}{\hat{\textbf{Z}}}

\title{\small{\hfill {\tt DESY 14-194}}\\\vspace*{1cm}
\LARGE{\textbf{Interference effects in BSM processes\\with a generalised narrow-width approximation}}}
\author{\vspace*{0.3cm}\\
\textsc{Elina Fuchs}\footnote{elina.fuchs@desy.de},~~
\textsc{Silja Thewes}\footnote{former address},~~
\textsc{Georg Weiglein}\footnote{georg.weiglein@desy.de} \vspace*{0.6cm}\\
\textit{DESY, Deutsches Elektronen-Synchrotron, Notkestr.\ 85, D-22607 Hamburg, Germany}}

\begin{document}
\maketitle
\thispagestyle{empty}
\begin{abstract}
A generalisation of the narrow-width approximation (NWA) is formulated
which allows for a consistent treatment of interference effects between
nearly mass-degenerate particles in the factorisation of a more
complicated process into production and decay parts. 
It is demonstrated that interference effects of this kind arising in BSM 
models can be very
large, leading to drastic modifications of predictions based on the
standard NWA. The application of the 
generalised NWA is demonstrated both at tree level and at one-loop order 
for an example process where the neutral Higgs bosons $h$
and $H$ of the MSSM are produced in the decay of a heavy neutralino and
subsequently decay into a fermion pair. The generalised
NWA, based on on-shell matrix elements or their approximations leading
to simple weight factors, is shown to produce UV- and IR-finite results
which are numerically close to the result of the full process at tree
level and at one-loop order, where an agreement of better than
$1\%$ is found for the considered process. The most accurate prediction
for this process based on the generalised NWA, taking into account also
corrections that are formally of higher orders, is briefly discussed.
\end{abstract}
\newpage
\thispagestyle{empty}
\tableofcontents
\thispagestyle{empty}
\newpage
\section{Introduction}
\setcounter{page}{1}
The description of the fundamental interactions of nature in terms of quantum
field theories that are evaluated perturbatively has been
extraordinarily successful in the context of elementary particle
physics. Nevertheless, this theoretical formulation is plagued by a
long-standing problem, since the asymptotic in- and outgoing states of
quantum field theories are defined at infinite times corresponding to
stable incoming and outgoing particles, while collider physics processes
usually involve numerous unstable particles. While in principle it would
be possible to perform calculations of the theoretical predictions for
the full process of stable incoming and outgoing particles, this is in
many cases not feasible in practice (and still leaves the problem of the
treatment of intermediate particles that can become resonant). Instead,
one often seeks to simplify the task of calculating a more complicated
process by separately treating the production of on-shell particles and
their decays, where the latter can happen in several separate steps,
each resulting in on-shell outgoing particles. Such an approach of
simplifying the task of computing a complicated process involving many
particles in the final state is in particular crucial in the context of
incorporating higher-order corrections.

The separation of a more complicated process into several sub-processes
involving on-shell particles as incoming and outgoing states is achieved with
the help of the ``narrow-width approximation'' (NWA) for particles having a
total width that is much smaller than their mass. The application of the NWA
is beneficial since the sub-processes can often be calculated at a higher loop
order than it would be the case for the full process, and it is also useful in
terms of computational speed. Indeed, many 
Monte-Carlo generators make use of the NWA.
An important condition
limiting the applicability of this approximation, however, is the requirement
that there should be no interference of the contribution of the intermediate
particle for which the NWA is applied with any other close-by resonance.
While within the Standard Model (SM) of particle physics
this condition is usually valid for relevant processes at
high-energy colliders such as the LHC or a future Linear Collider, many models
of physics beyond the SM (BSM) have mass spectra where two or more states can
be nearly mass-degenerate.
If the mass gap between two intermediate particles is
smaller than one of their total widths, the interference term between
the contributions from the two nearly mass-degenerate particles
may become large.

For instance, mass degeneracies can be encountered in the
Minimal Supersymmetric extension of the Standard Model
(MSSM)~\cite{Nilles:1983ge,Haber:1984rc,Barbieri:1987xf}. In particular,
the MSSM may contain approximately mass-degenerate first and second generation
squarks and sleptons. In the decoupling limit~\cite{Haber:1989xc},
the MSSM predicts a SM-like light Higgs boson, which can be compatible with
the signal discovered by 
ATLAS\,\cite{Aad:2012tfa} and CMS\,\cite{Chatrchyan:2012ufa} at 
a mass of about $M_h \simeq 125$\,GeV, and two further neutral Higgs bosons
and a charged Higgs boson $H^\pm$, which are significantly heavier and 
nearly mass-degenerate. 
While in the $\CP$-conserving case the heavy neutral Higgs bosons $H$ and $A$ are
$\CP$-eigenstates and therefore do not mix with
each other, $\CP$-violating loop contributions can induce sizable interference
effects, see e.g.\ Ref.~\cite{Fowler:2010eba}.
The compatibility of degenerate NMSSM Higgs
masses with the observed Higgs decay rate into two photons was recently
pointed out e.g.\ in Ref.\,\cite{Gunion:2012gc}. Another example are degenerate
Higgs bosons in (non-supersymmetric) two-Higgs doublet
models, see e.g.\ Refs.~\cite{Ferreira:2012nv,Drozd:2012vf}. 
Furthermore, degeneracies can also
occur in models of (universal) extra dimensions where the masses at one
Kaluza-Klein level are degenerate up to their SM masses and loop corrections,
see for example
Refs.~\cite{Appelquist:2000nn,Cheng:2002ab,Hooper:2007qk}.
On the other hand, models with new particles on various mass levels often
exhibit long cascade decays, so that there is a particular need in these cases 
for an approximation with which the complicated full process can be
simplified into smaller pieces that can be treated more easily.
However, several cases have been identified in the literature in which the NWA
is insufficient due to sizeable
interference effects, e.g.\ in the context of the MSSM in
Refs.~\cite{Reuter:2007me, Berdine:2007uv,Gigg:2008yc, Kalinowski:2008fk,
UhlemannDipl} and in the context of two- and multiple-Higgs models and in
Higgsless models in Ref.~\cite{Cacciapaglia:2009ic}. 

In the following we present a generalised NWA (gNWA), which extends the 
standard NWA (sNWA) by providing a
factorisation into on-shell production and decay while taking into
account interference effects.
In Ref.~\cite{Fowler:2010eba} such a method was introduced at the 
tree level and applied to interference effects in the MSSM Higgs sector. This
method was further extended in Ref.~\cite{ElinaMSc}, in particular by
incorporating partial loop contributions into an interference weight factor. 
A similar coupling-based estimation of an interference between new
heavy quarks at lowest order 
was suggested in Ref.~\cite{Barducci:2013zaa}.
In the present paper we formulate a gNWA based on an on-shell evaluation of
the interference contributions which is applicable at the loop level,
incorporating factorisable virtual and real corrections. We validate the
method for an example process by confronting the one-loop result within the
gNWA with the result of the full process at the one-loop level. We furthermore
investigate different levels of approximations, where we compare the
on-shell matrix elements in the interference term with possible further
simplifications based on interference weight factors. In the considered
example process we study
interference effects between the two neutral $\mathcal{CP}$-even MSSM Higgs
bosons $h$ and $H$ in the decay of a heavy neutralino and the
subsequent decay into a fermion pair. Besides the validation against the full
result for this process we also discuss additional improvements by the
incorporation of corrections that are formally of higher orders. The discussed
cases are meant to illustrate that the proposed method is applicable to a wide range
of possible processes in different models.

The paper is structured as follows. Sect.\,\ref{sect:sNWA} reviews the
standard NWA before introducing the inter\-ference-improved extension in two
different versions in Sect.\,\ref{sect:gNWA}. 
The notation of the parts of the MSSM that are needed for the phenomenological
discussion in the following sections
is defined in Sect.\,\ref{sect:MSSM}, with particular
emphasis on the mixing of Higgs bosons. 
In Sect.\,\ref{sect:example}, the gNWA
is applied at the tree level to the example process of Higgs production from
the decay of a heavier neutralino and its subsequent decay into a pair of
$\tau$-leptons. 
The numerical results for those contributions are discussed in
Sect.\,\ref{sect:numtree}.
In Sect.\,\ref{sect:IntNLO} the application of the gNWA at the loop level 
is demonstrated. 
For comparison, the full
one-loop calculation of the example process is performed in
Sect.\,\ref{sect:3bodyNLO}, including vertex, propagator, box and
bremsstrahlung corrections. The numerical comparison and accordingly the
validation of the gNWA at NLO is discussed in 
Sect.\,\ref{sect:ResultLoop}, where also the accuracy of the
gNWA is investigated. 
Sect.\,\ref{sect:conclusion} contains our conclusions.

\section{Standard narrow-width approximation}\label{sect:sNWA}
\begin{figure}[ht!]
\centering
 \includegraphics[width=8.2cm]{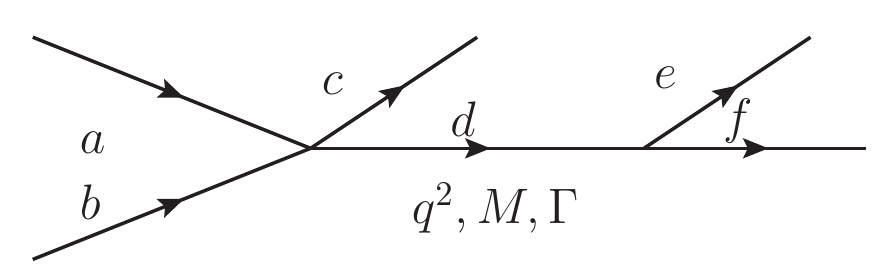}
 \includegraphics[width=8.2cm]{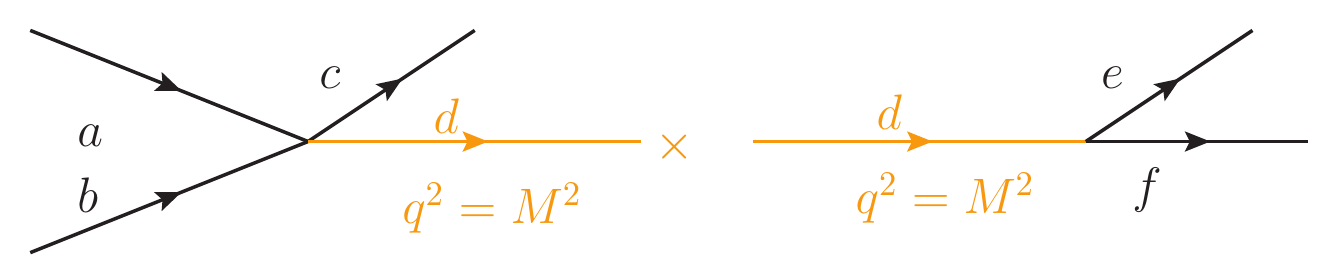}
\caption[Splitting a process into production and decay]{The resonant process $ab \rightarrow cef$ is split into the production $ab \rightarrow cd$ and decay $d\rightarrow ef$ with particle $d$ on-shell.}
\label{fig:prod_decay}
\end{figure}

The narrow-width approximation (NWA) is a useful way to simplify the
calculation of complicated processes involving the resonant contribution 
of an unstable particle.
The basic idea is to factorise the whole process into the on-shell
production and the subsequent decay of the resonant particle. 
The following picture in
Fig.\,\ref{fig:prod_decay} visualises this splitting using the example of an
arbitrary process $ab \rightarrow cef$ with an intermediate particle $d$.
\\
In the following, we focus on scalar propagators. Nonetheless, although the
production and decay are calculated independently, the spin of an intermediate
particle can be taken into account by means of spin
correlations\,\cite{Dicus:1984fu} 
giving rise to spin--density matrices. While we do not consider the non-zero
spin case explicitly, the formalism of spin--density matrices should be
applicable to the gNWA discussed below in the same way as for the standard NWA (sNWA).

\subsection{Unstable particles and the total decay width}\label{sect:unstable}
Since the total width $\Gamma$ plays a crucial role in resonant production and
decay, we will briefly discuss resonances and unstable particles, see e.g.\
Refs.~\cite{Wackeroth:1996hz,Freitas:2002ja}. 
While stable particles are associated with a
real pole of the S-matrix, 
for unstable particles the associated self-energy develops an imaginary part,
so that the pole of the propagator is located off the real axis within the
complex plane. For a single pole $M_c$, 
the scattering matrix as a function of the squared centre-of-mass energy $s$ can 
be written in the vicinity of the complex pole in a gauge-invariant way as
\begin{equation}
 \mathcal{M}(s) = \frac{R}{s-M_c^{2}}+F(s) \label{eq:gaugeinv},
\end{equation}
where $R$ denotes the residue and $F$ represents non-resonant contributions.
Writing the complex pole as $M_c^{2}=M^{2}-iM\Gamma$, the mass $M$ of an
unstable particle is obtained from the real part of the complex pole, while
the total width is obtained from the imaginary part. Accordingly, the
expansion around the complex pole leads to a Breit--Wigner propagator with a
constant width,
\begin{equation}
\Delta^{\text{BW}}(q^{2}):=\frac{1}{q^2 - M^2 + i M\Gamma} . \label{eq:BWdef} 
\end{equation}
In the following, we will use a Breit--Wigner propagator of this form
to express the contribution
of the unstable scalar $d$ with mass $M$ and total width $\Gamma$ in the
resonance region (a Breit--Wigner propagator with a running width can be
obtained from a simple reparametrisation of the mass and width appearing in 
Eq.~(\ref{eq:BWdef})).

The NWA is based on the observation that the on-shell contribution in 
Eq.~(\ref{eq:BWdef}) is strongly enhanced if the total width is much smaller than
the mass of the particle, 
$\Gamma \ll M$. 
Within its range of validity (see the discussion in the following section) the 
NWA provides an approximation of the cross section for
the full process 
in terms of the product of the production cross
section (or the previous step in a decay cascade) 
times the respective branching ratio:
\begin{equation}
 \sigma_{ab \rightarrow cef} \simeq \sigma_{ab \rightarrow cd} \times \text{BR}_{d\rightarrow ef}. \label{eq:NWAbasic}
\end{equation}

\subsection{Conditions for the narrow-width approximation}\label{sect:cond}
The NWA can only be expected to hold reliably if the following prerequisites
are fulfilled (see e.g.\ Refs.~\cite{Berdine:2007uv, Kauer:2007zc}):
\begin{itemize}
 \item A narrow mass peak is required in order to justify the on-shell
approximation. Otherwise off-shell effects may become large, cf.\ e.g.\,\cite{Gigg:2008yc,Uhlemann:2008pm}.
 \item Furthermore, the propagator needs to be separable from the matrix
element. However, loop contributions involving a particle exchange between
the initial and the final state give rise to non-factorisable corrections.
Hence, the application of the NWA beyond lowest order relies on the assumption
that the non-factorisable and non-resonant contributions are sufficiently
suppressed compared to the dominant contribution where the unstable particle
is on resonance. Concerning the incorporation of non-factorisable but resonant
contributions from photon exchange, see e.g.\ Ref.~\cite{Denner:2000bj}.
 \item Both sub-processes have to be kinematically allowed. For the production of the intermediate particle, this means that the centre of mass energy $\sqrt{s}$ must be well above the production threshold of the intermediate particle with mass $M$ and the other particles in the final state of the production process, i.e. $\sqrt{s}\gg M+m_c$ for the process shown in Fig.\,\ref{fig:prod_decay}. Otherwise, threshold effects must be considered \cite{Kauer:2007nt}.
 \item On the other hand, the decay channel must be kinematically open and
sufficiently far above the decay threshold, i.e.\ $M \gg \sum m_f$, where $m_f$
are the masses of the particles in the final state of the decay process, here
$m_e+m_f$. Off-shell effects 
can be enhanced if intermediate thresholds are present. This is the case for
instance for the decay of a Higgs boson with a mass of about 125~GeV into four
leptons. Since for an on-shell Higgs boson of this mass this process is far
below the threshold for on-shell $WW$ and $ZZ$ production, it suffers from a
significant phase-space suppression. Off-shell Higgs contributions above the
threshold for on-shell $WW$ and $ZZ$ production are therefore numerically more
important than one would expect just from a consideration of $\Gamma/M$.
 \item As another crucial condition, interferences with other resonant or
non-resonant diagrams have to be small because the mixed term would be
neglected in the NWA. The major part of the following chapters is dedicated to
a generalisation of the NWA for the inclusion of interference effects, see also Refs.~\cite{Fowler:2010eba,ElinaMSc}.
\end{itemize}

\subsection{Factorisation of the phase space and cross section} \label{sect:dlips}
In order to fix the notation used for the formulation of the gNWA in Sect.\,\ref{sect:gNWA}, we review some kinematic relations.
\paragraph{The phase space}
The phase space $\Phi$ is a Lorentz invariant quantity. Its differential is denoted as $dlips$ (differential Lorentz invariant phase space) or $d\Phi_n$. It is characterised by the number $n$ of particles in the final state\,\cite{Beringer:1900zz}
\begin{equation}
 d\Phi_n \equiv  dlips\left( P;p_1,...,p_n\right) 
	 =(2\pi)^4\delta^{(4)}(P-\sum\limits_{f=1}^np_f)\displaystyle\prod_{f=1}^n\frac{d^3p_f}{(2\pi)^32E_f}\label{eq:dPhi_n}.
\end{equation}
\paragraph{Factorisation}
Eq.\,(\ref{eq:NWAbasic}) is based on the property of the phase space and the matrix element to be factorisable into sub-processes. The phase space element $d\Phi_n$ with $n$ particles in the final state as in Eq.\,(\ref{eq:dPhi_n}) will now be expressed as a product of the $k$-particle phase space $\Phi_k$ with $k < n$ and the remaining $\Phi_{n-k+1}$ \cite{ByckKaja:72,Beringer:1900zz},
\begin{equation}
 d\Phi_n =  d\Phi_k  \frac{dq^2}{2\pi}  d\Phi_{n-k+1},\label{eq:phasespace}
\end{equation}
where $q$ denotes the momentum of the resonant particle.
Now $\Phi_k(q)$ can be interpreted as the \textit{production} phase space $P \rightarrow \left\lbrace p_1, ...,p_{k-1},q\right\rbrace $ and $\Phi_{n-k+1}(q)$ as the \textit{decay} phase space $q \rightarrow \left\lbrace p_k,...,p_n\right\rbrace$.
The factorisation of $d\Phi_n$ is exact, no approximation has been made so far. Next, we rewrite the amplitude with a scalar propagator as a product of the production (P) and decay (D) part. Beyond the tree level, this is only possible if non-factorisable loop-contributions are absent or negligible,
\begin{equation}
 \mathcal{M} = \mathcal{M}_{P}\frac{1}{q^2 - M^2 +iM\Gamma}\mathcal{M}_{D}\hspace*{0.5cm}
\Rightarrow |\mathcal{M}|^2 = |\mathcal{M}_{P}|^{2}\frac{1}{(q^2 - M^2)^2 +(M\Gamma)^2}|\mathcal{M}_{D}|^{2}. \label{eq:ampl_sq}
\end{equation}
One can distinguish two categories of processes. On the one hand, for a scattering process $a,b \rightarrow X$ to any final state $X$ (in particular $a,b\rightarrow c,e,f$ for the example in Fig.\,\ref{fig:prod_decay}), the flux factor is given by $F=2\lambda^{1/2}(s,m_a^{2},m_b^{2})$
with the kinematic function\,\cite{ByckKaja:72}
\begin{equation}
 \lambda(x,y,z) := x^{2}+y^{2}+z^{2}-2(xy+yz+zx)\label{eq:lambda}.
\end{equation}
On the other hand, for a decay process $a\rightarrow X$ (for example $a\rightarrow c,e,f$), the flux factor is determined by the mass of the decaying particle, $F=2m_a$. Then the full cross section is given as
\begin{equation}
 \sigma = \frac{1}{F}\int d\Phi|\mathcal{M}|^2 \label{eq:xsec}.
\end{equation}
For the decomposition into production and decay, we do not only factorise the
matrix elements as in Eq.\,(\ref{eq:ampl_sq}). Based on
Eq.\,(\ref{eq:phasespace}), also the phase space of the full process is
factorised into the production phase space $\Phi_P$ and the decay phase space
$\Phi_D$ (here defined for the example process in Fig.\,\ref{fig:prod_decay},
but they can be generalised to other external momenta), which depend on 
the momentum of the resonant particle:
\begin{align}
 d\Phi~~&= dlips(\sqrt{s}; p_c, p_e, p_f)\nonumber\\
 d\Phi_P &= dlips(\sqrt{s}; p_c, q)\nonumber\\
 d\Phi_D &= dlips(q; p_e, p_f)\label{eq:dlipsPD}.
\end{align}
Under the assumption of negligible
non-factorisable loop contributions, one can then express the cross section in
(\ref{eq:xsec}) as
\begin{align}
 \sigma &= \frac{1}{F}\int \frac{dq^2}{2\pi}\left(\int d\Phi_{P} \overline{|\mathcal{M}_{P}|^{2}}\right) \frac{1}{(q^2 - M^2)^2 +(M\Gamma)^2} \left(\int d\Phi_{D} \overline{|\mathcal{M}_{D}|^{2}}\right)\label{eq:sigma}.
\end{align}
In this analytical formula of the cross section, the production and decay matrix elements and the sub-phase spaces are separated from the Breit-Wigner propagator. However, the full $q^{2}$-dependence of the matrix elements and the phase space is retained. The off-shell production cross section of a scattering process with particles $a,\,b$ in the initial state and the production flux factor $F$
reads
\begin{equation}
 \sigma_P(q^{2}) =\frac{1}{F}\int d\Phi_P |\mathcal{M}_P(q^{2})|^{2} \label{eq:sigmaP}.
\end{equation}
The decay rate of the unstable particle, $d\rightarrow ef$, with energy $\sqrt{q^{2}}$ is obtained from the integrated squared decay matrix element divided by the decay flux factor $F_D=2\sqrt{q^{2}}$,
\begin{equation}
  \Gamma_D(q^{2}) = \frac{1}{F_D}\int d\Phi_D|\mathcal{M}_D(q^{2})|^{2} \label{eq:GammaD}.
\end{equation}
Hence one can rewrite the full cross section from Eq.\,(\ref{eq:sigma}) as
\begin{equation}
 \sigma = \int \frac{dq^{2}}{2\pi} \sigma_P(q^{2})\frac{2\sqrt{q^{2}}}{(q^2 - M^2)^2 +(M\Gamma)^2} \Gamma_D(q^{2}). \label{eq:sigmaofs}
\end{equation}
In the limit where $(\Gamma M)\rightarrow0$ 
the Dirac $\delta$-distribution emerges from the Cauchy
distribution,
\begin{equation}
\lim\limits_{(M\Gamma)\rightarrow 0}\frac{1}{(q^{2}-M^{2})^2+(M\Gamma)^2} = \delta(q^{2}-M^{2})\frac{\pi}{M\Gamma}\label{eq:BWcauchy}.
\end{equation}
For the integration of the $\delta$-distribution, the integral boundaries are
shifted from $q^{2}_{\rm{max}},\,q^{2}_{\rm{min}}$, i.e.\ the upper and lower bound on
$q^{2}$, respectively, to $\pm \infty$ because the contributions outside the
narrow resonance region are expected to be small. So this extension of the
integral should not considerably alter the result. The zero-width limit
implies that the production cross section, decay width
and the factor $\sqrt{q^{2}}$ are evaluated 
on-shell at $q^{2}=M^{2}$. This applies both to
the matrix elements and the phase space elements. The described approximation
leads to the 
well-known factorisation into the production cross section times the decay
branching ratio,
\begin{align}
 \sigma \stackrel{(M\Gamma)\rightarrow 0}{\rightarrow}&\int\limits_{-\infty}^{+\infty} \frac{dq^{2}}{2\pi}\sigma_P(q^{2})\,2\sqrt{q^{2}}\,\delta(q^{2}-M^{2})\frac{\pi}{M\Gamma}\,\Gamma_D(q^{2})
~=~\sigma_P(M^{2})\cdot\frac{\Gamma_D(M^{2})}{\Gamma} \equiv \sigma_P\cdot \text{BR}\label{eq:NWAproof},
\end{align}
with the branching ratio $\text{BR}=\Gamma_{D}/\Gamma$, where
$\Gamma_{D}$ denotes the partial decay width into the particles in the final
state of the considered process, and $\Gamma$ is the total decay width of the
unstable particle. While
Eq.~(\ref{eq:NWAproof}) has been obtained in the limit $(M\Gamma)\rightarrow 0$,
it is expected to approximate the result for non-zero $\Gamma$ up to terms of
$\mathcal{O}(\frac{\Gamma}{M})$. 

Going beyond the approximation of Eq.~(\ref{eq:NWAproof}) for the
treatment of finite width effects, the on-shell approximation can
be applied just to the matrix elements for production and decay 
while keeping a finite width in the
integration over the Breit-Wigner propagator in the form of
Eq.~(\ref{eq:sigmaofs}). This is motivated 
by the consideration
that the Breit-Wigner function is rapidly falling causing that only matrix
elements close to the mass shell $q^{2}=M^{2}$ contribute significantly.
It results in the \textit{finite-narrow-width approximation} improved for
off-shell effects, see e.g.\ Ref.\,\cite{Berdine:2007uv},
\begin{align}
 \sigma^{(ofs)} &= \sigma_P(M^{2})\left[\int \frac{dq^{2}}{2\pi} \frac{2\sqrt{q^{2}}}{(q^2 - M^2)^2 +(M\Gamma)^2} \right]\Gamma_D(M^{2})\label{eq:ofs}.
\end{align}

\section{Formulation of a generalised narrow-width approximation}\label{sect:gNWA}
\subsection{Cross section with interference term}
If all conditions in Sect.\,\ref{sect:cond} are met, the NWA is expected to
work reliably 
up to terms of $\mathcal{O}(\frac{\Gamma}{M})$. This
section addresses the issue of how to extend the NWA such that interference
effects can be 
included, leading to a generalised NWA (gNWA)~\cite{Fowler:2010eba,ElinaMSc}. 
Interference effects can be large if
there are several resonant diagrams whose intermediate particles are close in
mass compared to their total decay widths:
\begin{equation}
 |M_1 - M_2| \lesssim \Gamma_1,\Gamma_2. \label{eq:DMGamma}
\end{equation}
In these nearly mass-degenerate cases, the Breit-Wigner functions
$\Delta^{\text{BW}}_1(q^2),~ \Delta^{\text{BW}}_2(q^2)$ overlap significantly,
and an integral of the form
\begin{equation}
 \int\limits_{q^2_{\rm{min}}}^{q^2_{\rm{max}}}dq^{2}\Delta^{\text{BW}}_1(q^2)\Delta^{*\text{BW}}_2(q^2)\cdot f(\mathcal{M},p_i,...) \label{eq:overlap}
\end{equation}
is not negligible. The boundaries $q^{2}_{\rm{min}}, q^{2}_{\rm{max}}$ are the lower and upper limits of the kinematically allowed region of $q^{2}$, and $f$ summarises a possible dependence on matrix elements $\mathcal{M}$ and momenta $p_i$ in the phase space. Such interference effects might especially be relevant in models of new physics where an enlarged particle spectrum allows for more possibilities of mass degeneracies in some parts of the parameter space.\\
Let $h_1, h_2$ be two resonant intermediate particles, for example two Higgs
bosons, with similar masses occurring in a process $ab \rightarrow cef$, i.e.\
$ab \rightarrow c h_i, h_i \rightarrow ef$ (cf.\ Fig.\,\ref{fig:prod_decay}
with $d=h_1,h_2$).
If non-factorisable loop corrections can be neglected, the full matrix element
(dropping the $q^{2}$-dependence of the matrix elements to simplify the
notation) is given by
(as mentioned above, see Sect.~\ref{sect:sNWA},
we explicitly treat the case of scalar resonant particles; spin correlations
of intermediate particles with non-zero spin can be taken into account using
spin--density matrices)
\begin{align}
 \mathcal{M} &=\mathcal{M}_{ab\rightarrow c h_1}\frac{1}{q^2-M_1^2 + iM_1\Gamma_1}\mathcal{M}_{h_1\rightarrow ef} + 				\mathcal{M}_{ab\rightarrow c h_2}\frac{1}{q^2-M_2^2 + iM_2\Gamma_2}\mathcal{M}_{h_2\rightarrow ef}\label{eq:Mfull}.
\end{align}
The squared matrix element contains the two separate contributions of $h_1,\,h_2$ and in the second line of Eq.\,(\ref{eq:full}) the interference term,
\begin{align}
|\mathcal{M}|^2 &=\frac{|\mathcal{M}_{ab\rightarrow c h_1}|^2|\mathcal{M}_{h_1\rightarrow ef}|^2}{(q^2 - M_1^2)^2 + M_1^2 \Gamma_1^2} +\frac{|\mathcal{M}_{ab\rightarrow c h_2}|^2|\mathcal{M}_{h_2\rightarrow ef}|^2}{(q^2 - M_2^2)^2 + M_2^2 \Gamma_2^2}\nonumber\\ 
	&\hspace*{0.5cm}+2 \text{Re}\left\lbrace \frac{\mathcal{M}_{ab\rightarrow c h_1}\mathcal{M}^*_{ab\rightarrow c h_2}\mathcal{M}_{ h_1\rightarrow ef}\mathcal{M}^*_{h_2\rightarrow ef}}{(q^2-M_1^2 + iM_1\Gamma_1)(q^2-M_2^2 - iM_2\Gamma_2)} \right\rbrace\label{eq:full}.
\end{align}
So the full cross section from Eq.\,(\ref{eq:sigmaofs}) with the matrix element from Eq.\,(\ref{eq:full}) can be written as
\begin{align}
 &\sigma_{ab\rightarrow cef} =\int \frac{dq^{2}}{2\pi}\left[\frac{\sigma_{ab\rightarrow ch_1}(q^{2})~2\sqrt{q^{2}}~ \Gamma_{h_1\rightarrow ef}(q^{2})}{(q^2 - M_{h_1}^2)^2 +(M_{h_1}\Gamma_{h_1})^2}+ \frac{\sigma_{ab\rightarrow ch_2}(q^{2})~2\sqrt{q^{2}}~ \Gamma_{h_2\rightarrow ef}(q^{2})}{(q^2 - M_{h_2}^2)^2 +(M_{h_2}\Gamma_{h_2})^2}\right]\nonumber\\
 &~+\int \frac{dlips(s;p_c,q)dq^{2}dlips(q;p_e,p_f)}{2\pi\cdot2\lambda^{1/2}(s,m_a^{2},m_b^{2})} 2\text{Re}\left\lbrace\frac{\mathcal{M}_{ab\rightarrow c h_1}\mathcal{M}^*_{ab\rightarrow c h_2}\mathcal{M}_{ h_1\rightarrow ef}\mathcal{M}^*_{h_2\rightarrow ef}}{(q^2-M_1^2 + iM_1\Gamma_1)(q^2-M_2^2 - iM_2\Gamma_2)}\right\rbrace  \label{eq:fullxsec}.
\end{align}
We will use Eq.\,(\ref{eq:fullxsec}) as a starting point for approximations of the full cross section. The first two terms can again be approximated by the finite-narrow-width approximation according to Eq.\,(\ref{eq:ofs}), or by the usual narrow-width approximation in the limit of a vanishing width from Eq.\,(\ref{eq:NWAproof}) as $\sigma\times \text{BR}$. The interference term still consists of an integral over the $q^{2}$-dependent matrix elements, the product of Breit-Wigner propagators and the phase space.

\subsection{On-shell matrix elements}\label{sect:intMos}
Our approach is to evaluate the production ($\mathcal{P}$) and decay ($\mathcal{D}$) matrix elements 
\begin{equation}
 \Pp_i(q^{2}) \equiv \mathcal{M}_{ab\rightarrow ch_i}(q^{2}),
~~~\Dd_i(q^{2}) \equiv\mathcal{M}_{h_i\rightarrow ef}(q^{2}) \label{eq:PDdef}
\end{equation}
on the mass shell of the intermediate particle $h_i$ \cite{ElinaMSc}. This is motivated by the assumption of a narrow resonance region $[M_{h_i}-\Gamma_{h_i}, M_{h_i}+\Gamma_{h_i}]$ so that off-shell contributions of the matrix elements in the integral are suppressed by the non-resonant tail of the Breit-Wigner propagator. Then the interference term from the last line of Eq.\,(\ref{eq:fullxsec}) is approximated by
\begin{align}
 \sigma_{\rm{int}} &= \int \frac{d\Phi_P dq^{2} d\Phi_D}{2\pi F} \text{Re}\frac{\Pp_{1}(\qsq)\Pp^*_{2}(\qsq)\Dd_{1}(\qsq)\Dd^*_{2}(\qsq)}{(q^2-M_1^2 + iM_1\Gamma_1)(q^2-M_2^2 - iM_2\Gamma_2)}\label{eq:intexact}\\
 &= \frac{2}{F}\text{Re}\int \frac{dq^{2}}{2\pi}\Delta^{\text{BW}}_1(q^{2}) \Delta^{*\text{BW}}_2(q^{2})\left[\int d\Phi_P(\qsq)\Pp_{1}(q^{2})\Pp^*_{2}(q^{2})\right]\left[\int d\Phi_D(\qsq) \Dd_{1}(q^{2})\Dd^*_{2}(q^{2})\right] \nonumber\\
&\simeq \frac{2}{F}\text{Re}\int \frac{dq^{2}}{2\pi}\Delta^{\text{BW}}_1(q^{2}) \Delta^{*\text{BW}}_2(q^{2})\left[\int d\Phi_P(\qsq)\Pp_{1}(M_1^{2})\Pp^*_{2}(M_2^{2})\right]\left[\int d\Phi_D(\qsq) \Dd_{1}(M_1^{2})\Dd^*_{2}(M_2^{2}) \right]\label{eq:masterformula}.
\end{align}
Eq.~(\ref{eq:masterformula}) represents our master formula for the
interference contribution. At this stage, we
have only evaluated the matrix elements on the mass shell of the particular
Higgs boson by setting $\qsq=M_{h_i}^{2}$ (this is also important for ensuring
the gauge invariance of the considered contributions). 
So the on-shell matrix elements can
be taken out of the $\qsq$-integral. But the dependence of the matrix elements
on further invariants and momenta is kept. For 2-body decays, it is possible
to carry out the phase space integration without referring to the specific
form of the matrix elements. In general, however, the matrix elements are
functions of the phase space integration variables.

The approximation in Eq.\,(\ref{eq:masterformula}) is a simplification of the
full expression in Eq.\,(\ref{eq:intexact}) since the integrand of the
$\qsq$-integral is simplified. We will use Eq.\,(\ref{eq:masterformula}) in
the numerical calculation of an example process in Sect.\,\ref{sect:example}.

We will furthermore investigate additional approximations of the integral
structure in Eq.\,(\ref{eq:masterformula}), which would simplify the
application of the gNWA. This issue is discussed at the tree level in the
following section.

\subsection{On-shell phase space and tree level interference weight factors}\label{sect:intR}

The following discussion,
which focuses on the tree-level case, 
concerns a technical
simplification of the master formula in Eq.\,(\ref{eq:masterformula}). It will
be numerically applied in Fig.\,\ref{fig:sgNWA_tree} below and extended to the
1-loop level in Sect.\,\ref{sect:IntNLO}. As a possible further simplification
on top of the
on-shell approximation for matrix elements, one can also evaluate the
production and decay phase spaces on-shell. This is based on the same argument
as for the on-shell evaluation of the matrix elements because off-shell phase
space elements are multiplied with the non-resonant tail of Breit-Wigner
functions. Now the $\qsq$-independent matrix elements and phase space
integrals can be taken out of the $\qsq$-integral,
\begin{align}
 \sigma_{\rm{int}} \simeq \frac{2}{F}\text{Re}\left\lbrace\left[\int d\Phi_P\Pp_{1}(M_1^{2})\Pp^*_{2}(M_2^{2})\right]\left[\int d\Phi_D \Dd_{1}(M_1^{2})\Dd^*_{2}(M_2^{2})\right] \int \frac{dq^{2}}{2\pi}\Delta^{\text{BW}}_1(q^{2})\Delta^{*\text{BW}}_2(q^{2})\right\rbrace\label{eq:intPSos}.
\end{align}
The choice at which mass, $M_1$ or $M_2$, to evaluate the production and decay
phase space regions is not unique. We thus introduce a weighting factor between
the two possible processes, as an ansatz based on their production cross
sections and branching ratios:
\begin{equation}
 w_i := \frac{\sigma_{P_i}\,\text{BR}_i}{\sigma_{P_1}\,\text{BR}_1+\sigma_{P_2}\,\text{BR}_2}\label{eq:wi}.
\end{equation}
Then we define the on-shell phase space regions as
\begin{equation}
 d\Phi_{P/D} := w_1 d\,\Phi_{P/D}(q^{2}=M_1^{2}) + w_2\,d\Phi_{P/D}(q^{2}=M_2^{2}) \label{eq:wdPhi}.
\end{equation}
In Eq.\,(\ref{eq:intPSos}), a universal integral over the Breit-Wigner propagators emerges:
\begin{align}
 I:=&\int\limits_{q^{2}_{\rm{min}}}^{q^{2}_{\rm{max}}} \frac{dq^{2}}{2\pi}\Delta^{\text{BW}}_1(q^2)\,\Delta^{*\text{BW}}_2(q^2)\label{eq:defI},                                                                                                                                                                                                                                                                                                                                          
\end{align}
which is analytically solvable,
\begin{footnotesize}
 \begin{equation}
  I = \left[\frac{\text{arctan}\left[\frac{\Gamma_1 M_1}{M_1^2-\qsq}\right]+\text{arctan}\left[\frac{\Gamma_2 M_2}{M_2^2-\qsq}\right]+\frac{i}{2}\left(\ln\left[\Gamma_1^2 M_1^2+\left(M_1^2-\qsq\right)^2\right]-\ln\left[\Gamma_2^2 M_2^2+\left(M_2^2-\qsq\right)^2\right]\right)}{2\pi i\left(M_1^2-M_2^2-i(M_1 \Gamma_1 + M_2 \Gamma_2)\right)}\right]_{q^{2}_{\rm{min}}}^{q^{2}_{\rm{max}}}.\label{eq:Ianalyt}
 \end{equation}
\end{footnotesize}In the limit of equal masses and widths, $M=M_1=M_2$ and $\Gamma=\Gamma_1=\Gamma_2$, the product of Breit-Wigner propagators would become the absolute square, and the integral is reduced to
\begin{equation}
 I(M,\Gamma) = \int\limits_{q^{2}_{\rm{min}}}^{q^{2}_{\rm{max}}} d\qsq \frac{1}{(\qsq-M^{2})^{2} + (M\Gamma)^{2}} = \left[-\frac{1}{M\Gamma}\,\text{arctan}\left[\frac{M^{2}-\qsq}{M\Gamma}\right]\right]_{q^{2}_{\rm{min}}}^{q^{2}_{\rm{max}}}\label{eq:Iem}.
\end{equation}
This absolute square of the Breit-Wigner function is also present in the usual NWA in Eq.\,(\ref{eq:BWcauchy}), and for vanishing $\Gamma$ it can be approximated by a $\delta$-distribution. Here, however, we allow for different masses and widths from the two resonant propagators. We evaluate only the matrix elements and differential phase space on-shell, but we do not perform a zero-width approximation. This approach is analogous to the finite-narrow-width approximation without the interference term in Eq.\,(\ref{eq:ofs}).\\
Under the additional assumption of equal masses, the interference part can be approximated in terms of cross sections, branching ratios and couplings in order to avoid the explicit calculation of the product of unsquared amplitudes and their conjugates. This will also avoid the phase space integrals in the interference term as in Eq.\,(\ref{eq:intPSos}).\\
For this purpose, each matrix element is written as the coupling of the particular production or decay process, $C_{P_i}$ or $C_{D_i}$, times the helicity part $p(M_i^{2})$ or $d(M_i^{2})$, respectively,
\begin{equation}
 \Pp_{i}(M_i^{2}) = C_{P_i}\,p(M_i^{2}),~~~~~~	\Dd_{i}(M_i^{2}) = C_{D_i}\, d(M_i^{2})\label{eq:gK}.
\end{equation}
The on-shell calculation of helicity matrix elements is demonstrated in
Sect.\,\ref{sect:Helicity} where also left- and right-handed couplings are
distinguished. Here we use the schematic notation of Eq.\,(\ref{eq:gK}), but
it could directly be replaced by the L/R-sum as in Eq.\,(\ref{eq:PLR}) below.

If we then make the additional assumption $M_1\simeq M_2$, the helicity matrix elements coincide, $p(M_1^{2})\simeq p(M_2^{2})$, $d(M_1^{2})\simeq d(M_2^{2})$, thus the matrix elements differ just by fractions of their couplings,
\begin{equation}
 \Pp_{2}(M_2^{2}) \simeq \frac{C_{P_2}}{C_{P_1}}\Pp_{1}(M_1^{2}),~~~~~~	\Dd_{2}(M_2^{2}) \simeq \frac{C_{D_2}}{C_{D_1}}\Dd_{1}(M_1^{2})\label{eq:gfrac}.
\end{equation}
This enables us to replace the products of an amplitude involving the resonant particle 1 with a conjugate amplitude of resonant particle 2 by absolute squares of amplitudes as follows, where $i,j \in \{1,2\}$, $i\neq j$, and no summation over indices is implied:
\begin{align}
 \sigma_{\rm{int}} &\stackrel{(\ref{eq:intPSos})}{\simeq} 2\text{Re}\left\lbrace\left[\frac{1}{F}\int d\Phi_P\Pp_{1}\Pp^*_{2}\right]\left[\frac{1}{2M_i}\int d\Phi_D \Dd_{1}\Dd^*_{2}\right]\,2M_i \int \frac{dq^{2}}{2\pi}\Delta^{\text{BW}}_1(q^{2})\Delta^{*\text{BW}}_2(q^{2})\right\rbrace\label{eq:intMi}\\
&\stackrel{(\ref{eq:gK})}{\simeq} 2\text{Re}\left\lbrace\left[\frac{1}{F}\int d\Phi_P|\Pp_{i}|^{2}\frac{C_{P_j}^{*}}{C_{P_i}^{*}}\right]\left[\frac{1}{2M_i}\int d\Phi_D |\Dd_{i}|^{2}\frac{C_{D_j}^{*}}{C_{D_i}^{*}}\right]\,2M_i\int \frac{dq^{2}}{2\pi}\Delta^{\text{BW}}_1(q^{2})\Delta^{*\text{BW}}_2(q^{2})\right\rbrace\label{eq:intMsqi}\\
&\stackrel{(\ref{eq:sigmaP}, \ref{eq:GammaD})}{=}\sigma_{P_i}\, \Gamma_{D_i}\cdot 2M_i \cdot 2\text{Re}\left\lbrace \frac{C_{P_j}^{*}C_{D_j}^{*}}{C_{P_i}^{*}C_{D_i}^{*}}\,\int \frac{dq^{2}}{2\pi}\Delta^{\text{BW}}_1(q^{2})\Delta^{*\text{BW}}_2(q^{2})  \right\rbrace\label{eq:intijlong}\\
&~~~~=~~~\sigma_{P_i}\, \text{BR}_i\cdot 2M_i\Gamma_i \cdot 2\text{Re}\left\lbrace x_i\cdot I\right\rbrace\label{eq:intij}.
\end{align}
In the last step, we divided and multiplied by the total width $\Gamma_i$ to
obtain the branching ratio $\text{BR}_i = \frac{\Gamma_{D_i}}{\Gamma_i}$. The
universal integral $I$ over the overlapping Breit-Wigner propagators is given
in Eq.\,(\ref{eq:defI}). Furthermore, we defined a scaling factor as the ratio
of couplings\,\cite{Fowler:2010eba,ElinaMSc,Barducci:2013zaa},
\begin{equation}
 x_i := \frac{C_{P_j}^{*}C_{D_j}^{*}}{C_{P_i}^{*}C_{D_i}^{*}} = \frac{C_{P_i}C_{P_j}^{*}C_{D_i}C_{D_j}^{*}}{|C_{P_i}|^{2}|C_{D_i}|^{2}}\label{eq:xi}.
\end{equation}
Using Eq.\,(\ref{eq:intij}) and the scaling factor $x_i$ with $i=1,~ j=2$ or vice versa allows to express $\sigma_{\rm{int}}$ alternatively in terms of the cross section, branching ratio, mass and width of either of the resonant particle 1 or 2. Since no summation over $i$ or $j$ is implied in Eq.\,(\ref{eq:intij}), both contributions are accounted for by the weighting factor $w_i\in [0,1]$ from Eq.\,(\ref{eq:wi}).\\
Next, we summarise the components of $\sigma_{\rm{int}}$ apart from $\sigma_{P_i}$ and $\text{BR}_i$, which also occur in the usual NWA, in an interference weight factor
\begin{align}
 R_i &:= 2M_i \Gamma_i w_i\cdot 2\text{Re}\left\lbrace x_i I \right\rbrace \label{eq:R}.
\end{align}
Hence, in this approximation of on-shell matrix elements and production and decay phase spaces with the additional condition of equal masses, the interference contribution can be written as the weighted sum
\begin{align}
 \sigma_{\rm{int}} &\simeq \sigma_{P_1}\, \text{BR}_1\cdot R_1 + \sigma_{P_2}\, \text{BR}_2\cdot R_2 \label{eq:intR},
\end{align}
or in terms of only one of the resonant particles,
\begin{align}
 \sigma_{\rm{int}} &\simeq \sigma_{P_i}\, \text{BR}_i\cdot \tilde{R}_i \label{eq:intRtilde},\\
\tilde{R}_i &:= 2M_i \Gamma_i \cdot \text{Re}\left\lbrace x_i I \right\rbrace \equiv \frac{R_i}{2w_i} \label{eq:Rtilde}.
\end{align}
Finally, we are able to express the cross section of the complete process in this $R$-factor approximation, comprising the exchange of the resonant particles 1 and 2 as well as their interference, in the following compact form
\begin{align}
 \sigma &\simeq \sigma_{P_1}\, \text{BR}_1\cdot (1+R_1) + \sigma_{P_2}\, \text{BR}_2\cdot (1+R_2) \label{eq:sigmaR}\\
	&\simeq \sigma_{P_i}\, \text{BR}_i\cdot \,(1+2\tilde{R}_i) + \sigma_{P_j}\, \text{BR}_j \label{eq:sigmaRtilde}
\end{align}
Furthermore, it is possible to replace the term $\sigma_i\,\text{BR}_i$ in the two separate processes without the interference term by the finite-width integral from Eq.\,(\ref{eq:ofs}).

\subsection{Discussion of the steps of approximations}
In the previous sections, we presented two levels of approximations for the
interference term with two resonant particles. 
The first approximation in Sect.\,\ref{sect:intMos} represents our main
result. It relies only on the on-shell evaluation of the matrix elements,
justified by a narrow resonance region, but no further assumptions are
implied. Different masses and finite widths are taken into account. This
version requires the explicit calculation of unsquared on-shell amplitudes,
preventing the use of e.g.\ convenient spinor trace rules. Furthermore, the
phase space integration depends on $\qsq$ so that the universal,
process-independent Breit-Wigner integral $I$ from Eq.\,(\ref{eq:defI}) does
not appear here.

The second approximation in Sect.\,\ref{sect:intR} has been 
formulated only at tree level so far.
It is based on the additional approximation, motivated by
the same argument as for the matrix elements, of setting the differential
Lorentz invariant phase spaces on-shell at either mass, scaled by a weighting
factor. This makes the $\qsq$-integration easier because only the universal
integral $I$ is left.
Furthermore, it avoids the unusual calculation of
on-shell amplitudes in an explicit representation by expressing the
interference part as an interference weight factor $R$ in terms of cross
sections, branching ratios, masses and widths, which are already needed in the
simple NWA, plus the universal integral $I$ and a scaling factor $x$ which
consists of the process-specific couplings.
Yet, this approximation holds only for equal masses. As discussed in the
context of Eq.\,(\ref{eq:overlap}), the interference term is largest if the
Breit-Wigner shapes overlap significantly due to the relation $\Delta M
\lesssim \Gamma_i$. Nevertheless, the masses are not necessarily equal in the
interference region. Instead, the overlap criterion in Eq.\,(\ref{eq:DMGamma})
can as well be satisfied if one of the widths is relatively large. In this
respect, the equal-mass condition is more restrictive than
the overlap criterion.
However, the equal-mass constraint is just applied on the matrix elements and
phase space, whereas different masses and widths are distinguished in the
Breit-Wigner integral. The $R$-factor method is technically easier to handle
because the constituents of $R$ can be obtained by standard routines in the
program packages such as \texttt{FormCalc}\,\cite{Hahn:1998yk, Hahn:1999wr,
Hahn:2000jm, Hahn:2006qw, Hahn:2006zy} and
\texttt{FeynHiggs}\,\cite{Heinemeyer:1998np, Heinemeyer:1998yj,
Degrassi:2002fi, Heinemeyer:2007aq} that we use in the numerical computation.
For one example process, this is done in Sect.\,\ref{sect:example}. An
extension of the generalised narrow-width approximation to the 1-loop level
is discussed in Sect.\,\ref{sect:IntNLO}.

\section{Particle content and mixing in the MSSM}\label{sect:MSSM}

Before we discuss the 
application of the gNWA to an example process within the MSSM, we briefly summarise 
here the different particle sectors of the MSSM 
in order to clarify the notation and conventions.

\subsection{Propagator mixing in the Higgs sector}

\paragraph{Higgs sector at tree level}
The MSSM contains two Higgs doublets,
\begin{align}
 \mathcal{H}_1 = \begin{pmatrix}H_{11}\\H_{12}\end{pmatrix}
                  =\bpm v_1+\frac{1}{\sqrt{2}}(\phi_1^{0}-i\chi_1^{0})\\ -\phi_1^{-}\epm, \hspace*{1cm}
 \mathcal{H}_2 = \begin{pmatrix}H_{21}\\H_{22}\end{pmatrix}
                  = \bpm \phi_2^{+}\\v_2+\frac{1}{\sqrt{2}}(\phi_2^{0}+i\chi_2^{0})\epm \label{eq:Hdoublets},
\end{align}
with the vacuum expectation values $v_1, v_2$, respectively, whose ratio
$\tan\beta\equiv\frac{v_2}{v_1}$ determines together with $M_{H^{\pm}}$ 
(or $M_A$)
the MSSM Higgs sector at tree level. In principle, complex parameters can enter through loops, but in this methodical study we consider the MSSM with real parameters. The neutral fields $\phi_1^{0},\phi_2^{0},\chi_1^{0},\chi_2^{0}$ are rotated into the mass eigenstates $h,H,A,G$, where $h$ and $H$ are neutral $\CP$-even Higgs bosons (rotated from $\phi_1^{0},\phi_2^{0}$ by the mixing angle $\alpha$), $A$ is the neutral $\CP$-odd Higgs boson and $G$ denotes the neutral Goldstone boson. Besides, there are the charged Higgs and Goldstone bosons, $H^{\pm}, G^{\pm}$. 

\paragraph{Mixing in the MSSM Higgs sector}\label{sect:Hmix}
Higher-order corrections have a crucial impact on the phenomenology in the
Higgs sector. We adopt the 
renormalisation scheme in the Higgs sector from
Refs.\,\cite{Frank:2006yh,Williams:2011bu}, where $M_A$ (or $M_{H^{\pm}}$) is
renormalised on-shell while a $\overline{\text{DR}}$-renormalisation is used
for the Higgs fields
and $\tan\beta$. For the prediction of the considered process we incorporate 
important higher-order corrections from the Higgs sector already at the Born
level by using for the Higgs-boson masses and total decay widths the
predictions from \texttt{FeynHiggs}~\cite{Heinemeyer:1998np, Heinemeyer:1998yj,
Degrassi:2002fi, Heinemeyer:2007aq}, which contain the full one-loop and
dominant two-loop contributions. 

Furthermore, because of the presence of off-diagonal self-energies like 
$\hat{\Sigma}_{ij}$ with $i,j=h, H, A$, the propagators of the neutral Higgs bosons mix with
each other and in general also with contributions from the gauge and
Goldstone bosons, see e.g.\ Refs.\,\cite{Frank:2006yh,Williams:2011bu}. The
Higgs-boson masses therefore have to be determined from the complex poles
of the Higgs propagator matrix. 

For correct on-shell properties of external Higgs bosons the residues of the
propagators have to be normalised to one. This is achieved by finite wave
function normalisation factors, which can be collected in a matrix
$\hat{\textbf{Z}}$, such that for a one-particle irreducible
vertex $\hat{\Gamma}_i$ with an external Higgs boson $i$ the effect of Higgs
mixing amounts to
\begin{align}
 \hat{\Gamma}_i &\rightarrow \hat{\textbf{Z}}_{ih}\hat{\Gamma}_h
+\hat{\textbf{Z}}_{iH}\hat{\Gamma}_H  +\hat{\textbf{Z}}_{iA}\hat{\Gamma}_A
+ \ldots , \label{eq:1PI}
\end{align}
where the ellipsis indicates the mixing with the Goldstone
and $Z$-bosons, which are not comprised in the $\hat{\textbf{Z}}$-factors, but
have to be calculated explicitly. 
The (non-unitary) matrix $\hat{\textbf{Z}}$ can be written as 
(see Ref.~\cite{Williams:2011bu})
\begin{align}
 \hat{\textbf{Z}} &= \bpm
\sqrt{\hat{Z}_h} & \sqrt{\hat{Z}_h}\hat{Z}_{hH} &\sqrt{\hat{Z}_h}\hat{Z}_{hA}\\
\sqrt{\hat{Z}_H}\hat{Z}_{Hh} & \sqrt{\hat{Z}_H} &\sqrt{\hat{Z}_H}\hat{Z}_{HA}\\
\sqrt{\hat{Z}_A}\hat{Z}_{Ah} & \sqrt{\hat{Z}_A}\hat{Z}_{AH} &\sqrt{\hat{Z}_A}
\epm,\label{eq:Zmatrix}
\end{align}
with
\begin{equation}
 \hat{Z}_{i} = \frac{1}{\frac{\partial}{\partial \psq}\frac{i}{\Delta_{ii}}}\bigg\vert_{\psq=M_{c_a}^{2}},
\hspace*{1.5cm}
 \hat{Z}_{ij} = \frac{\Delta_{ij}(\psq)}{\Delta_{ii}(\psq)}\bigg\rvert_{\psq=M_{c_a}^{2}}\label{eq:Zij},
\end{equation}
where the wave function normalisation factors are evaluated at the complex
poles 
\begin{align}
 M_{c_a}^{2} &= M_{h_a}^{2} - i M_{h_a}\Gamma_{h_a}\label{eq:HComplexPole}
\end{align}
for $a=1,2,3$. We choose $a=1$ for $i=h$, $a=2$ for $i=H$ and $a=3$ for $i=A$. $M_{h_a}$ is the loop-corrected mass and $\Gamma_{h_a}$ the total width of $h_a$. In the $\CP$-conserving case, the $\hat{\textbf{Z}}$-matrix is reduced to the $2\times2$ mixing between $h$ and $H$.

An amplitude involving resonant Higgs-boson propagators therefore needs to
incorporate in general the full loop-corrected propagator matrix
(and also the mixing contributions with the gauge and Goldstone bosons). It
can be shown~\cite{Fowler:2010eba,HiggsMix:InPrep} that in the vicinity of the
resonance the full propagator matrix contribution can be approximated by
\begin{equation}
 \sum_{i,j}\gh{i}^{\textrm{A}}\Delta_{ij}(\psq)\gh{j}^{\textrm{B}}
\simeq
\sum_{\alpha,i,j}\gh{i}^{\textrm{A}}\Zb_{\alpha i}\Delta^{\rm{BW}}_{\alpha}(\psq) \Zb_{\alpha j}\gh{j}^{\textrm{B}},\label{eq:ZBW}
\end{equation}
involving the Breit-Wigner propagator $\Delta_{\alpha}^{\text{BW}}(p^{2})$ 
as given in Eq.~(\ref{eq:BWdef}), where $\hat{\Gamma}_{i}^{A}, \hat{\Gamma}_{j}^{B}$ are the one-particle irreducible vertices $A,B$ of the Higgs bosons $h_i,\,h_j$, and $i,j,\alpha= h,H,A$ are summed over.

\subsection{The neutralino and chargino sector}\label{sect:neucha}
The superpartners of the neutral gauge and Higgs bosons mix into the four
neutralinos $\ci$, $i=1,2,3,4$, as mass eigenstates whose mass matrix is
determined by the bino, wino and higgsino mass parameters $M_1, M_2, \mu$ and
the parameter $\tan\beta$,
 \begin{align}
 M_{\tilde{\chi}^{0}}&= \left(\begin{matrix}
	M_1 & 0 & -M_Z \cb s_W & M_Z \sinb s_W\\
	0 & M_2 & M_Z \cb c_W & -M_Z \sinb c_W \\
	-M_Z \cb s_W & M_Z \cb c_W & 0 &-\mu\\
	M_Z \sinb s_W & -M_Z \sinb c_W &  -\mu &0\\
	     \end{matrix} \right)\label{eq:Y}.
\end{align}
The admixture of gauginos and higgsinos in each neutralino can be determined from the components of the matrix $N$ which diagonalises $M_{\tilde{\chi}^{0}}$ by $N^{*}M_{\tilde{\chi}^{0}}N^{-1}$.\\
The charginos $\tilde{\chi}_i^{\pm},~i=1,2$, as mass eigenstates are superpositions of the charged wino and higgsino,
\begin{equation}
 M_{\tilde{\chi}^{\pm}}
= \bpm M_2 & \sqrt{2}M_W\sib\\ \sqrt{2}M_W \cb & \mu \epm\label{eq:X}.
\end{equation}
The chosen example process requires the couplings at the Higgs-neutralino-neutralino and the Higgs-fermion-fermion vertices. For the neutralinos $\ci,\cj$ with $i,j=1,2,3,4$ and the neutral Higgs bosons $h_{k}=\{h,H,A,G\}$ for $k=1,2,3,4$, the right-handed $C^{R}$ and left-handed $C^{L}$ neutralino-Higgs couplings are given by
\begin{align}
 C_{R}^{ijk}&=-C_{L*}^{ijk} = \frac{ie}{2\cw\,\sw}c^{ijk}, ~\text{with}\label{eq:CijkGeneric}\\
c^{ijk} &=  \left\{\begin{matrix}
                    (-\sa N_{i3}  -\ca N_{i4})(\sw N_{j1}-\cw N_{j2}) + (i\leftrightarrow j),~&k&=1,\nonumber\\
                    (+\ca N_{i3}  -\sa N_{i4})(\sw N_{j1}-\cw N_{j2}) + (i\leftrightarrow j),~&k&=2,\nonumber\\
                    (+i\sib N_{i3}  -i\cb N_{i4})(\sw N_{j1}-\cw N_{j2}) + (i\leftrightarrow j),~&k&=3,\nonumber\\
                    (-i\cb N_{i3}  -i\sib N_{i4})(\sw N_{j1}-\cw N_{j2}) + (i\leftrightarrow j),~&k&=4,
                   \end{matrix} \right.
\end{align}
where $\sa\equiv \sin\alpha, \ca\equiv\cos\alpha$ and likewise for $\beta$. With the left-/right-handed projection operators $\omega_{R/L}\equiv\frac{1}{2}(1\pm\gamma^{5})$, the 3-point function of the neutralino-Higgs vertex is at tree level composed of
\begin{equation}
 \Gamma^{\text{tree}}_{\co,\chi_j,h_k} = \pr C^{ijk}_{R} \pm \pl C^{ijk}_{L},\label{eq:3point}
\end{equation}
where the + applies to the $\mathcal{CP}$-even Higgs bosons $h$ and H, whilst
the $-$ appears for the $\mathcal{CP}$-odd Higgs boson $A$ and the Goldstone boson $G$. 
Mixing of Higgs bosons is taken into account by a linear combination of
the couplings and Z-factors (see Sect.\,\ref{sect:Hmix}). In the
following, the couplings $C_{h_kXY}$ always mean the mixed couplings
(for $k,l,m = 1,2,3$; no summation over indices implied),
\begin{equation}
 C_{h_kXY} \rightarrow  \hat{\textbf{Z}}_{h_k h_k} C_{h_kXY} + \hat{\textbf{Z}}_{h_k h_l} C_{h_lXY} + \hat{\textbf{Z}}_{h_k h_m} C_{h_mXY}.\label{eq:Cmix}
\end{equation}
For the calculation of higher-order corrections to the neutralino-chargino
sector, it is essential to identify a stable renormalisation scheme according
to the gaugino parameter hierarchy of $M_1, M_2$ and $\mu$ as it was pointed
out in Refs.\,\cite{Fowler:2010eba,Chatterjee:2011wc,Bharucha:2012nx}. 
Choosing the external neutralinos and charginos in the
considered process to be on-shell does not necessarily lead to the most stable
renormalisation scheme. Among the four neutralinos and two charginos, three
can be renormalised on-shell in relation to the three gaugino parameters $M_1,
M_2$ and $\mu$. The most bino-, wino- and higgsino-like states should be
chosen on-shell so that the three parameters are sufficiently constrained by
the renormalisation conditions. Otherwise, an unstable choice of the input
states can lead to unphysically large values for the parameter counterterms
and mass corrections. On-shell 
renormalisation conditions were derived in
Refs.\,\cite{Lahanas:1993ib,Pierce:1993gj,Pierce:1994ew,Eberl:2001eu,Fritzsche:2002bi,Oller:2003ge,Drees:2006um,Chatterjee:2011wc}
for the MSSM with real parameters and in
Refs.\,\cite{Fowler:2009ay,Bharucha:2012nx,Bharucha:2012re} for the complex
case. Schemes with two charginos and one neutralino on-shell are referred to
as NCC, with one chargino and two neutralinos as NNC and with three
neutralinos input states as NNN. In view of our example process (see
Sect.\,\ref{sect:example}) and the gaugino hierarchy of the scenario, we will
comment on our choice of a renormalisation scheme in
Sect.\,(\ref{sect:VirtNeutralino}).

\subsection{The sfermion and the gluino sectors}
The mixing of sfermions $\tilde{f}_L, \tilde{f}_R$ within one generation into mass eigenstates $\tilde{f}_1, \tilde{f}_2$ is parametrised  by the matrix
\begin{align}
 M_{\tilde{f}}&=\left(\begin{matrix}
                         M_{\tilde{f}_L}^{2}+m_f^{2}+M_Z^{2}\cos 2 \beta (I_3^{f}-Q_f\sw^{2}) & m_f X_f^{*}\\
			 m_f X_f & M_{\tilde{f}_R}^{2}+m_{f}^{2}+M_Z^{2}\cos2\beta Q_f\sw^{2}
                        \end{matrix} \right),\label{eq:Msf}\\
 X_f &:= A_f - \mu^{*}\cdot \left\{ \begin{matrix}
                               \cot \beta,~f = \text{up-type}~~~~\\\tan \beta,~f = \text{down-type}.
                               \end{matrix} \right.
\end{align}
The trilinear couplings $A_f$ as well as $\mu$ can be complex. Their phases
enter the Higgs sector via sfermion loops, but as mentioned above
here we only take real parameters into account. The sfermion masses at tree level are the eigenvalues of $M_{\tilde{f}}$. In the considered example process with $h,H$ decaying into $\tau^{+}\tau^{-}$, the couplings of Higgs bosons to $\tau$-leptons are involved,
\begin{eqnarray}
C_{h\tau\tau}^{\rm{tree}} = +\frac{igm_{\tau}s_{\alpha}}{2M_W c_{\beta}},\hspace*{2cm}
C_{H\tau\tau}^{\rm{tree}} = -\frac{igm_{\tau}c_{\alpha}}{2M_W c_{\beta}} \label{eq:GHtt}.
\end{eqnarray}
The mass of the gluino $\tilde{g}$ is given by $|M_3|$.

\section{\texorpdfstring{Generalised narrow-width approximation at leading order: example process $\cf \rightarrow\co\,h/H\rightarrow \co\,\tau^{+}\tau^{-}$}{gNWA at LO: example process}}\label{sect:example}
\begin{figure}[ht!]
\centering
\subfigure[3-body decay]{\includegraphics[width=8cm]{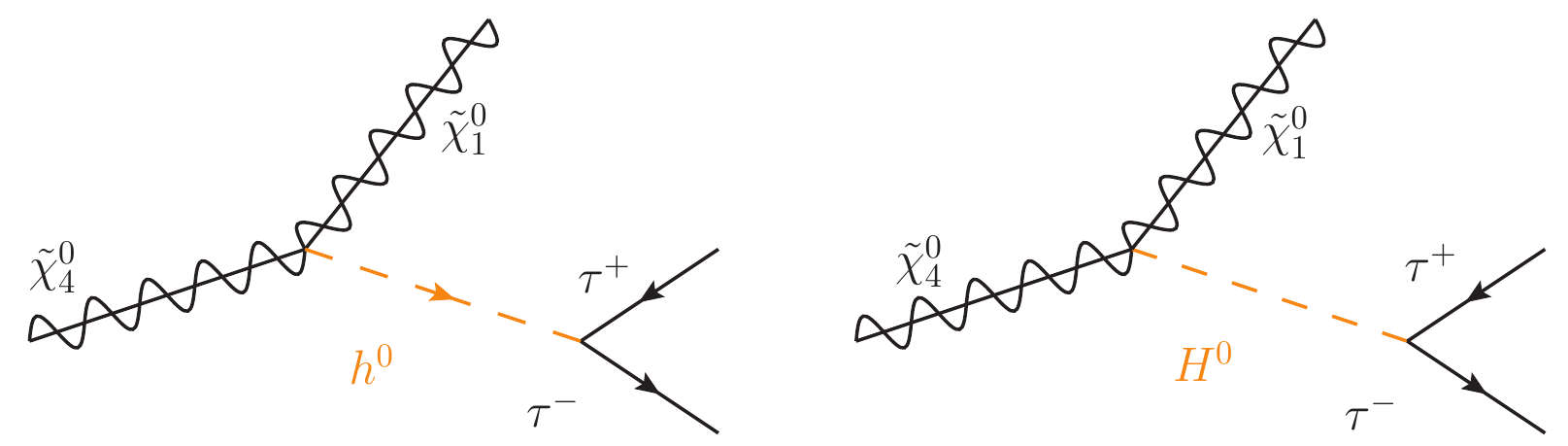}\label{fig:chi04_1to3}}\hspace*{0.2cm}
\subfigure[2-body decays]{\includegraphics[width=8cm]{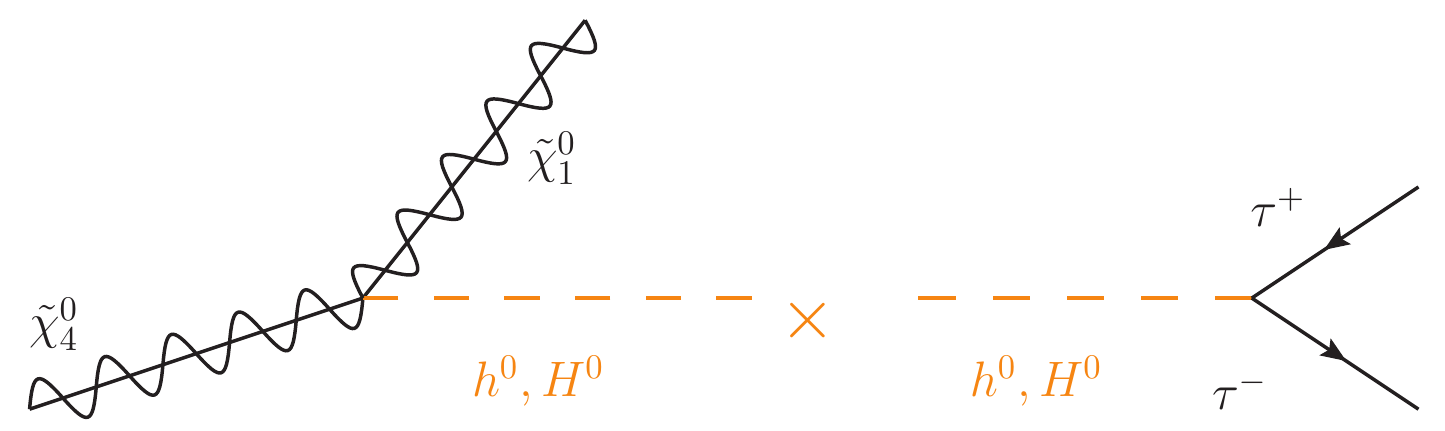}\label{fig:chi04_1to2}}
\caption[3-body decay of $\co$ split into 2-body decays]{$\cf \rightarrow \co \tau^+ \tau^-$ with $h$ or $H$ as intermediate particle in the two interfering diagrams. The decay process is either considered as \textbf{(a)} one 3-body decay or \textbf{(b)} decomposed in two 2-body decays.}
\label{fig:chi04decay}
 \end{figure}

The gNWA will be validated for a simple example process. The focus lies on providing a test case for the method rather than on the phenomenology of the process itself. For a comparison with the gNWA, we choose a process which can be calculated also at the 1-loop level without the on-shell approximation.\\
In the following, we will consider Higgs production from the decay of the heaviest neutralino and its subsequent decay into a pair of $\tau$-leptons, $\cf \rightarrow \co\,h/H\rightarrow \co\,\tau^{+}\tau^{-}$, which is a useful example process because it is computable as a full 3-body decay and it can be decomposed into two simple 2-body decays, see Fig.\,\ref{fig:chi04decay}. 
Moreover, the intermediate particles are scalars. 
Thus, for this process
the treatment of interference effects can be trivially disentangled from
any spin correlations between production and decay.
Due to the
neutralinos in the initial state and in the first decay step, soft
bremsstrahlung only appears in the final state, and there is no photon
exchange between the initial and final state. 
Restricting this test case to the MSSM with real parameters, only the two $\mathcal{CP}$-even
states $h, H$ mix due to $\mathcal{CP}$-conservation, instead of the
$3\times3$ mixing of $h, H, A$ in the complex case. We neglect
non-resonant diagrams from sleptons, 
which is a good approximation for the case of heavy sleptons.
Slepton contributions to
neutralino 3-body decays have been analysed in Ref.\,\cite{Drees:2006um}. As a
first step, we also neglect the exchange of an intermediate pseudoscalar $A$, Goldstone boson $G$ and $Z$-boson for 
the purpose of 
a pure comparison of the factorised and the full Higgs contribution.
For the most accurate prediction within the gNWA, which will be
discussed in Sect.\,\ref{sect:best}, 
we will add the tree-level
$A,G$- and $Z$-exchange, but they do not interfere with $h$ and $H$ in the case of real parameters.

The decay width
will be calculated using \texttt{FeynArts}\,\cite{Kublbeck:1990xc,
Denner:1992vza, Kublbeck:1992mt, Hahn:2000kx, Hahn:2001rv} and
\texttt{Form\-Calc}\,\cite{Hahn:1998yk, Hahn:1999wr,
Hahn:2000jm, Hahn:2006qw, Hahn:2006zy}\footnote{We used \texttt{FeynArts-3.7, FormCalc-7.4,
LoopTools-2.8} and \texttt{FeynHiggs-2.9.3}.} both as a 3-body decay
with the full matrix element and in the narrow-width approximation as a
combination of two 2-body decays - with and without the 
interference term. 
In this and the following section, the gNWA will be applied at the tree level. The application at the loop level will follow conceptually in Sect.\,\ref{sect:3bodyNLO} and numerically in Sect.\,\ref{sect:ResultLoop}.
\subsection{3-body decays: leading order matrix element}
In order to compare the gNWA to the unfactorised LO result, we calculate
the amplitude $\Mm_{h_k}$ of the 3-body decay via $h_k=h,H$. From
the matrix element of the form
\begin{align}
 \mathcal{M}_{h_k} &= i C_{h_k \ci \cj}C_{h_k \tau \tau}
\bar{u}(p_4,s_4)v(p_3, s_3) \frac{1}{q^2 - M_{h_k}^2 + iM_{h_k}
\Gamma_{h_k}}\bar{u}(p_2,s_2)u(p_1,s_1) \label{eq:amplMhk}
\end{align}
we obtain the spin-averaged, squared amplitude consisting of 
the separate $h, H$ contributions and the interference contribution,
\begin{align}
 \overline{|\mathcal{M}|^2} 
 &=8(p_1p_2 + m_{\tilde \chi_1^0}m_{\tilde \chi_4^0})(p_3p_4-m^2_{\tau})
 \left(\frac{|C_{h \co \cf}|^2|C_{h \tau \tau}|^2}{(q^2-m_h^2)^2 + m_h^2\Gamma_h^2}
	+\frac{|C_{H \co \cf}|^2 |C_{H \tau \tau}|^2}{(q^2-m_H^2)^2 + m_H^2\Gamma_H^2}\right.\nonumber \\
 &\hspace*{0.5cm}+\left. 2 \text{Re}\left[ C_{h \co \cf}C^*_{H \co \cf}C_{h \tau \tau}C^*_{H \tau \tau} \cdot \Delta_h^{\text{BW}}(q^2) \Delta_H^{*\text{BW}}(q^2) \right] \right) ,\label{eq:Msq_3body}
\end{align}
where the momenta and masses are labelled as $p_1\rightarrow p_2,p_3,p_4$ with $m_1 \equiv \mf, m_2\equiv \mo, m_3=m_4\equiv m_{\tau}$.
In order to calculate the decay width in one of the Gottfried-Jackson frames\,\cite{Gottfried:1964nx}, the products of momenta are rewritten in terms of two combined invariant masses, here e.g. $m_{23},m_{24}$:
\begin{eqnarray}
 p_1\cdot p_2 
	      &=& \frac{1}{2}(m_{23}^2+m_{24}^2)-m_{\tau}^{2},\hspace*{1cm}
 p_3\cdot p_4 
              = \frac{1}{2}\left(m_1^2 + m_2^2 -m_{23}^2 - m_{24}^2  \right)\nonumber,\\
q^2 &=& (p_1-p_2)^2 = m_1^2 + m_2^2 -m_{23}^2-m_{24}^2\label{eq:mij}\,.
\end{eqnarray}
This yields the partial 
decay width for the 3-body decay~\cite{Beringer:1900zz},
\begin{equation}
\Gamma =
\frac{1}{(2\pi)^3}\frac{1}{32m_{\cf}^3}\int|\mathcal{M}|^2dm_{23}dm_{24}\label{eq:dGamma3_2324}
\end{equation}
which we will use for a comparison with the gNWA.

\subsection{Decomposition into 2-body decays}\label{sect:dec2body}
In this section, we calculate the 2-body decay widths of the subprocesses needed in the NWA. The matrix element for the production of $h_k=h,H$ is
\begin{align}
 \mathcal{M}_{\cf \co h_k} &= i\bar{u}_2C_{h_k\cf \co}u_1,\\
 |\mathcal{M}_{\cf \co h_k}|^2 
 & = |C_{h_k\cf \co}|^2 2 (p_1p_2 +\mf \mo)\,.
\end{align}
In the rest frame of $\cf$ we have $p_1p_2=m_1 E_2$ with 
\begin{equation}
 E_2 = \frac{m_1^2 +m_2^2 -M_{h_k}^2}{2m_1}\label{eq:Eb}\,.
\end{equation}
Then the decay width of $\cf \rightarrow \co h_k$ for the production of $h_k=\left\lbrace h, H \right\rbrace$ equals
\begin{equation}
 \Gamma(\cf \rightarrow \co h_k) = \frac{|C_{h_k\cf \co}|^2}{16 \pi \mf^3}\left((\mf + \mo)^2 - M_{h_k}^2\right) \sqrt{(\mf^2-\mo^2-M_{h_k}^2)^2 - 4\mo^2M_{h_k}^2}\,. \label{eq:GammaPh}
\end{equation}
Summing over spins in the final states, the partial decay widths of $h$ and $H$ into a pair of $\tau$-leptons and the branching ratios are at tree level, improved by 2-loop Higgs masses and total widths from \texttt{FeynHiggs}\,\cite{Heinemeyer:1998np, Heinemeyer:1998yj,
Degrassi:2002fi, Heinemeyer:2007aq},
\begin{align}
\Gamma(h_k\rightarrow \tau\tau) &= \frac{1}{\pi}|C_{h_k\tau \tau}|^2 \frac{\left[\frac{M_{h_k}^2}{4}-m_{\tau}^2 \right]^{3/2}}{M_{h_k}^2},\hspace*{1cm}
\text{BR}_k = \frac{\Gamma(h_k\rightarrow \tau^+\tau^-)}{\Gamma_{h_k}^{\rm{tot}}}\label{eq:BR},
\end{align}
where $\Gamma_{h_k}^{\rm{tot}}$ is the total width. Loop-corrections to
the partial decay widths of these subprocesses are calculated with
\texttt{FormCalc}\,\cite{Hahn:1998yk, Hahn:1999wr,
Hahn:2000jm, Hahn:2006qw, Hahn:2006zy} in Sect.\,\ref{sect:ResultLoop_2body}.

\subsection{Unsquared matrix elements}\label{sect:Helicity}
For the calculation of the interference term according to Eq.\,(\ref{eq:masterformula}), we need the on-shell matrix elements of the production and decay part. Instead of evaluating absolute values of squared, spin-averaged matrix elements by applying spinor traces, we now aim at expressing the unsquared matrix elements explicitly in order to evaluate them on the appropriate mass shell. Therefore, we need to represent spin wave functions in terms of energy and mass. Following Ref.\,\cite{Haber:1994pe}, a Dirac spinor with an arbitrary helicity can be written as
\begin{align}
 u(p) &= \left(\begin{matrix} \sqrt{E+m}~\chi\\ \sqrt{E-m}~\vec{\sigma}\cdot\vec{p}~\chi
               \end{matrix} \right),
\end{align}
where $\chi$ is a two-component spinor. The eigenstates $\chi$ of the helicity operator $\vec{\sigma}\cdot\vec{p}$ with eigenvalues $\lambda=\pm\frac{1}{2}$ satisfy
\begin{align}
 \left[\frac{1}{2} \vec{\sigma}\cdot\vec{p}\right]\chi_{\lambda} =\lambda \chi_{\lambda}.\label{eq:helicityEigen}
\end{align}
For the unit vector $\hat{p}$ in the direction parametrised by the polar angle $\theta$ and azimuthal angle $\phi$ relative to the $z$-axis, the two-component spinors are expressed as
\begin{align}
\chi_{+1/2}(\hat{p}) = \left(\begin{matrix}
                                 \cos\frac{\theta}{2}\\ e^{i\phi}\sin\frac{\theta}{2} \end{matrix} \right),
\hspace*{2cm}\chi_{-1/2}(\hat{p}) = \left(\begin{matrix}
                               -e^{-i\phi}\sin\frac{\theta}{2}\\
				\cos\frac{\theta}{2}\end{matrix} \right).\label{eq:2spinor}
\end{align}
For the specific choice of $\vec{p}\propto\hat{e}_z$ we have $\theta=0$ and $\phi$ is arbitrary so that it can be set to $0$. Thus, the 2-spinors take the simpler form
\begin{equation}
 \chi_{1/2}(\hat{p}=e_z) = e_1\equiv\left(\begin{matrix}
                                        1\\0 \end{matrix} \right), \hspace{2cm}
\chi_{-1/2}(\hat{p}=e_z) = e_2\equiv\left(\begin{matrix}
                                        0\\1 \end{matrix} \right).\label{eq:2spinorPlus}
\end{equation}
We label the unit vectors in space as $\left\lbrace e_x, e_y, e_z\right\rbrace$ whereas the basis of the 2-spinors is denoted by $\left\lbrace e_1, e_2\right\rbrace$. The two-component spinors in the opposite momentum direction $-\hat{p}=-\hat{e}_z$ are constructed using
\begin{equation}
 \chi_{-\lambda}(-\hat{p}) = \xi_{\lambda}\chi_{\lambda}(\hat{p}) \label{eq:chiMinus}
\end{equation}
from Ref.\,\cite{Haber:1994pe} with $\xi_{\lambda}=1$ in the Jacob-Wick convention for a second particle spinor \cite{Jacob:1959at}, 
resulting in 
\begin{align}
  \chi_{+1/2}(-e_z) =e_2, \hspace{4cm}
\chi_{-1/2}(-e_z) =e_1.
\end{align}
Defining $\epsilon_+ := \sqrt{E+m}$ and $\epsilon_- := \sqrt{E-m}$ for a simpler notation, we can rewrite the particle and antiparticle four-component spinors as
\begin{align}
 u_{\lambda}(p) = \bpm \epsilon_+ \chi_{\lambda}(\hat{p})\\
		      2\lambda\,\epsilon_- \chi_{\lambda}(\hat{p}) \epm
		 = \bpm \rho^{\lambda} \\ \psi^{\lambda}\epm,
\hspace*{2cm}
 v_{\lambda}(p) = \bpm \epsilon_- \chi_{-\lambda}(\hat{p})\\
		      -2\lambda\,\epsilon_+ \chi_{-\lambda}(\hat{p}) \epm
		  = \bpm \sigma^{\lambda} \\ \varphi^{\lambda}\epm
\label{eq:uv}.
\end{align}
Here we introduced the nomenclature $\rho/ \psi$ for the upper/lower 2-spinor within a particle 4-spinor $u$ and likewise $\sigma/ \varphi$ for an antiparticle $v$. For later use, we now list the combinations of $\lambda=\pm\frac{1}{2}$ and $\hat{p}\pm e_z$ explicitly:
\begin{align}
&u_+(e_z)=\bpm~~\epsp e_1\\ ~~\epsm e_1 \epm,
&&u_-(e_z)=\bpm~~\epsp e_2\\ -\epsm e_2 \epm,
&&&u_+(-e_z)=\bpm~~\epsp e_2\\ ~~\epsm e_2 \epm,
&&&&u_-(-e_z)=\bpm~~\epsp e_1\\ -\epsm e_1 \epm,\nonumber\\
&v_+(e_z)=\bpm~~\epsm e_2\\ -\epsp e_2 \epm,
&&v_-(e_z)=\bpm~~\epsm e_1\\ ~~\epsp e_1 \epm,
&&&v_+(-e_z)=\bpm~~\epsm e_1\\ -\epsp e_1 \epm,
&&&&v_-(-e_z)=\bpm~~\epsm e_2\\ ~~\epsp e_2 \epm.\label{eq:uvAllExpl}
\end{align}
In the following, we will apply this formalism to Higgs production and decay within our example process.

\paragraph{Higgs production}
As illustrated in Fig.\,\ref{fig:chi04_1to2}, the incoming spinor $u_1$ (in the example case $\cf$) decays into $u_2$ ($\co$) and a scalar ($h/H$). The matrix element $\Pp$ of this production process is decomposed into a right- and left-handed part,
\begin{align}
 \Pp = \bar{u}_2 C_R\omega_R u_1 + \bar{u}_2 C_L\omega_L u_1 \label{eq:PLR},
\end{align}
where $C_{R/L}$ are form factors.
Using $\gamma^{0}, \gamma^{5}$ in the Dirac representation, and the 2-spinor notation introduced in Eq.\,(\ref{eq:uv}),
we calculate the spinor chains with arbitrary helicity of $\lambda_1, \lambda_2=\pm\frac{1}{2}$,
\begin{align}
 p_R &:= \bar{u}_2\omega_R u_1= \frac{1}{2}(\rho_2-\psi_2)(\rho_1+\psi_1),\nonumber\\
 p_L &:= \bar{u}_2\omega_L u_1= \frac{1}{2}(\rho_2+\psi_2)(\rho_1-\psi_1)\label{eq:pLR}.
\end{align}
Given the 2-body decay in the rest frame of particle 1, it follows that $E_1=m_1$ and consequently $\epsm=0,~\psi_1=0$. In order to obtain the helicity matrix elements $p_{R/L}^{\lambda_2\lambda_1}$, we insert the explicit spinors from Eq.\,(\ref{eq:uvAllExpl}) into the generic Eq.\,(\ref{eq:pLR}):
\begin{align}
p_R^{++} &= \bar{u}_{2+} \omega_R u_{1+}
	 = \frac{1}{2}(\epsilon_{2+}-\epsilon_{2-})\epsilon_{1+}~e_1\cdot e_1\nonumber\\
	&= \frac{1}{2}\left(\sqrt{E_2+m_2}-\sqrt{E_2-m_2}\right)\sqrt{2m_1},\nonumber\\
p_L^{++} &= \frac{1}{2}\left(\sqrt{E_2+m_2}+\sqrt{E_2-m_2}\right)\sqrt{2m_1},\nonumber\\
p_R^{--} &= p_L^{++},~~
p_L^{--} = p_R^{++},\nonumber\\
p_{R/L}^{+-} &= p_{R/L}^{-+} \propto e_1\cdot e_2 \equiv 0 \label{eq:pLRpm}.
\end{align}
Since the helicity matrix elements are real, their complex conjugates 
$p_{R/L}^{*}=\bar{u}_1\omega_{L/R}u_2$ are equal to the results in Eq.\,(\ref{eq:pLRpm}). The products of matrix elements are summed over all helicity combinations (but no averaging is done yet),
 with $i,j\in\left\lbrace R,L\right\rbrace$, leading to\footnote{These helicity matrix elements correspond to the \texttt{FormCalc-HelicityME}s via $A_{ij} = 4\cdot\textrm{MatF}(i,j)$. The factor of $4$ arises because the \texttt{FormCalc} expressions are multiplied later on by $2$ for each external fermion.}
\begin{align}
 A_{ij} &:= \sum_{\lambda_1,\lambda_2=\pm1/2} p_i\,p_j^{*}\label{eq:Aij},\\
 A_{RR} &= A_{RR}^{++} + A_{RR}^{--} = 2m_1E_2 = m_1^{2}+m_2^{2}-M^{2}, \nonumber\\
 A_{LL} &= A_{LL}^{++} + A_{LL}^{--} = A_{RR},\nonumber\\
 A_{RL} &= A_{RL}^{++} + A_{RL}^{--} = 2m_1m_2, \nonumber\\
 A_{LR} &= A_{LR}^{++} + A_{LR}^{--} = A_{RL} \label{eq:ARL},
\end{align}
where the energy relation of a 2-body decay with $m_1\rightarrow \left\lbrace m_2, M\right\rbrace $ was applied:
\begin{equation}
 E_2 = \frac{m_1^{2}+m_2^{2}-M^{2}}{2m_1} \label{eq:E2body}.
\end{equation}
Finally, the squared production matrix element is constructed as
\begin{align}
 \Pp \Pp^{*}&=\sum_{i,j=R,L}C_i C_j^{*} A_{ij} \nonumber\\
      &= (|C_R|^{2}+|C_L|^{2})(m_1^{2}+m_2^{2}-M^{2})+(C_R C_L^{*}+C_L C_R^{*})\,2m_1 m_2 \label{eq:PpPp}.
\end{align}
If the left- and right-handed form factors coincide ($C_L=C_R\equiv C$), Eq.\,(\ref{eq:PpPp}) is reduced to
\begin{align}
 \left(\Pp \Pp^{*}\right)_C &= 2|C|^{2}\left((m_1+m_2)^{2}-M^{2}\right).\label{eq:PpPpC}
\end{align}
However, in the interference term we need the product $\Pp_h \Pp_H^{*}$ with different Higgs masses in $E_2$ from Eq.\,(\ref{eq:E2body}). This distinction leads to
\begin{align}
 A_{ij} &= \sum_{\lambda_1,\lambda_2=\pm1/2}p_i^{h}p_j^{H*},\label{eq:AijhH}\\
 A_{RR} &= A_{LL} = m_1\left(\epsilon_{2+}^{h}\,\epsilon_{2+}^{H}+\epsilon_{2-}^{h}\epsilon_{2-}^{H}\right),\label{eq:ARRLLhH}\\
 A_{RL} &= A_{LR} = m_1\left(\epsilon_{2+}^{h}\,\epsilon_{2+}^{H}-\epsilon_{2-}^{h}\epsilon_{2-}^{H}\right)\label{eq:ARLhH}.
\end{align}
As before, we give the resulting product of matrix elements for the independent $C_{R/L}$ and for simpler use in the special case of $C_{R/L}\equiv C$,
\begin{align}
 \Pp_h\Pp_H^{*} = &(C_R^{h}C_R^{H*}+C_L^{h}C_L^{H*})m_1\left(\epsilon_{2+}^{h}\,\epsilon_{2+}^{H}+\epsilon_{2-}^{h}\epsilon_{2-}^{H}\right) + (C_R^{h} C_L^{H*}+C_L^{h} C_R^{H*})m_1\left(\epsilon_{2+}^{h}\,\epsilon_{2+}^{H}-\epsilon_{2-}^{h}\epsilon_{2-}^{H}\right)\label{eq:PpPphH}\\
 \stackrel{C}{\longrightarrow} &4C^{h}C^{H*} m_1 \epsilon_{2+}^{h}\epsilon_{2+}^{H}  = 2C^{h}C^{H*}\sqrt{(m_1+m_2)^{2}-M_h^{2}}\,\sqrt{(m_1+m_2)^{2}-M_H^{2}}\label{eq:PphHC}.
\end{align}
Eq.\,(\ref{eq:PpPphH}) shows that the method of on-shell matrix elements enables us to distinguish between different masses of the intermediate particles, in this example $M_h$ and $M_H$.

\paragraph{Higgs decay} In the decay of a Higgs boson into a pair of fermions, the representation of antiparticle spinors from Eq.\,(\ref{eq:uvAllExpl}) is also needed. Furthermore, the fermions are generated back to back in the rest frame of the decaying Higgs boson. So if we align the momentum direction of the particle spinor $u_4$ with the $z$-axis, $\hat{p}_4=e_z$, the momentum of the antiparticle spinor $v_3$ points into the direction of $\hat{p}_3=-e_z$.

Analogously to Eq.\,(\ref{eq:PLR}), the decay matrix element is in general composed of a left- and right-handed part,
\begin{align}
 \Dd &= \bar{u}_4 C_R\omega_R v_3 + \bar{u}_4 C_L\omega_L v_3 \label{eq:DLR},\\
 d_R &:= \bar{u}_4(e_z) C_R\omega_R v_3(-e_z) = \frac{1}{2}(\rho_4-\psi_4)(\sigma_3+\varphi_3)\label{eq:dR},\\
 d_L &:= \bar{u}_4(e_z) C_L\omega_R v_3(-e_z) = \frac{1}{2}(\rho_4+\psi_4)(\sigma_3-\varphi_3)\label{eq:dL}.
\end{align}
With the mass $M$ of the decaying Higgs boson, the fermion masses $m_3=m_4\equiv m$ and the resulting $E_3=E_4\equiv \frac{M}{2}$, the spinor chains $d_R, d_L$ are now calculated for all helicity configuration of $\lambda_3, \lambda_4 =\pm\frac{1}{2}$,
\begin{align}
 d_R^{++} &= d_L^{--} = \sqrt{E^{2}-m^{2}}-E, \nonumber\\
 d_L^{++} &= d_R^{--} = \sqrt{E^{2}-m^{2}}+E, \hspace*{1cm}
 d_{R/L}^{+-} = d_{R/L}^{-+} =0\label{eq:dRL}.
\end{align}
Summing over all helicity combinations, we obtain
\begin{align}
 A_{RR} = A_{LL} = M^{2}-2m^{2}, \hspace{2cm}
 A_{RL} = A_{LR} = -2m^{2}\label{eq:ARLdec}.
\end{align}
So the product of on-shell decay matrix elements results in
\begin{equation}
 \Dd\Dd^{*}= \left(|C_R|^{2}+|C_L|^{2}\right)(M^{2}-2m^{2})-\left(C_RC_L^{*}+C_LC_R^{*}\right)2m^{2} \label{eq:Dd}.
\end{equation}
In case of identical left- and right-handed couplings $C$ of the decay vertex, Eq.\,(\ref{eq:Dd}) simplifies to
\begin{equation}
 \Dd\Dd^{*} = 4|C|^{2} (M^{2}-4m^{2}).\label{eq:DdC}
\end{equation}
As in the production case, we are interested in the contribution to the on-shell interference term, so we distinguish between $E_h= \frac{M_h}{2}$ and $E_H= \frac{M_H}{2}$,
\begin{align}
 A_{RR} &= A_{LL} =2 \left(\sqrt{(E_h^{2}-m^{2})(E_H^{2}-m^{2})} +E_hE_H\right),\nonumber\\
 A_{RL} &= A_{LR} =2 \left(\sqrt{(E_h^{2}-m^{2})(E_H^{2}-m^{2})} -E_hE_H\right)\label{eq:ALRhHdec}.
\end{align}
Finally, the product of decay matrix elements with different masses reads
\begin{align}
 \Dd_h\Dd_H^{*} &= 2\left(C_R^{h}C_R^{H*}+C_L^{h}C_L^{H*}\right) \left(\sqrt{(E_h^{2}-m^{2})(E_H^{2}-m^{2})} +E_hE_H\right) \nonumber\\
  &+ 2\left(C_R^{h}C_L^{H*}+C_L^{h}C_R^{H*}\right) \left(\sqrt{(E_h^{2}-m^{2})(E_H^{2}-m^{2})} -E_hE_H\right)\label{eq:DdhH}\\
 &\stackrel{C}{\longrightarrow}8C^{h}C^{H*}\sqrt{\left(\frac{M_h^{2}}{4}-m^{2}\right)\left(\frac{M_H^{2}}{4}-m^{2}\right)},
\end{align}
where the last line applies for identical L/R form factors.

The outcome of the explicit spinor representations in the context of factorising a longer process into production and decay is the possibility to express the interference term with on-shell matrix elements depending on the mass of the intermediate particle. The method was here introduced in a generic way and then applied to the example of Higgs production and decay with two external fermions in each subprocess in the rest frames of the decaying particles.

\section{Numerical evaluation at lowest order}
\label{sect:numtree}
\subsection{Modified \texorpdfstring{$M_h^{\rm{max}}$}{Mhmax} scenario}\label{sect:Mhmax}
In order to apply the gNWA on the example process of $\cf \rightarrow \co\,h/H\rightarrow \co\,\tau^{+}\tau^{-}$ numerically, we specify a scenario. In this study, we restrict the MSSM parameters to be real so that there is no new source of $\mathcal{CP}$-violation compared to the SM and only the two $\mathcal{CP}$-even neutral Higgs bosons, $h$ and $H$, mix and interfere with each other. The aim here is not to determine the parameters which are currently preferred by recent limits from experiments, but to provide a setting in which interference effects between $h$ and $H$ become large in order to investigate the performance of the generalised narrow-width approximation for this simple example process.

\begin{figure}[ht!]
 \begin{center}
  \subfigure[Higgs masses.]{\includegraphics[width=8.4cm]{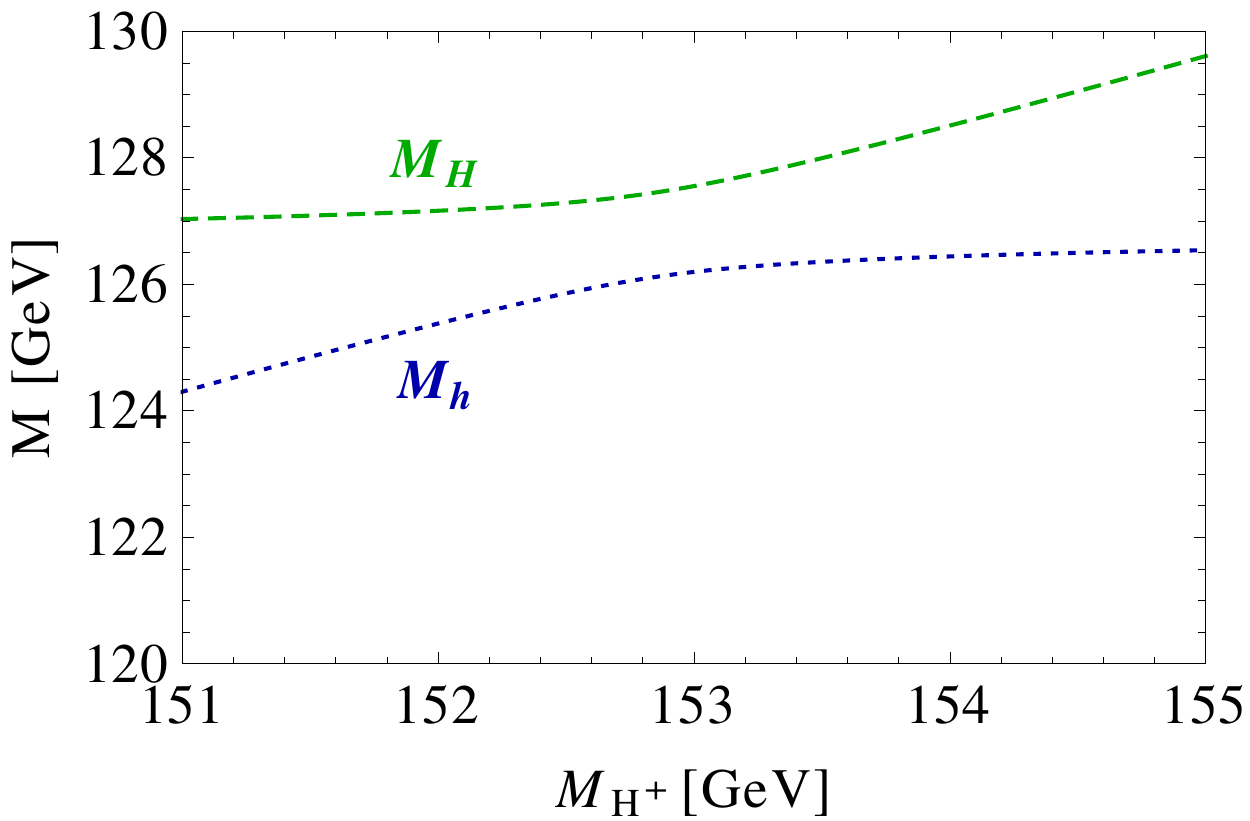}\label{fig:MhH}}
  \subfigure[Mass difference and total widths.]{\includegraphics[width=8.15cm]{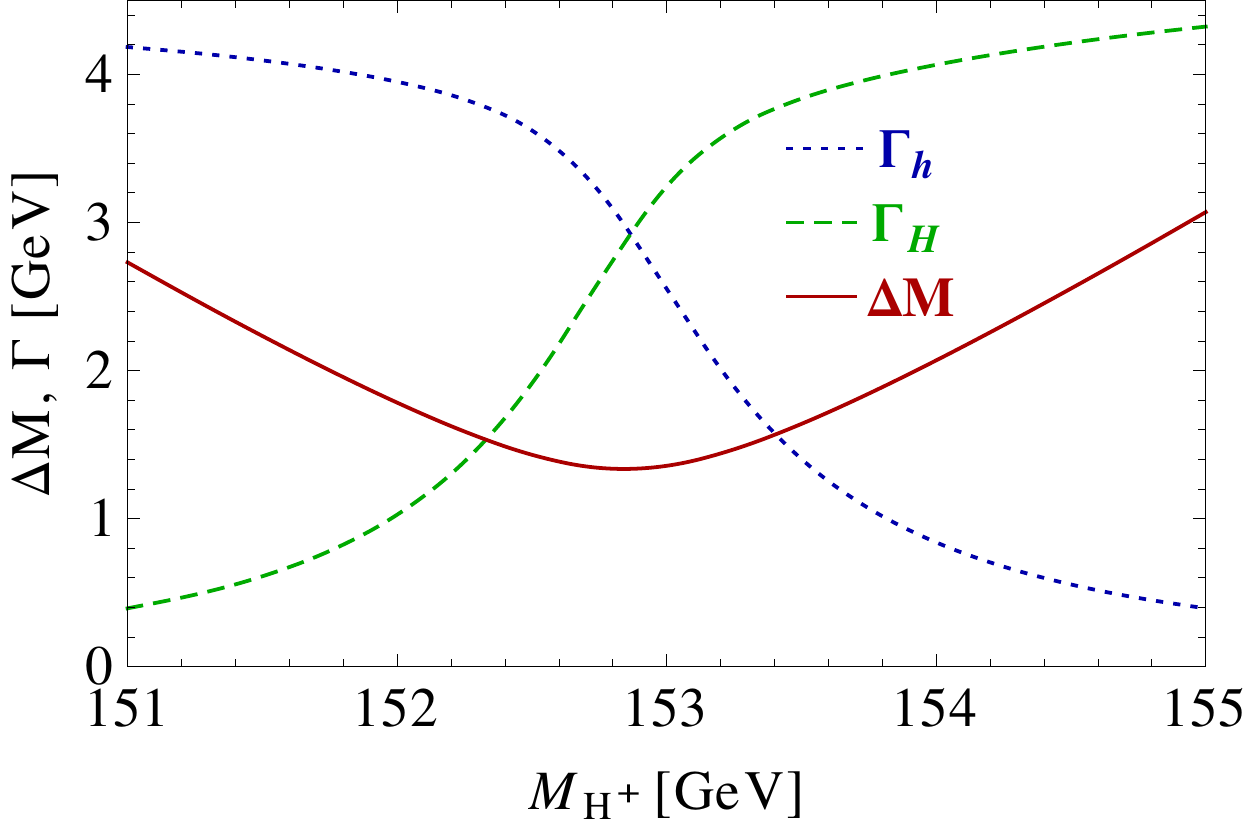}\label{fig:GM}\label{fig:DMGamma}}\\
  \subfigure[Ratio of mass difference and total widths.]{\includegraphics[width=8.3cm]{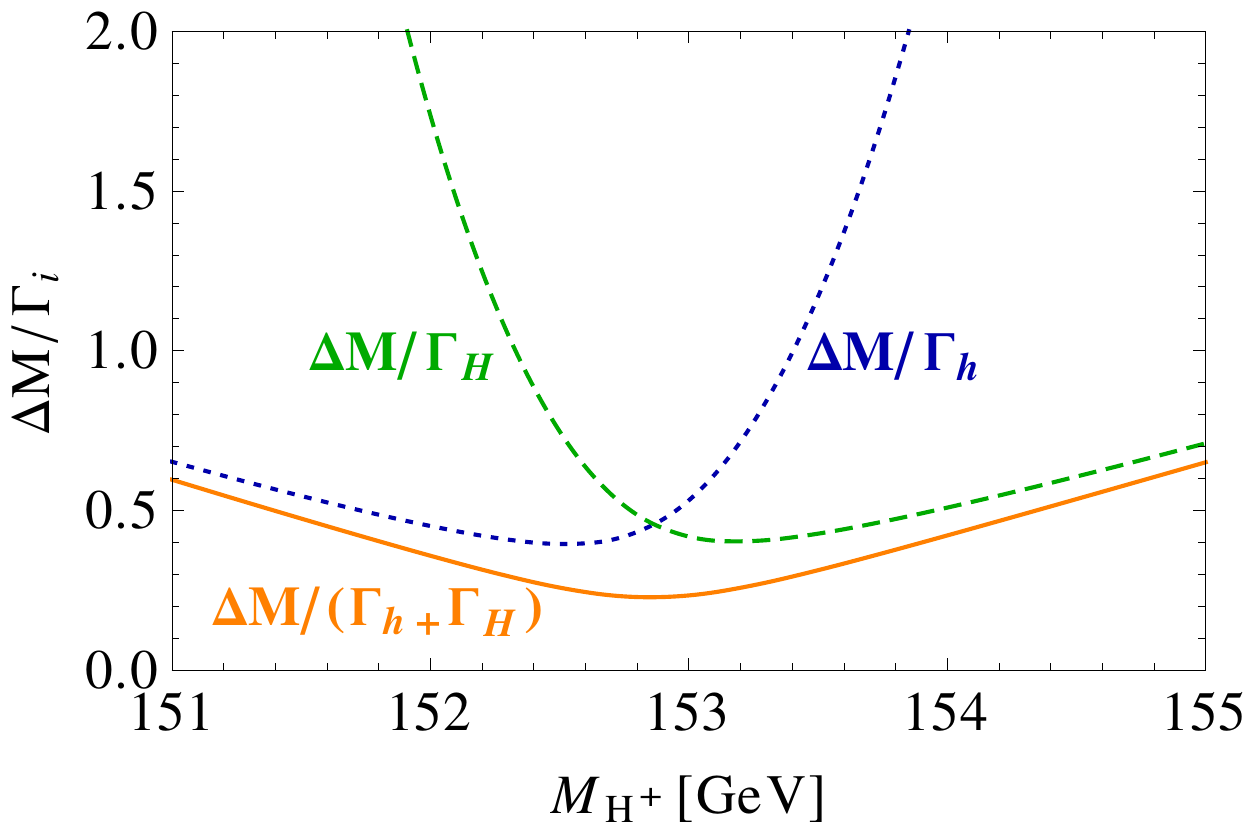}\label{fig:dmg}}
  \subfigure[Ratio of total widths and masses.]{\includegraphics[width=8.3cm]{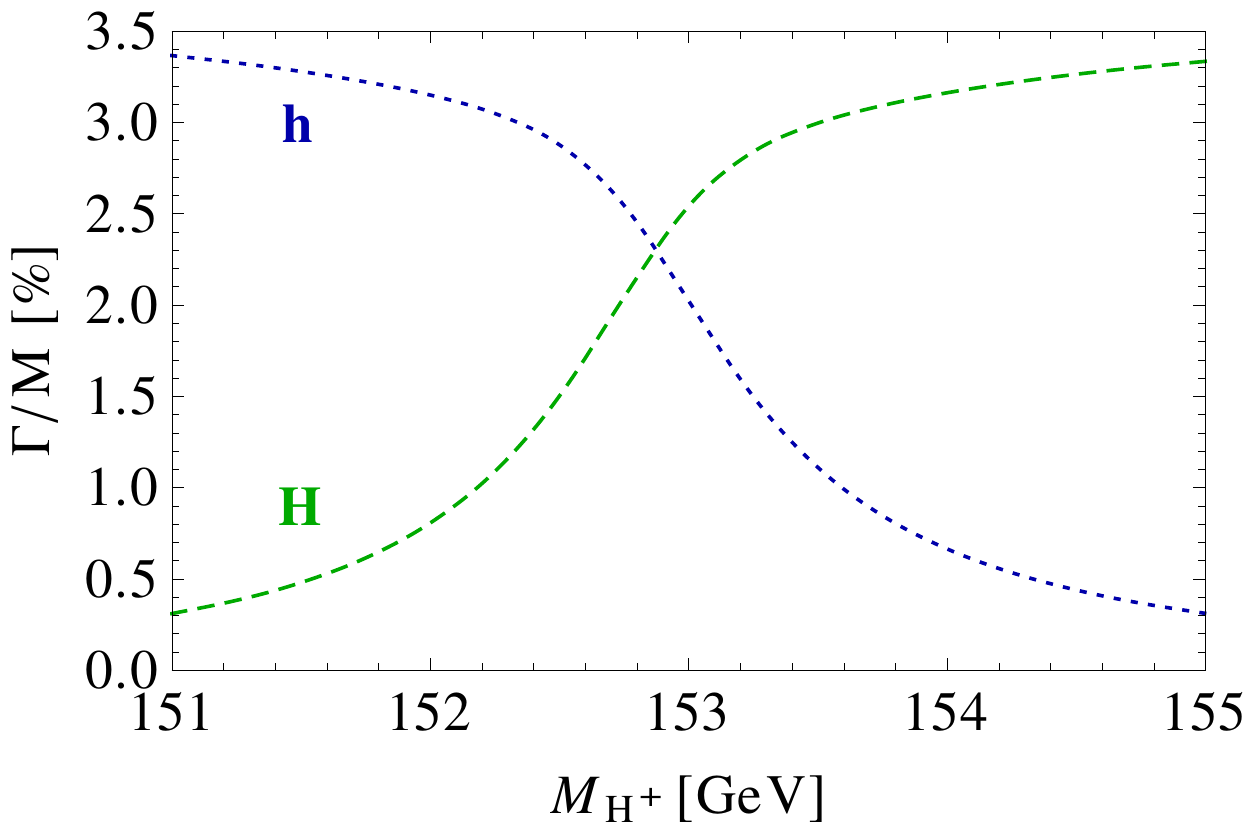}\label{fig:GoM}}
\caption{Higgs masses and widths from \texttt{FeynHiggs}\,\cite{Heinemeyer:1998np, Heinemeyer:1998yj,
Degrassi:2002fi, Heinemeyer:2007aq} including dominant 2-loop corrections in the modified $M_{h}^{\rm{max}}$ scenario.\textbf{(a):} Higgs masses $M_h$ (blue, dotted) and $M_H$ (green, dashed). \textbf{(b):} Mass difference $\Delta M\equiv M_H-M_h$ (red) compared to total widths $\Gamma_h$ (blue, dotted) and $\Gamma_H$ (green, dashed). \textbf{(c):} Mass difference $\Delta M$ divided by total width of $h$ (blue, dotted), $H$ (green, dashed) and sum of both widths (orange). \textbf{(d):} Ratio $\Gamma_i/M_i$ for $h$ (blue, dotted) and $H$ (green, dashed).}
\label{fig:GammaDeltaM}
 \end{center}
\end{figure}

The $M_h^{\rm{max}}$ scenario\,\cite{Carena:1999xa,Carena:2002qg} is defined
such that the loop corrections to the mass $M_h$ reach their maximum for
fixed $\tan\beta,~M_A$ and $M_{\rm{SUSY}}$. This requires a large stop
mixing, i.e.\ a large off-diagonal element $X_t$ of the stop mixing
matrix in Eq.\,(\ref{eq:Msf}). A small mass difference $\Delta M \equiv
M_H-M_h$ requires a rather low value of $M_A$, or equivalently
$M_{H^{\pm}}$, and a high value of $\tan\beta$. On the other hand, $\tan
\beta$ must not be chosen too large because otherwise the bottom Yukawa
coupling would be enhanced to an non-perturbative value.
We modify the $M_h^{\rm{max}}$ scenario such that $M_h$ is not maximised, but
the mass difference $\Delta M$ is reduced by raising $X_t$. As one of
the Higgs sector input parameters, we choose $M_H^{\pm}$ for a later
extension to $\mathcal{CP}$-violating mixings instead of $M_A$, which is
more commonly used in the MSSM with real parameters. The charged Higgs
mass is scanned over the range $\mhp\in$[151\,GeV, 155\,GeV]. The other
parameters are defined in Tab.\,\ref{tab:scenario}, and we assume
universal trilinear couplings $A_f=A_t$.
\begin{table}[ht!]
 \begin{center}
\caption[Parameters in the numerical analysis]{Parameter settings of the modified $M_h^{\rm{max}}$ scenario in the numerical analysis. A value in brackets indicates that the parameter is varied around this central value.}
  \begin{tabular}{|c|c|c|c|c|c|c|c|}
   \hline
$M_1$& $M_2$& $M_3$& $M_{\rm{SUSY}}$& $X_t$ & $\mu$&$\tb$& $M_{H^{\pm}}$ \\ \hline
100\,GeV& 200\,GeV & 800\,GeV & 1\,TeV & 2.5\,TeV & 200\,GeV& 50& (153\,GeV) \\ \hline
  \end{tabular}
\label{tab:scenario}
 \end{center}
\end{table}\\
Under variation of the input Higgs mass $\mhp$, the resulting masses and
widths of the interfering neutral Higgs bosons $h, H$ change as shown in
Fig.\,\ref{fig:GammaDeltaM} with results from \texttt{FeynHiggs}\,\cite{Heinemeyer:1998np, Heinemeyer:1998yj,
Degrassi:2002fi, Heinemeyer:2007aq}
including dominant 2-loop corrections. 
Fig.\,\ref{fig:MhH} displays the
dependence of the masses of $h$ (blue, dotted) and $H$ (green, dashed)
on $\mhp$. Within the analysed parameter range of $\mhp=
151...155\,$GeV, their mass difference $\Delta M$ (red) in
Fig.\,\ref{fig:DMGamma} is around its minimum at $\mhp\simeq 153\,$GeV
below both total widths $\Gamma_h$ (blue, dotted) and $\Gamma_H$ (green,
dashed). While $\Gamma_h$ decreases, $\Gamma_H$ increases with
increasing $\mhp$. This is caused by a change of the predominantly
diagonal or off-diagonal structure of the $\Zbf$-matrix which has a
cross-over around $\mhp\simeq 153$\,GeV in this scenario. Since both
widths contribute to the overlap of the two resonances, the ratio
$R_{M\Gamma}=\Delta M/(\Gamma_h+\Gamma_H)$ gives a good 
indication of the parameter region of most significant interference. This is visualised (in orange) in Fig.\,\ref{fig:dmg} and compared to the ratios $\Delta M/\Gamma_h$ (blue, dotted) and $\Delta M/\Gamma_H$ (green, dashed), which only take one of the 
widths into account and are therefore a less suitable criterion for the
importance of the interference term. Fig.\,\ref{fig:GoM} presents the
ratio $\Gamma_i/M_i$ for $i=h$ (blue, dotted) and $H$ (green, dashed) as
a criterion for a \textit{narrow} width. Both ratios lie in the range of
about $0.5\%$ to $3.5\%$, and this represents the expected order of the
NWA uncertainty.

\subsection{Results for tree level process \texorpdfstring{$\cf \rightarrow\co\,h/H\rightarrow \co\,\tau^{+}\tau^{-}$}{}}\label{sect:ResultTree}
In order to understand the possible impact of interference terms, we
confront the prediction of the standard NWA (sNWA) with the 3-body decay
width of our example process $\cf\rightarrow\co\tau^{+}\tau^{-}$ at the
tree level (improved by 2-loop predictions for the masses, widths and
$\hat{\textbf{Z}}$-factors) in the modified $M_h^{\rm{max}}$ scenario.

\begin{figure}[ht!]
 \begin{center}
  \includegraphics[width=12.5cm]{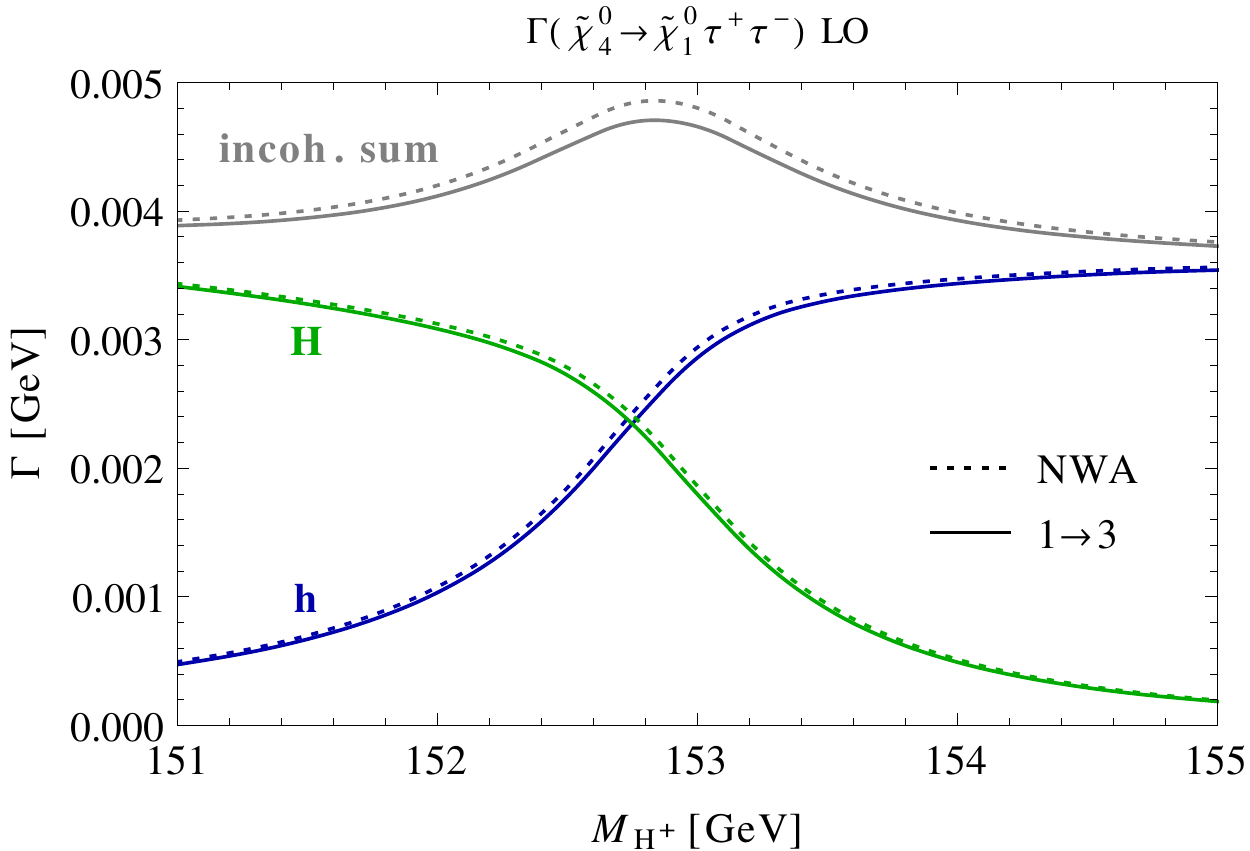}
\caption{The 1$\rightarrow$3 decay width (solid) of  $\cf \rightarrow \co \tau^{+}\tau^{-}$ at tree level with separate contributions from $h$ (blue), $H$ (green) and their incoherent sum (grey) confronted with the sNWA (dotted).}
\label{fig:sNWA_intrinsic}
 \end{center}
\end{figure}

First of all, we verify that the other conditions from
Sect.\,\ref{sect:cond} for the NWA are met. The widths of the involved
Higgs bosons do not exceed $3.5\%$ of their masses, hence they can be
considered \textit{narrow} (see Fig.\,\ref{fig:GoM}). At tree level,
there are no unfactorisable contributions so that the scalar propagator
is separable from the matrix elements. Besides, our scenario is far away
from the production and decay thresholds since $M_{h_k}\gg 2m_{\tau}$
holds independently of the parameters, and with neutralino masses of
$m_{\cf} \simeq 264.9\,$GeV and $m_{\co}\simeq 92.6\,$GeV, also 
$m_{\cf} - (m_{\co}+M_{h_k})> 32\,$GeV does not violate the threshold condition.
The neutralino masses are independent of $\mhp$. 
Thus, the NWA is applicable for the individual contributions of $h$ and $H$, so the factorised versions
\begin{equation}
\Gamma^{i}_{NWA}:= \Gamma_{P_i}(\cf\rightarrow \co h_i)\,\text{BR}_i(h_i\rightarrow \tau^{+}\tau^{-}) \label{eq:NWAi}
\end{equation} 
should agree with the separate terms of the 3-body decays via the exchange of only one of the Higgs bosons, $h_i$,
\begin{equation}
\Gamma^{i}_{1\rightarrow3}:=\Gamma(\cf \stackrel{h_i}{\rightarrow}\co \tau^{+}\tau^{-})
\end{equation}
within the uncertainty of $\mathcal{O}\left(\frac{\Gamma_{h_i}}{M_{h_i}}\right)$. This is tested in Fig.\,\ref{fig:sNWA_intrinsic}. The blue lines compare $\Gamma^{h}_{1\rightarrow3}$ (solid) with the factorised process $\Gamma^{h}_{NWA}$ (dotted), the green lines represent the corresponding expressions for $H$. The standard narrow-width approximation (sNWA) is composed of the \textit{incoherent} sum of both factorised processes, i.e., 
\begin{equation}
 \Gamma_{sNWA}= \Gamma_{P_h}\,\text{BR}_h + \Gamma_{P_H}\,\text{BR}_H\label{eq:sNWA}.
\end{equation}
This is confronted with the incoherent sum of the 3-body decays which are only $h$-mediated or $H$-mediated. For a direct comparison with the sNWA, the interference term is not included,
\begin{equation}
\Gamma^{incoh}_{1\rightarrow3} = \Gamma^{h}_{1\rightarrow3}+\Gamma^{H}_{1\rightarrow3}\label{eq:incoh}.
\end{equation}
The sNWA (dotted) and the incoherent sum of the 3-body decay widths are both shown in grey. Their relative deviation of $0.8-3.3\%$ is of the order of the ratio $\Gamma/M$ from Fig.\,\ref{fig:GoM}. Consequently, the NWA is applicable to the terms of the separate $h/H$-exchange within the expected uncertainty.

However, the fifth condition in Sect.\,\ref{sect:cond} concerns the absence of
a large interference with other diagrams. But with $\Delta
M<\Gamma_h+\Gamma_H$ throughout the analysed parameter range (see
Fig.\,\ref{fig:dmg}), we expect a sizeable interference effect in this
scenario owing to a considerable overlap of the Breit-Wigner propagators and a sizeable mixing between $h$ and $H$.
Since the masses and widths of the interfering Higgs bosons depend on $\mhp$,
the size of the interference term varies with the input charged Higgs mass.
Based on the minimum of the ratio $R_{\Gamma M}=\Delta M/(\Gamma_h+\Gamma_H)$ and a significant mixing between $h$ and $H$,
we expect the most significant interference contribution near
$\mhp=153$\,GeV.

\begin{figure}[ht!]
 \begin{center}
  \includegraphics[width=14cm]{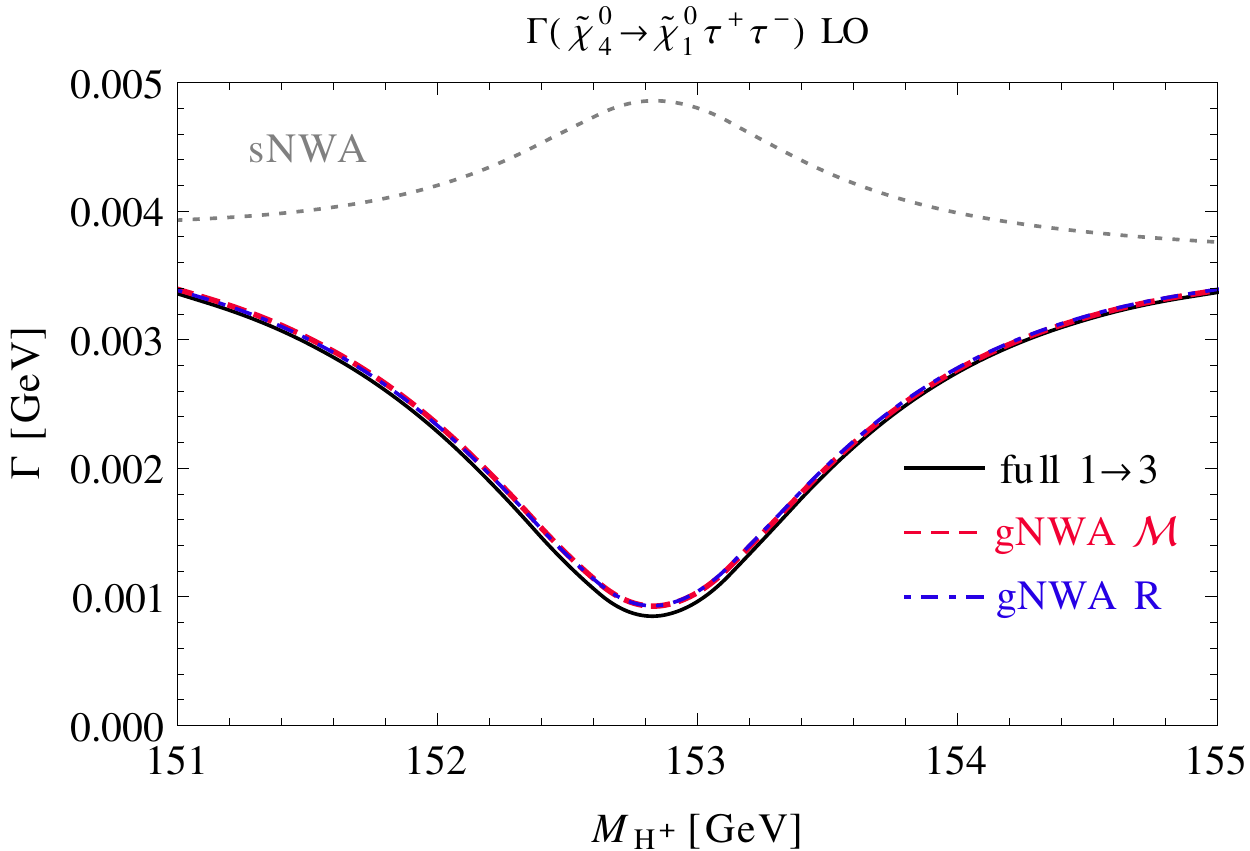}
\caption{The 1$\rightarrow$3 decay width of  $\cf \rightarrow \co \tau^{+}\tau^{-}$ at tree level with contributions from $h, H$ including their interference (black) confronted with the NWA: sNWA without the interference term (grey, dotted), gNWA including the interference term based on on-shell matrix elements denoted by $\Mm$ (red, dashed) and on the R-factor approximation denoted by R (blue, dash-dotted).}
\label{fig:sgNWA_tree}
 \end{center}
\end{figure}

Fig.\,\ref{fig:sgNWA_tree} presents the partial decay width
$\Gamma(\cf\rightarrow \co \tau^{+}\tau^{-})$ in dependence of the input Higgs
mass $M_{H^{\pm}}$. In the sNWA (grey), the interference term is absent. In
contrast, the full 3-body decay\footnote{In this section, the \textit{full}
tree level refers to the sum of $h$- and $H$-mediated 3-body decays including
the interference term (but without $A$- and $Z$-boson exchange or non-resonant
propagators) at the improved Born level, i.e.\ including Higgs masses, total
widths and $\hat{\textbf{Z}}$-factors at the leading 2-loop level from
\texttt{FeynHiggs}\,\cite{Heinemeyer:1998np, Heinemeyer:1998yj,
Degrassi:2002fi, Heinemeyer:2007aq}.} (black) takes both $h-$ and $H-$ propagators and their
interference into account.
Comparing the prediction of the sNWA with the full 3-body decay width reveals
an enormous discrepancy between both results, especially in the region of the
smallest ratio $R_{\Gamma M}$ around $\mhp\simeq153\,$GeV, due to a large
negative interference term. Consequently, the NWA in its standard version is
insufficient in this parameter scenario.

In the generalised narrow-width approximation, on the other hand,
the sNWA is extended by incorporating the on-shell interference term.
The red line indicates the prediction
of the complete process in the gNWA using the on-shell evaluation of unsquared
matrix elements in the interference term as derived conceptually in
Eq.\,(\ref{eq:masterformula}) and explicitly in Sect.\,\ref{sect:Helicity}.
Furthermore, the blue line demonstrates the result of the gNWA using the
additional approximation of interference weight factors $R$ defined in
Eq.\,(\ref{eq:R}). While the sNWA overestimates the full result by a factor of
up to $5.5$ on account of the neglected destructive interference, both
variants of the gNWA result in a good approximation of the full 3-body decay
width.

The slight relative deviation between either form of the gNWA and the full
result amounts to $\left(\Gamma_{gNWA}-\Gamma_{1\rightarrow
3}\right)/\Gamma_{sNWA}\simeq  0.4\%-1.7\%$ if normalised to the sNWA and to
$\left(\Gamma_{gNWA}-\Gamma_{1\rightarrow 3}\right)/\Gamma_{1\rightarrow
3}\simeq  0.5\%-9.2\%$ if normalised to the 3-body decay width. The largest
relative deviation between $\Gamma_{gNWA}$ and $\Gamma_{1\rightarrow 3}$
arises in the region where the reference value $\Gamma_{1\rightarrow 3}$
itself is very small so that a small deviation has a 
pronounced relative
effect. This uncertainty, however, is not intrinsically introduced by the
approximated interference term, but it stems from the factorised constituents
$\Gamma_{NWA}^{h},~\Gamma_{NWA}^{H}$ already present in the sNWA, see
Fig.\,\ref{fig:sNWA_intrinsic}. 

\section{Application of the gNWA to the loop level}\label{sect:IntNLO}
Motivated by the good performance of the gNWA at the tree level, in this
section we investigate the application of
the generalised narrow-width approximation at the loop level by
incorporating 1-loop corrections of the production and
decay part into the predictions. Before treating the full
3-body decay width at the next-to leading order (NLO) in
Sect.\,\ref{sect:3bodyNLO}, we will start with the method of on-shell matrix
elements in Sect.\,\ref{sect:M1loop} and turn to the R-factor approximation in
Sect.\,\ref{sect:R1loop}. 

At the 1-loop level we write the 
product of the
production cross-section times partial decay width in the standard NWA 
as
\begin{equation}
 \sigma_P \cdot \text{BR} \longmapsto \frac{\sigma_P^{1}\Gamma_D^{0} + \sigma_P^{0}\Gamma_D^{1}}{\Gamma^{\rm{tot}}},\label{eq:sigBR1}
\end{equation}
where the total width is obtained from \texttt{FeynHiggs}\,\cite{Heinemeyer:1998np, Heinemeyer:1998yj,
Degrassi:2002fi, Heinemeyer:2007aq} incorporating
corrections up to the 2-loop level
as in the definition of the branching ratio and in the Breit-Wigner
propagator. While restricting the numerator of Eq.\,(\ref{eq:sigBR1}) formally
to one-loop order to enable a consistent comparison with the full process, at
the end (in Sect.\,\ref{sect:best}) all constituents of the NWA will be used
at the highest available precision, i.e.\ $\sigma_P^{\rm{best}} \cdot
\text{BR}^{\rm{best}}$ for the most advanced prediction with the branching
ratio obtained from \texttt{FeynHiggs}.

\subsection{On-shell matrix elements at 1-loop order}\label{sect:M1loop}
In analogy to the procedure in Sect.\,\ref{sect:intMos} at the tree level,
on-shell matrix elements are used here in the 1-loop expansion. Special
attention must be paid to the cancellation of infrared (IR) divergences from
virtual photons (or gluons) in 1-loop matrix elements and real photon (gluon)
emission off charged external legs. In preparation for the example process
$\cf \rightarrow \co\,h/H\rightarrow \co\,\tau^{+}\tau^{-}$ (see
Sect.\,\ref{sect:example}), we focus on IR-divergences from photons in loops
of the decay part and soft final state photon radiation.

The aim is to approximate only the 1-loop contribution, but to keep  the full momentum dependent expression at the Born level with $\Mm_i^{0} = \Mm_i^{0}(q^{2})$, 
\begin{align}
 |\Mm^{0}|^{2} &= |\Mm_h^{0}|^{2} + |\Mm_H^{0}|^{2} + 2\text{Re}\left[\Mm_h^{0}\Mm_H^{0*}\right]. \label{eq:Mtree}
\end{align}
In contrast, the 1-loop matrix elements are factorised into the on-shell
production and decay parts times the momentum dependent Breit-Wigner
propagator $\Delta_i^{\text{BW}}\equiv \Delta_i^{\text{BW}}(q^{2})$. The
squared matrix elements are expanded up to the 1-loop order.
Since the emission of soft real photons is
proportional to the Born contribution, the virtual contribution is
supplemented by the 
absolute value squared of the tree-level matrix element,
multiplied by the QED-factor $\delta_{\text{SB}}$ of soft
bremsstrahlung \cite{Denner:1991kt,Hahn:FCmanual}, 
\begin{align}
2\rm{Re}\left[\Mm^{0}\Mm^{1*}\right]
 + \delta_{\text{SB}} |\Mm^{0}|^{2} &\simeq 2\text{Re}\left[\left(\Pp_h^{1}\Dd_h^{0}+ \Pp_h^{0}\Dd_h^{1} + \delta_{\text{SB}} \Pp_h^{0}\Dd_h^{0}\right)\Pp_h^{0*}\Dd_h^{0*} \cdot|\Delta_h^{BW}|^{2}\right]\nonumber\\
&\,+2\text{Re}\left[\left(\Pp_H^{1}\Dd_H^{0}+ \Pp_H^{0}\Dd_H^{1} + \delta_{\text{SB}} \Pp_H^{0}\Dd_H^{0}\right)\Pp_H^{0*}\Dd_H^{0*}\cdot|\Delta_H^{\text{BW}}|^{2} \right]\nonumber\\
 &\,+2\text{Re}\left[\left\lbrace\left(\Pp_h^{1}\Dd_h^{0}+\Pp_h^{0}\Dd_h^{1}\right)\Pp_H^{0*}\Dd_H^{0*}+\Pp_h^{0}\Dd_h^{0}\left(\Pp_H^{1*}\Dd_H^{0*}+ \Pp_H^{0*}\Dd_H^{1*}\right) \right. \right.\nonumber\\
&\hspace{1.4cm} +\left.\left. \delta_{\text{SB}}\, \Pp_h^{0}\Dd_h^{0}\Pp_H^{0*}\Dd_H^{0*} \right\rbrace \cdot \Delta_h^{\text{BW}}\Delta_H^{\text{BW}*}\right]. \label{eq:Mloopsoft}
\end{align}
The first line of Eq.\,(\ref{eq:Mloopsoft}) represents the pure
contribution from $h$, factorised into production and decay, the second
line accordingly for $H$. The third and fourth lines constitute the
1-loop and bremsstrahlung interference term as the product of $h$- and
$H$-matrix elements and Breit-Wigner propagators. For a consistent
comparison with the full 1-loop result, each term is restricted to
1-loop corrections in only one of the matrix elements.

The 1-loop prediction of the full process in the approximation of
on-shell matrix elements consists --- besides the Born cross section
without an approximation\footnote{If the full Born cross section cannot
be calculated, this term can be replaced by the gNWA at the Born level.}
--- of the squared contribution of $h$ and $H$ and the interference term
$\sigma_{\Mm}^{\rm{int}1}$ at the strict 1-loop level\footnote{With
\textit{strict 1-loop} we refer to the expansion of the products of
matrix elements whereas 2-loop Higgs masses, total widths and wave
function renormalisation factors are employed.},
\begin{align}
 \sigma_{\Mm}^{1} &= \sigma_{\rm{full}}^{0} + \frac{\sigma_{P_h}^{1}\Gamma_{D_h}^{0}+\sigma_{P_h}^{0}\Gamma_{D_h}^{1}}{\Gamma_h^{\rm{tot}}}+ \frac{\sigma_{P_H}^{1}\Gamma_{D_H}^{0}+\sigma_{P_H}^{0}\Gamma_{D_H}^{1}}{\Gamma_H^{\rm{tot}}}
+\sigma_{\Mm}^{\rm{int}1}\label{eq:M1strict},\\
\sigma_{\Mm}^{\rm{int}1} &=\frac{2}{F}\text{Re}\left\lbrace\int \frac{dq^{2}}{2\pi}\Delta^{\text{BW}}_h(q^{2}) \Delta^{*\text{BW}}_H(q^{2}) \right. \nonumber\\
&\hspace*{1.5cm} \left(\left[\int d\Phi_P(\qsq)(\Pp_{h}^{1}\Pp_{H}^{0*}+\Pp_{h}^{0}\Pp^{1*}_{H})\right]
\left[\int d\Phi_D(\qsq) \Dd_{h}^{0}\Dd^{0*}_{H}\right]\right. \nonumber\\
&\hspace*{1.5cm} \left.\left.+\left[\int d\Phi_P(\qsq)\Pp_{h}^{0}\Pp^{0*}_{H}\right]
 \left[\int d\Phi_D(\qsq) (\Dd_{h}^{1}\Dd^{0*}_{H}+\Dd_{h}^{0}\Dd^{1*}_{H}+\delta_{\text{SB}}\Dd_{h}^{0}\Dd^{0*}_{H})\right]\right) \right\rbrace.\label{eq:Mloop}
\end{align}
For the prediction with the most precise constituents, we use 2-loop branching ratios, $\rm{BR}_i^{\rm{best}}$. We include also the products of 1-loop matrix elements. Their contribution to the interference term is denoted by $\sigma_{\Mm}^{\rm{int}+}$,
\begin{align}
 \sigma_{\Mm}^{\rm{int}+} &= \frac{2}{F}\text{Re}\left\lbrace \int \frac{dq^{2}}{2\pi}\Delta^{\text{BW}}_h(q^{2}) \Delta^{*\text{BW}}_H(q^{2})\right.\nonumber\\
&\hspace*{1.5cm}\left.\left[\int d\Phi_P(\qsq) (\Pp_h^{1}\Pp_H^{0*}+\Pp_h^{0}\Pp_H^{1*})\right]
\left[\int d\Phi_D(\qsq) (\Dd_h^{1}\Dd_H^{0*}+\Dd_h^{0}\Dd_H^{1*}+\delta_{\text{SB}}\Dd_h^{0}\Dd_H^{0*})\right]   \right\rbrace. \label{eq:Mint+}
\end{align}
The approximation of the whole process based on on-shell matrix elements and incorporating higher-order corrections wherever possible is denoted by $\sigma_{\Mm}^{\text{best}}$, which reads then
\begin{align}
 \sigma_{\Mm}^{\text{best}} &= \sigma_{\rm{full}}^{0}+\sum_{i=h,H}\left(
\sigma_{P_i}^{\text{best}}\text{BR}_i^{\text{best}}-\sigma_{P_i}^{0}\text{BR}_i^{0}\right)+
 \sigma_{\Mm}^{\rm{int}1}+\sigma_{\Mm}^{\rm{int}+}\label{eq:Mbest}.
\end{align}
The \textit{best} production cross section $\sigma_{P_i}^{\text{best}}$ and branching ratios $\text{BR}_i^{\text{best}}$ mean the sum of the tree level, strict 1-loop and all available higher-order contribution to the respective quantity. Therefore, the products of tree level production cross sections and branching ratios are subtracted because their unfactorised counterparts are already contained in the full tree level term $\sigma_{\rm{full}}^{0}$. If a more precise result of the production cross sections is available, it can be used instead of the explicit 1-loop calculation that was performed in our example process.

\subsubsection{IR-finiteness of the factorised matrix elements}
\paragraph{On-shell evaluation}
The UV-divergences of the virtual corrections are cancelled by the same
counterterms as in the full process at 1-loop order. Although it would
be technically possible in most processes to compute the full
brems\-strahlung term without the NWA, i.e.\
$\delta_{\text{SB}}\,|\Mm_{\rm{full}}^{0}|^{2}$, the IR-divergences from
the on-shell decays need to be exactly cancelled by those from the real
photon emission. But the IR-singularities in the sum of the factorised
(on-shell) virtual corrections and the momentum-dependent real ones
would not match each other. Consequently, the tree level matrix elements
are also factorised, and the IR-divergent parts of the 1-loop decay matrix elements $\Dd_h^{1}(M_h^{2}, \overline{M}^{2}), \Dd_H^{1}(M_H^{2}, \overline{M}^{2})$ and the soft QED-factor $\delta_{\text{SB}}(\overline{M}^{2})$ have to be calculated at the same mass $\overline{M}=M_h$ or $M_H$. The LO matrix elements are evaluated at their mass-shell, i.e. $\Dd_i^{0}(M_{h_i}^{2})$. The NLO matrix elements are split into the 
part containing loop integrals on the one hand and the helicity matrix elements on the other hand. While the individual Higgs masses can be inserted into the finite helicity matrix elements (see Sect.\,\ref{sect:Helicity}
), the loop integrals have to be evaluated at the same mass
$\overline{M}^{2}$ as in $\delta_{\text{SB}}$ to preserve the
IR-cancellations. Hence, a choice must be made whether to define
$\overline{M}=M_h$ or $M_H$. We evaluate the numerical difference in
Sect.\,\ref{sect:treatphoton}.

The production matrix elements are completely evaluated on their
respective mass-shells, $\Pp^{0}_i(M_{h_i}^{2})$ and
$\Pp^{1}_i(M_{h_i}^{2})$. This is possible because the initial state in
this example contains only neutral particles. But the calculation can be
directly generalised to charged initial states according to the
procedure described for the decay matrix elements. The IR-singularities
in the product of initial and final state radiation are then cancelled
by those from a virtual photon connecting charged legs of the initial
and final state. Such non-factorisable contributions can be treated in
a pole approximation in analogy to the 
double-pole approximation (DPA) that has been used for instance for the
process
$e^{+}e^{-}\rightarrow W^{+}W^{-}\rightarrow 4$\,leptons, see 
Ref.\,\cite{Denner:1997ia}. An alternative approach for the treatment of
IR-singularities is formulated in
Refs.\,\cite{Grunewald:2000ju,Denner:2000bj}. There, the singular parts
from the real photon contribution are extracted, and the DPA is only
applied for those terms which exactly match the 
singularities from the virtual photons. In our calculation, 
we do not split up the real corrections in this way, but employ instead
the procedure described above.
We discuss a possibility of splitting the diagrams with virtual photons
into an IR-singular and a finite subgroup in
Sect.\,\ref{sect:SpecialTreatment}.

\paragraph{Cancellation of IR-divergences}
According to the Kinoshita-Lee-Nauenberg (KLN) theorem\,\cite{Kinoshita:1962ur, Lee:1964is}, the IR-divergence from a virtual photon is cancelled by the emission of a real photon off a charged particle from the initial or final state, i.e., in our example process as soft bremsstrahlung in the final state of a Higgs decay. We will derive the IR-finiteness of the on-shell matrix elements in analogy to the cancellation of the IR divergencies for the
full 3-body decay. Writing the momentum-dependent 3-body matrix elements
with the resonant particle either $h_i=h$ or $H$ as the sum of the
tree level ($\Mm_{h_i}^{0}$) and virtual ($\Mm_{h_i}^{v}$) 
contributions,
\begin{align}
 \Mm_{h_i}(q^{2}) = \Mm_{h_i}^{0}(q^{2})+\Mm_{h_i}^{v}(q^{2}) \label{eq:Mh0v},
\end{align}
and adding to the squared matrix element the corresponding contribution
from real soft photon
($\Mm_{h_i}^{\rm{Br}}$) radiation, we find
\begin{align}
 |\Mm_{h}+\Mm_H|^{2}+|\Mm_{h}^{\rm{Br}}+\Mm_H^{\rm{Br}}|^{2}
&= \sum_{h_i=h,H}\left(|\Mm_{h_i}|^{2}+|\Mm_{h_i}^{\rm{Br}}|^{2}\right)
 + 2\text{Re}\left[\Mm_{h}\Mm_{H}^{*}+\Mm_{h}^{\rm{Br}}\Mm_H^{\rm{Br}*}\right].\label{eq:MhHBr}
\end{align}
Because the complete sum in Eq.\,(\ref{eq:MhHBr}) and the individual $h$-
and $H$-terms are IR-finite, the interference term must be IR-finite by
itself. With the proportionality of the bremsstrahlung contribution
to the tree level
term,
\begin{equation}
\Mm_{h}^{\rm{Br}}(q^{2})\Mm_H^{\rm{Br}*}(q^{2}) = \delta_{\text{SB}}(q^{2})\Mm_h^{0}(q^{2})\Mm_H^{0*}(q^{2}),\label{eq:Brems}
\end{equation}
and keeping only the terms of $\mathcal{O}(\alpha)$ relative to the lowest order, the interference term $\text{Int}^{\alpha}(q^{2})$ results in
\begin{align}
 \text{Int}^{\alpha}(q^{2}) &= 2\text{Re}\left[\left.\Mm_h(q^{2})\Mm_H^{*}(q^{2})\right|_{\alpha}+\Mm_{h}^{Br}(q^{2})\Mm_H^{Br*}(q^{2})\right]\label{eq:Intalpha}\\
&= 2\text{Re}\left[\Mm_h^{v}(q^{2})\Mm_H^{0*}(q^{2})+\Mm_h^{0}(q^{2})\Mm_H^{v*}(q^{2})+\delta_{\text{SB}}(q^{2})\Mm_h^{0}(q^{2})\Mm_H^{0*}(q^{2})  \right]\label{eq:Intq2}.
\end{align}
As described above, the on-shell evaluation is performed at the individual mass $M_{h_i}$ in all production and tree level matrix elements and the helicity elements, whereas the soft photon factor $\delta_{\text{SB}}$ and the 1-loop form factors of the decay are evaluated at the same mass $\overline{M}$ in the on-shell interference term $\text{Int}_{\text{os}}^{\alpha}$ of $\mathcal{O}(\alpha)$ relative to the lowest order,
\begin{align}
  \text{Int}_{\text{os}}^{\alpha}
&= 2\text{Re}\left[\Mm_h^{v}(M_h^{2},\overline{M}^{2})\Mm_H^{0*}(M_H^{2})+\Mm_h^{0}(M_h^{2})\Mm_H^{v*}(M_H^{2},\overline{M}^{2})+\delta_{\text{SB}}(\overline{M}^{2})\Mm_h^{0}(M_h^{2})\Mm_H^{0*}(M_H^{2})  \right]\label{eq:IntosMbar2}\\
&= 2\text{Re}\left[\left\lbrace\left(\Pp_h^{v}(M_h^{2})\Dd_h^{0}(M_h^{2})+\Pp_h^{0}(M_h^{2})\Dd_h^{v}(M_h^{2},\overline{M}^{2})\right)\cdot\Pp_H^{0*}(M_H^{2})\Dd_H^{0*}(M_H^{2})\right.\right.\nonumber\\
&\left.\left.\hspace{1.3cm}+\Pp_h^{0}(M_h^{2})\Dd_h^{0}(M_h^{2})\cdot\left(\Pp_H^{v*}(M_H^{2})\Dd_H^{0*}(M_H^{2})+\Pp_H^{0*}(M_H^{2})\Dd_H^{v*}(M_H^{2},\overline{M}^{2})\right)\right.\right., \nonumber\\
&\hspace{1.3cm}\left.\left.
+\delta_{\text{SB}}(\overline{M}^{2})\,\Pp_h^{0}(M_h^{2})\Dd_h^{0}(M_h^{2})\,\Pp_H^{0*}(M_H^{2})\Dd_H^{0*}(M_H^{2}) \right\rbrace\Delta_h(q^{2})\Delta_H^{*}(q^{2}) \right].\label{eq:IntosAlpha}
\end{align}
Since the virtual production matrix elements are IR-finite in our example process, we can drop the first term in each of the brackets in the first and second line of Eq.\,(\ref{eq:IntosAlpha}) for the discussion of IR-singularities, which are contained in $\left.\text{Int}_{\text{os}}^{\alpha}\right|_{\text{IR}}$,
\begin{align}
   \left.\text{Int}_{\text{os}}^{\alpha}\right|_{\text{IR}} 
&= 2\rm{Re}\biggl[\Pp_h^{0}(M_h^{2})\Pp_H^{0*}(M_H^{2})\cdot\Delta_h(q^{2})\Delta_H^{*}(q^{2})\cdot\biggr.\nonumber\\
&\hspace*{0.75cm}\biggl.\left(\Dd_h^{v}(M_h^{2},\overline{M}^{2})\Dd_H^{0*}(M_H^{2})+\Dd_h^{0}(M_h^{2})\Dd_H^{v*}(M_H^{2},\overline{M}^{2})+\delta_{\text{SB}}(\overline{M}^{2})\Dd_h^{0}(M_h^{2})\Dd_H^{0*}(M_H^{2})\right)\biggr]\label{eq:IntosAlphaIR}.
\end{align}
Moreover, the $M_{h_i}^{2}$-dependent helicity matrix elements $d_{h_i}(M_{h_i}^{2})$ from Sect.\,(\ref{sect:Helicity}) can be factored out by $\Dd_{h_i}=C_{h_i}d_{h_i}$ so that the IR-singularities from $\left.\text{Int}_{\text{os}}^{\alpha}\right|_{\text{IR}}$ can be further extracted:
\begin{align}
  \left.\text{Int}_{\text{os}}^{\alpha}\right|_{\text{IR}} 
&= 2\text{Re}\left[\Pp_h^{0}(M_h^{2})\Pp_H^{0*}(M_H^{2})\cdot\Delta_h(q^{2})\Delta_H^{*}(q^{2})\cdot d_h(M_h^{2})\,d_H^{*}(M_H^{2})\right.\nonumber\\
&\hspace*{0.7cm}\left.\left(C_h^{v}(\overline{M}^{2})C_H^{0*}+C_h^{0}C_H^{v*}(\overline{M}^{2})+\delta_{\text{SB}}(\overline{M}^{2})C_h^{0}C_H^{0*}\right)\right]\label{eq:IRextract}.
\end{align}
Compared to Eq.\,(\ref{eq:Intq2}) which can also be factorised into
$q^{2}$-dependent form factors and helicity matrix elements, the
structure of the IR-singularities is the same. In
Eq.\,(\ref{eq:IRextract}), all of those contributions
are just evaluated at $\overline{M}^{2}$ instead of $q^{2}$. Hence the cancellation works analogously so that Eq.\,(\ref{eq:IntosMbar2}) is an IR-finite formulation of the factorised interference term. Because the $\hat{\textbf{Z}}$-factors can be factored out in the same way for the on-shell approximation as for the full matrix elements, their inclusion preserves the cancellations of IR-divergences.

\subsubsection{Separate calculation of photon diagrams}\label{sect:SpecialTreatment}
As an alternative to the method described above, it is possible to reduce the number of diagrams whose loop integrals need to be evaluated at the common mass $\overline{M}$ instead of their on-shell mass $M_i$ by splitting the 1-loop decay matrix elements into an IR-finite and an IR-divergent part,
\begin{equation}
 \Dd_i^{1} = \Dd_i^{1,\rm{no}\gamma} + \Dd_i^{1,\gamma}\label{eq:split}.
\end{equation}
Both subgroups of diagrams are rendered UV-finite by the corresponding counterterms. Since the diagrams without any photon are already IR-finite, their loop integrals can safely be calculated on-shell, $\Dd_i^{1,\rm{no}\gamma}(M_{h_i}^{2})$. Hence, only the loop-integrals of the photon contribution need to be evaluated at a fixed mass $\overline{M}$, resulting in $\Dd_i^{1,\gamma}(M_{h_i}^{2}, \overline{M}^{2})$ and $\delta_{\text{SB}}(\overline{M}^{2})$.

If the fixed Higgs mass were inserted into both the loop integrals and
the helicity matrix elements, the IR-cancellation would work in the same 
way as for the unfactorised process, just with the special choice of
$q^{2}=\overline{M}^{2}$. In our approach, the helicity matrix elements
are determined at the specific masses $M_{h_i}$ as it is demonstrated in
Eqs.\,(\ref{eq:PpPphH}) and (\ref{eq:DdhH}). Furthermore, those mass
values
are equal in the matrix elements at lowest and higher orders
as loop-corrected masses are used also at the improved Born level. Because the $M_{h_i}$-dependent helicity matrix elements can be factored out, the IR-singularities cancel in the decay contribution to the interference term of $\mathcal{O}(\alpha)$ relative to the lowest order, with $\Dd_i^{0}$ at $M_{h_i}^{2}$,
\begin{align}
 \left(\Dd_h\Dd_H^{*}\right)^{\alpha}
&=\Dd_h^{1,\gamma}(M_h^{2},\overline{M}^{2})\Dd_H^{0*}+\Dd_h^{0}\Dd_H^{1,\gamma*}(M_H^{2},\overline{M}^{2}) + \delta_{\text{SB}}(\overline{M}^{2})\Dd_h^{0}\Dd_H^{0*}.
\end{align}
On the one hand, this approach requires the separate calculation of
purely photonic and non-photonic contributions. On the other hand, it
enables the on-shell evaluation of IR-finite integrals and is thus
closer to the full result. However, in case of a virtual photino
contribution one needs to be careful
not to break supersymmetry by treating the photon differently than
its superpartner. Thus, the possibility of such a separate treatment of
the photon diagrams, whose numerical impact is small in the studied
example process, should be considered in view of the investigated model and its
particle content.

\subsection{Interference weight factors at 1-loop order}\label{sect:R1loop}
In the previous section, we derived how to include virtual and real
contributions in the product of factorised matrix elements in a UV- and
IR-finite way. However, special attention is needed to ensure the
correct treatment of the on-shell matrix elements of the interference
contribution.

We now discuss additional approximations with which the R-factor method
introduced in Sect.\,\ref{sect:intR} can be extended beyond the tree
level. We develop 
a method that facilitates an
approximation of the interference term based on higher-order cross
sections and decay widths, but only tree level couplings. This
technically simpler
treatment comes at the price of the further assumption, as in
the tree level version of the interference weight factor, that
both Higgs masses be equal. Thus, the method presented in this section is
an optional, additional approximation with respect to
Eq.\,(\ref{eq:Mloopsoft}).

Under the assumption of equal masses, the product of unsquared matrix
elements for the production and decay of $h$ and $H$ can 
be re-expressed at the tree level in terms of either $h$ or $H$ 
with the help of
Eq.\,(\ref{eq:xi}). Hence, one can choose to keep the 1-loop matrix
elements and to replace only the tree level ones so that only
lowest-order couplings will be present in the $x$-factor. We will now
apply this prescription to
the third term in Eq.\,(\ref{eq:Mloopsoft})
containing the 1-loop virtual corrections to the interference term
$\text{Int}^{v}$:
\begin{align}
 \text{Int}^{v}&=2\text{Re}\left[\left\lbrace \left(\Pp_h^{1}\Dd_h^{0}+ \Pp_h^{0}\Dd_h^{1}\right) \Pp_H^{0*} \mathcal{D}_H^{0*}
+ \Pp_h^{0}\Dd_h^{0} \left(\Pp_H^{1*} \mathcal{D}_H^{0*}+\Pp_H^{0*} \Dd_H^{1*} \right) \right\rbrace \Delta_h^{\text{BW}}\Delta_H^{\text{BW}*} \right]\nonumber\\
&\simeq2\text{Re}\left[ \left(\Pp_h^{1}\Dd_h^{0}+ \Pp_h^{0}\Dd_h^{1}\right) \mathcal{P}_h^{0*} \mathcal{D}_h^{0*}\cdot \frac{C_{P_H}^{0*}}{C_{P_h}^{0*}} \frac{C_{D_H}^{0*}}{C_{D_h}^{0*}}\cdot \Delta_h^{\text{BW}}\Delta_H^{\text{BW}*}\right]\nonumber\\
&\hspace*{0.5cm}+2\text{Re}\left[\left\lbrace\mathcal{P}_H^{0}\mathcal{D}_H^{0}\cdot \frac{C_{P_h}^{0}}{C_{P_H}^{0}} \frac{C_{D_h}^{0}}{Cc_{D_H}^{0}} \left(\Pp_H^{1*} \mathcal{D}_H^{0*}+\Pp_H^{0*} \Dd_H^{1*} \right)  \Delta_h^{\text{BW}}\Delta_H^{\text{BW}*}\right\rbrace^{*} \right]\nonumber\\
&=2\text{Re}\left[\left(\Pp_h^{1}\mathcal{P}_h^{0*}|\Dd_h^{0}|^{2} + |\Pp_h^{0}|^{2}\Dd_h^{1}\mathcal{D}_h^{0*}\right) x_h^{0}\cdot \Delta_h^{\text{BW}}\Delta_H^{\text{BW}*}\right] \nonumber\\
&\hspace*{0.5cm}+2\text{Re}\left[\left(\mathcal{P}_H^{1}\Pp_H^{0*}|\mathcal{D}_H^{0}|^{2} + |\mathcal{P}_H^{0}|^{2}\mathcal{D}_H^{1}\mathcal{D}_H^{0*}\right) x_H^{0}\cdot \Delta_H^{\text{BW}}\Delta_h^{\text{BW}*}\right]. \label{eq:2Rexh0}
\end{align}
Hence we exploited the choice of expressing the product of $h$- and $H$-matrix elements either in a weighted sum of both or in terms of one of them. The latter choice, as selected in Eq.\,(\ref{eq:2Rexh0}), has the advantage that the matrix elements containing loop contributions of $h$ and only tree level contributions of $H$ are transformed in terms of $h$ and vice versa.
Including the flux factor and the phase space integrals as in Eq.\,(\ref{eq:intijlong}), adding soft bremsstrahlung according to the last line of Eq.\,(\ref{eq:Mloopsoft}) and keeping in mind that
\begin{align}
 \frac{1}{F}\int d\Phi_P 2\text{Re}\left[\Pp_i^{1}\Pp_i^{0*}\right] = \sigma_{P_i}^{1}, \hspace*{1cm}
 \frac{1}{2M_i}\int d\Phi_D \left(2\text{Re}\left[\Dd_i^{1}\Dd_i^{0*}\right]+\delta_{\text{\text{SB}}}|\Dd_i^{0}|^{2}\right) = \sigma_{D_i}^{1}, \label{eq:2ReLoop}
\end{align}
the expressions from Eq.\,(\ref{eq:2Rexh0}) lead to
\begin{align}
 \sigma_{\rm{int}}^{1,R} 
 &= \frac{\sigma_{P_h}^{1}\Gamma_{D_h}^{0}+\sigma_{P_h}^{0}\Gamma_{D_h}^{1}}{\Gamma_h^{\rm{tot}}} \tilde{R}_h
  +\frac{\sigma_{P_H}^{1}\Gamma_{D_H}^{0}+\sigma_{P_H}^{0}\Gamma_{D_H}^{1}}{\Gamma_H^{\rm{tot}}} \tilde{R}_H \label{eq:int1R0},
\end{align}
where $\tilde{R}_i$ has been defined in Eq.\,(\ref{eq:Rtilde}). Eq.\,(\ref{eq:int1R0}) is meant for the consistent comparison with the full result in the strict one-loop expansion. Using the most precise predictions of all components and the unfactorised tree level result leads to the final prediction:
\begin{align}
\sigma_{R}^{\rm{best}} &= \sigma_{\rm{full}}^{0}+\sum_{i=h,H}\left(
\sigma_{P_i}^{\text{best}}\text{BR}_i^{\text{best}}-\sigma_{P_i}^{0}\text{BR}_i^{0}\right)+
 \sigma_{R}^{\rm{int}1}+\sigma_{R}^{\rm{int}+}\label{eq:intbestR0},\\
\sigma_{R}^{\rm{int}1} &= \left(\sigma_{P_h}^{1}\text{BR}_h^{0}+\sigma_{P_h}^{0}\text{BR}_h^{1}\right)\tilde{R}_h
+\left(\sigma_{P_H}^{1}\text{BR}_H^{0}+\sigma_{P_H}^{0}\text{BR}_H^{1}\right)\tilde{R}_H\label{eq:Rint1},\\
\sigma_{R}^{\rm{int}+} &= \frac{1}{2}\sigma_{P_h}^{1}\left(\text{BR}_h^{1}\tilde{R}_h+\text{BR}_H^{1}\tilde{R}_{hH}\right)
+\frac{1}{2}\sigma_{P_H}^{1}\left(\text{BR}_H^{1}\tilde{R}_H+\text{BR}_h^{1}\tilde{R}_{Hh}\right)\label{eq:Rint+},
\end{align}
where $\sigma_{R}^{\rm{int}1}$ denotes the contribution to the interference term for which the product of production cross sections and partial decay widths is restricted to the 1-loop level, but the branching ratios are at all levels normalised to the 2-loop total width from \texttt{FeynHiggs}\,\cite{Heinemeyer:1998np, Heinemeyer:1998yj,
Degrassi:2002fi, Heinemeyer:2007aq}. In addition, $\sigma_{R}^{\rm{int}+}$ contains terms beyond the 1-loop level.
In Eq.\,(\ref{eq:Rint+}), we introduced the generalised interference
weight factors $\tilde{R}_{ij}$,
\begin{align}
 \tilde{R}_{ij} &=2M_j\Gamma_j\text{Re}\left\lbrace x_{ij} I
\right\rbrace \label{eq:Rij},
\end{align}
involving the scaling factors $x_{ij}$,
\begin{align}
x_{ij} &= \frac{C_{P_h}C_{P_H}^{*}C_{D_h}C_{D_H}^{*}}{|C_{P_i}|^{2}|C_{D_j}|^{2}} \label{eq:xij},
\end{align}
to account for the product of 1-loop production and decay matrix elements in Eq.\,(\ref{eq:Mint+}). For the most precise prediction, the 1-loop branching ratios in Eqs.\,(\ref{eq:Rint1}, \ref{eq:Rint+}) can additionally be replaced by $\text{BR}_i^{\rm{best}}-\text{BR}_i^{0}$ which is beyond the $\Mm$-method in Eq.\,(\ref{eq:Mint+}). As in Eq.\,(\ref{eq:Mbest}) for the $\Mm$-method, the products of tree level production cross section and branching ratios have to be subtracted because their contribution is already accounted for by $\sigma_{\text{full}}^{0}$. The most precise branching ratios can be obtained from \texttt{FeynHiggs}\,\cite{Heinemeyer:1998np, Heinemeyer:1998yj,
Degrassi:2002fi, Heinemeyer:2007aq} including full 1-loop and leading 2-loop corrections.

\section{Full 3-body decay at the one-loop level}\label{sect:3bodyNLO}
The numerical validation of the gNWA at the next-to-leading order
requires the calculation of the process $\cf \rightarrow
\co\,\tau^{+}\tau^{-}$ 
with intermediate $h$ and $H$ 
as the full 3-body
decay including virtual and real corrections.

Ref.\,\cite{Drees:2006um} provides a 1-loop calculation of the decay of the next-to-lightest neutralino $\tilde{\chi}_2^{0}$ into $\co$ and a pair of leptons, thus a similar process, but with a dominant contribution from an on-shell slepton, while the Higgs propagators are treated as non-resonant. In the following, we focus on the diagrams contributing to resonant intermediate Higgs bosons, as well as box-diagrams with and without Higgs bosons. The 1-loop integrals are computed with \texttt{LoopTools}\,\cite{Hahn:1998yk,Hahn:2010zi}.

\subsection{Contributing diagrams}
\subsubsection{Virtual corrections at the neutralino-Higgs vertex}\label{sect:VirtNeutralino}
\begin{figure}[ht!]
 \begin{center}
  \includegraphics[width=3.2cm]{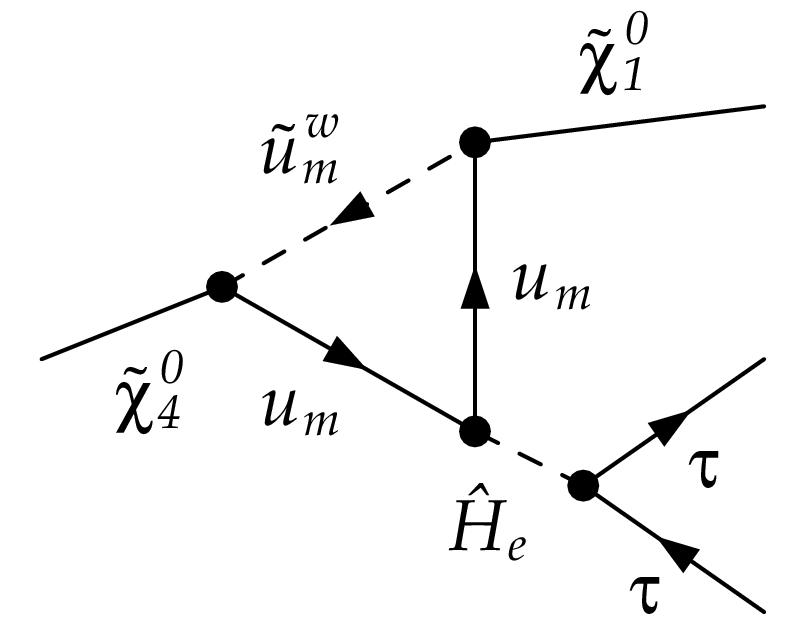}
  \includegraphics[width=3.2cm]{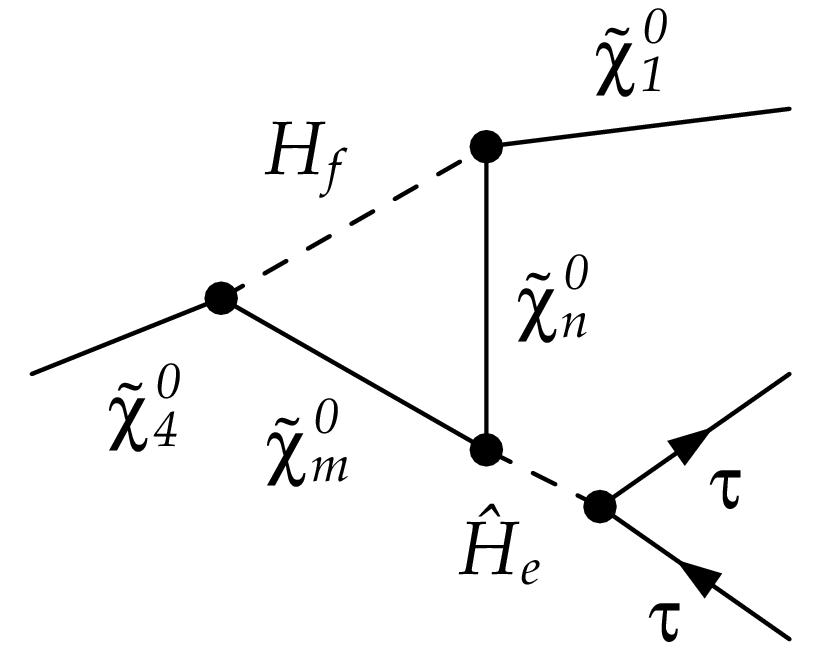}
  \includegraphics[width=3.2cm]{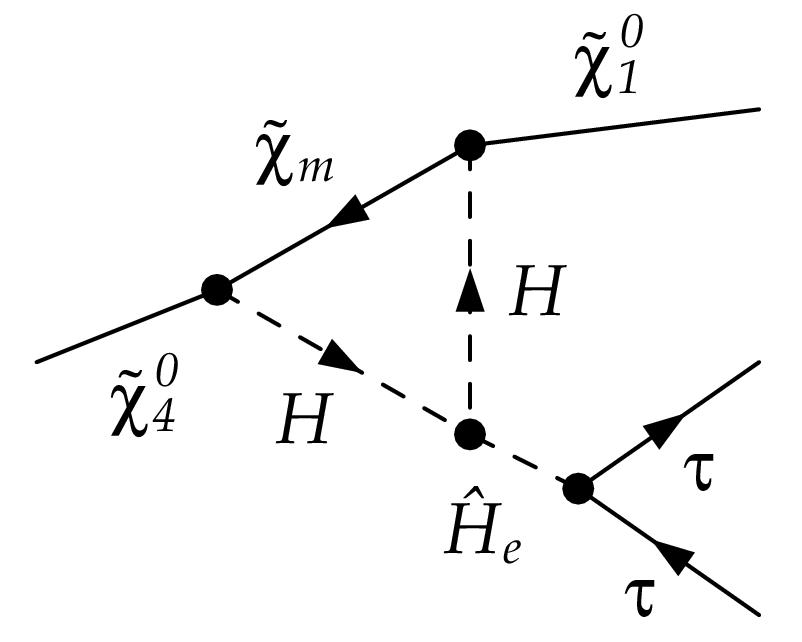}
  \includegraphics[width=3.2cm]{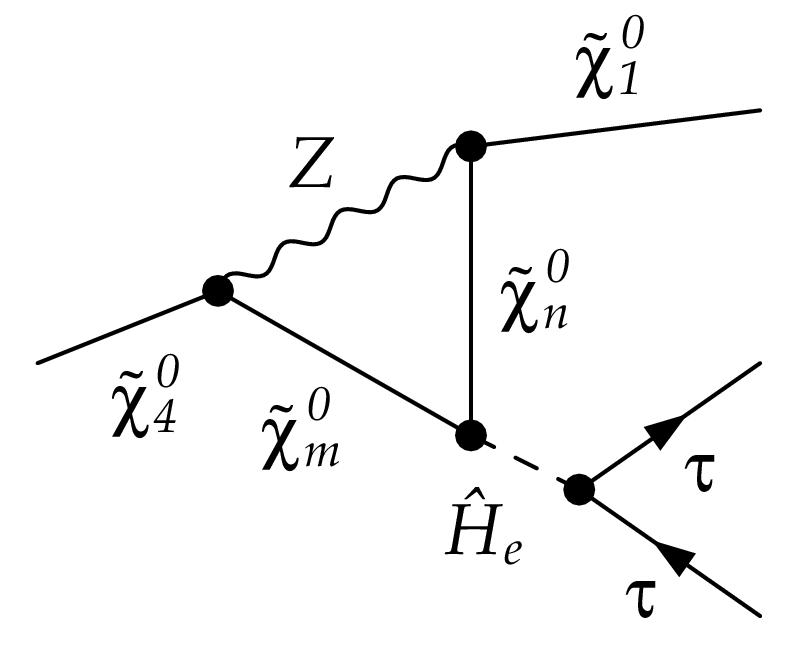}
  \includegraphics[width=3.2cm]{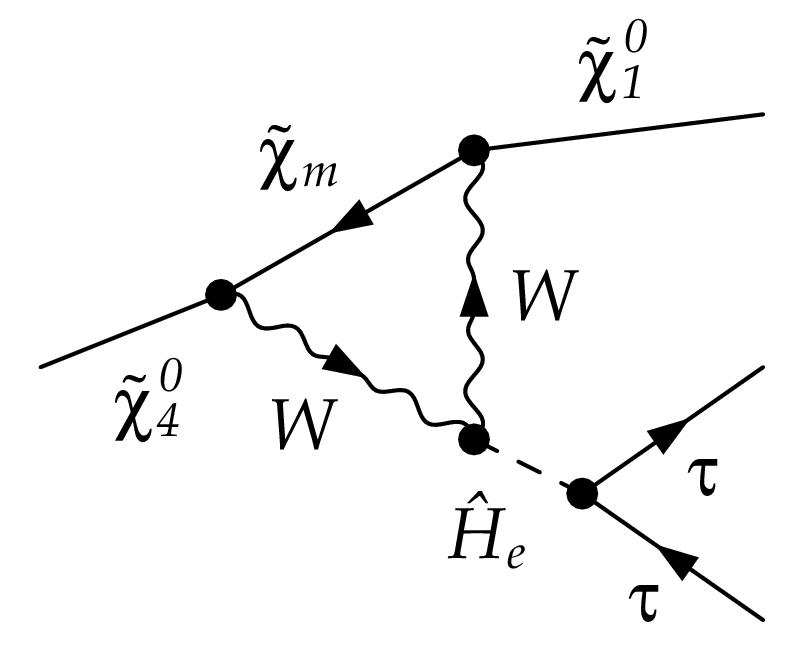}
\caption{Example triangle diagrams of the 3-body decay
$\cf\rightarrow\co\tau^{+}\tau^{-}$ with 1-loop corrections at the
$\cf\co \hat{H}_e$-vertex, where $\hat{H}_e$ denotes a Higgs boson mixed
by $\hat{\textbf{Z}}$-factors, $H_f$ an internal Higgs boson (see text)
and $H\equiv H^{\pm}$. $u$ and $\tilde{u}$ represent the up-type (s)quarks,  $\tilde{\chi}^{0}$ are the neutralinos and $\tilde{\chi}$ the charginos.}
\label{fig:Vert13body}
 \end{center}
\end{figure}
Virtual SM and MSSM particles contribute to the correction of the $\ci
\cj h_k$-vertex. A selection of diagrams is displayed in
Fig.\,\ref{fig:Vert13body}. 
We treat here the intermediate Higgs bosons $\hat{H}_e$ 
appearing outside of the vertex loop contribution as ``external'', 
while $H_f$ denotes an internal Higgs boson within the loop ($e,f=1,2,3$). 
Furthermore, 
$H\equiv H^{\pm}$ denotes the charged Higgs bosons. The neutralinos
are labelled by $\tilde{\chi}^{0}_n,~n=1,2,3,4$ and the charginos by
$\tilde{\chi}_m,~m=1,2$. The first example diagram contains up-type
quarks and a squark of generation $m=1,2,3$.

For $\hat{H}_e$ the
mixing with $\hat{\textbf{Z}}$-factors is taken into account, i.e., 
Eq.\,(\ref{eq:1PI}) is applied for both vertices of $\hat{H}_e$.
This treatment has been applied in order to enable a comparison with the factorised
production and decay contributions in the gNWA. 
The appearance of  $\hat{\textbf{Z}}$-factors in external Higgs boson lines 
is related to the fact that we use a renormalisation scheme without 
on-shell conditions for the Higgs-boson fields. In such a case, like the 
$\overline{\text{DR}}$ renormalisation of the Higgs fields employed here,
the $\mathbf{\hat{Z}}$-factors are
introduced to ensure correct on-shell properties of \textit{external} Higgs
bosons\,\cite{Chankowski:1992er,Heinemeyer:2001iy}. In the NWA, the Higgs
bosons appear as external particles in the on-shell production and decay,
and we therefore treat the intermediate Higgs bosons of resonant propagators
in the full 3-body decay in the same way for comparison purposes. 

The triangle corrections appearing at the $\ci \cj h_k$-vertex are renormalised by the counterterm
\begin{align}
 \delta C_{ijk}^{R/L}~=~&\frac{e}{2\cw \sw}\delta c^{(*)}_{ijk} + \left(\delta Z_e - \frac{\delta \sw}{\sw}- \frac{\delta \cw}{\cw}\right) C_{ijk}^{R/L}\nonumber\\
  &+ \frac{1}{2}\sum_{l=1}^{4}(\delta Z_{li}^{R/L}\,C_{ljk}^{R/L}+\delta
\bar{Z}_{jl}^{L/R}\,C_{ilk}^{R/L} +\delta
Z_{h_kh_l}C_{ijk}^{R/L})\label{eq:deltaCijk}
\end{align}
in the on-shell scheme, see Ref.\,\cite{Fowler:2009ay} and references therein.
In Eq.\,(\ref{eq:deltaCijk}), $h_l=\left\lbrace h,H,A,G\right\rbrace$ for
$l=1,2,3,4,$ denote the neutral Higgs and Goldstone bosons. The parameters
$M_1,~M_2,~\mu$ are related to the choice of the three electroweakinos which
are renormalised on-shell and thus define the choice for the 
on-shell renormalisation scheme
for the neutralino-chargino sector, as mentioned in Sect.\,\ref{sect:neucha}.
In our scenario, we identify $\co$ as the most bino-like, $\cth$ as the most
higgsino-like and $\cf$ as the most wino-like state and hence renormalise these
three neutralinos on-shell. By this choice of an NNN scheme, we avoid large
mass corrections to the remaining neutralino and the charginos. Alternatively,
$\ctw$ instead of $\cf$ could
be identified as the most wino-like state because
the two corresponding elements in the matrix $N$, which diagonalises the
neutralino mass matrix (see Sect.\,\ref{sect:neucha}), have nearly the same 
magnitude. 
Thus, this alternative choice would lead to a comparable sensitivity to the
three parameters of this sector and thereby also to a stable renormalisation
scheme. But since $\cf$ is involved in our process as an external particle, we
prefer to set it on-shell. The 1-loop effect on the 2-body decay widths
$\Gamma(\cf\rightarrow \co h/H)$ is shown in Fig.\,\ref{fig:2body}.

\subsubsection{Virtual corrections at the \texorpdfstring{Higgs- $\tau^{+}\tau^{-}$}{Higgs-tau} vertex and real photon emission}
\begin{figure}[ht!]
 \begin{center}
  \includegraphics[width=3.2cm]{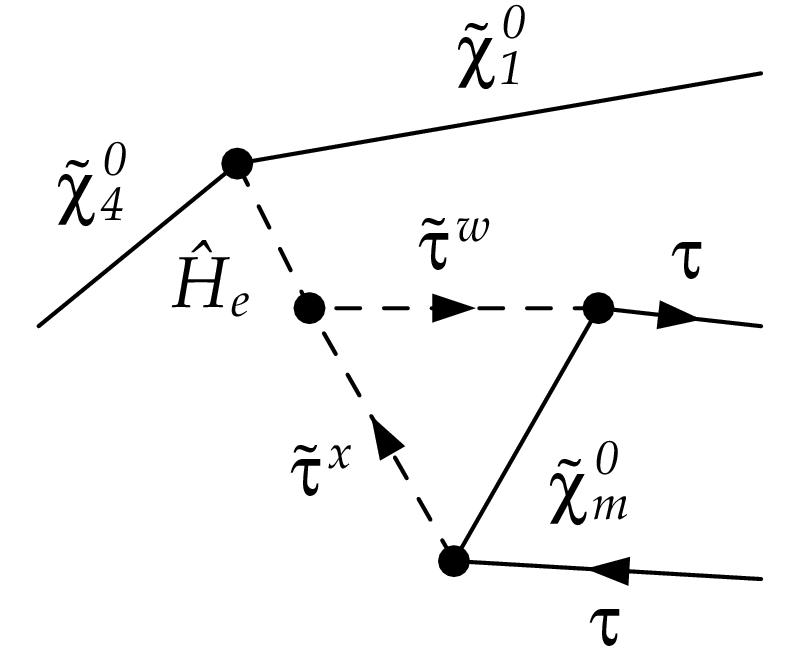}
  \includegraphics[width=3.2cm]{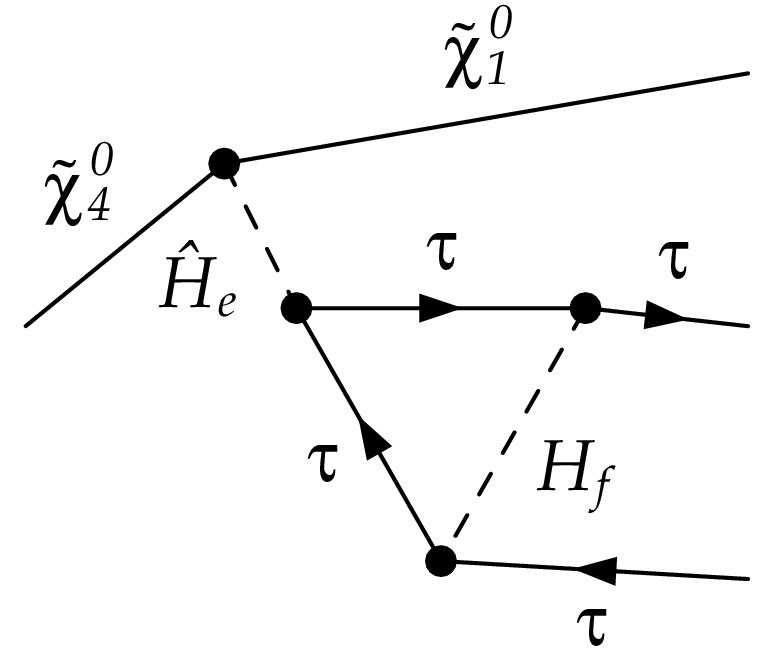}
  \includegraphics[width=3.2cm]{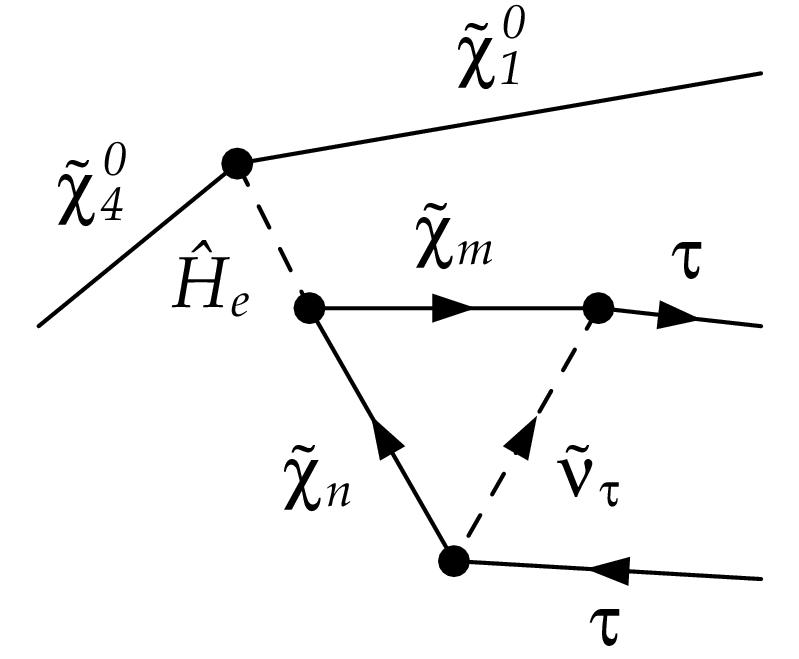}
  \includegraphics[width=3.2cm]{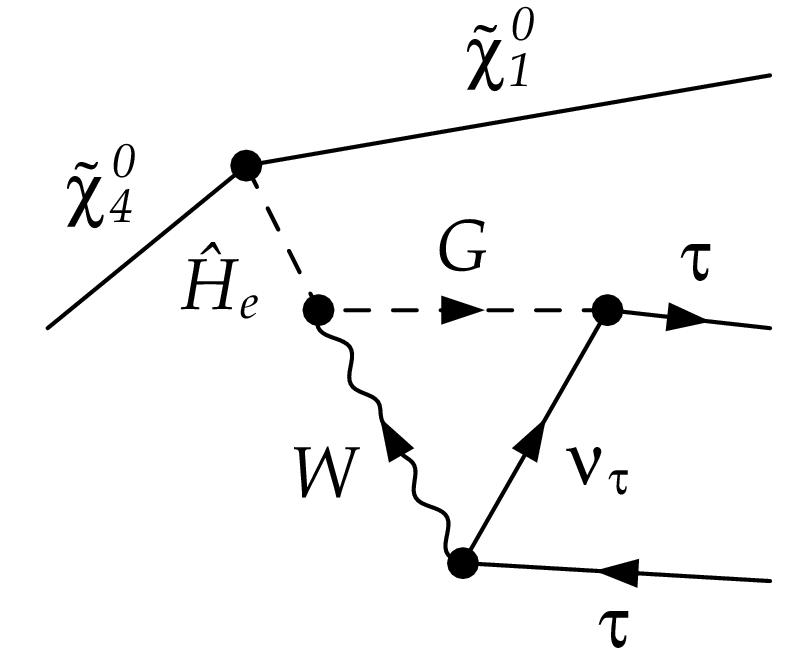}
  \includegraphics[width=3.2cm]{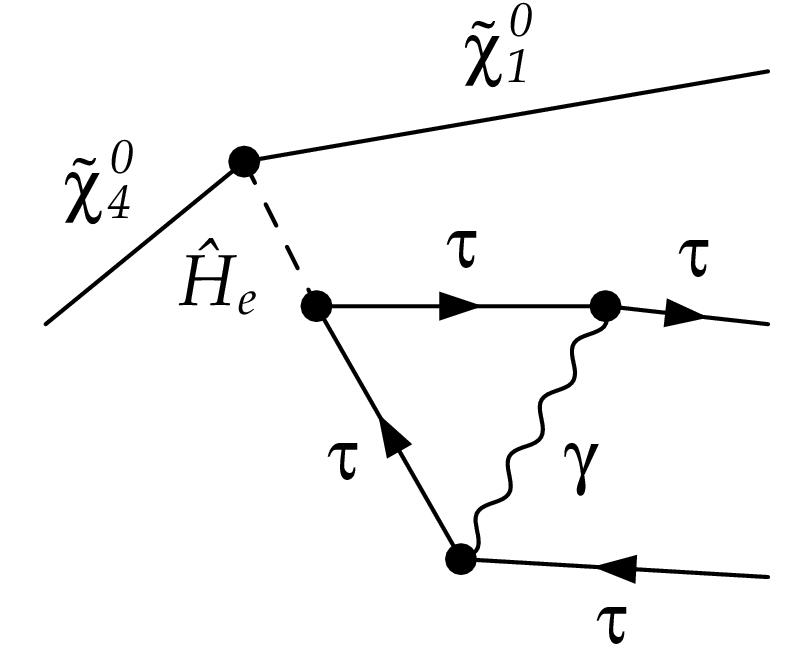}
\caption{Example triangle diagrams of the 3-body decay $\cf\rightarrow\co\tau^{+}\tau^{-}$ with 1-loop corrections at the $\hat{H}_e\tau^{+}\tau^{-}$-vertex, where the particles are labelled as is Fig.\,\ref{fig:Vert13body}.}
\label{fig:Vert23body}
 \end{center}
\end{figure}
Furthermore, the $h_k\tau^{+}\tau^{-}$-vertex diagrams shown in
Fig.\,\ref{fig:Vert23body} are UV-divergent, and the last diagram is also IR-divergent due to the virtual photon. The UV-divergences are cancelled by the counterterm, analogous to the SM, 
$\delta C_{h_k\tau^{+}\tau^{-}} = \delta C_{h_k\tau\tau}^L\omega_L+\delta C_{h_k\tau\tau}^R\omega_R$, with \cite{Denner:1991kt,KarinaPhD}
\begin{align}
 \delta C_{h_k\tau^{+}\tau^{-}}^{L/R}&=
C_{h_k\tau^{+}\tau^{-}}^{\rm{tree}}\cdot\left(\delta Z_e+\frac{1}{2}\delta Z_{h_k h_k}+\frac{1}{2}\delta Z_{hH}\frac{C_{h_l \tau\tau}^{\rm{tree}}}{C_{h_k \tau\tau}^{\rm{tree}}} 
-\frac{\delta M_W^{2}}{2M_W^{2}} -\frac{\delta \sw}{\sw} + \sinb^{2}\delta \tb \right.\nonumber\\
&\hspace*{2.5cm}\left.+\frac{\delta m_{\tau}}{m_{\tau}}+\frac{1}{2}\left\lbrace \delta Z_{\tau}^{L/R}+\delta Z_{\tau}^{R/L\dagger} \right\rbrace \right),
\end{align}
where $k,l=h,H$ and $\delta Z_{\tau}^{L/R}$ are the left-/right-handed field
renormalisation constants of the $\tau$-lepton. The tree-level couplings
$C_{h_k\tau^{+}\tau^{-}}^{\rm{tree}}$ are given in Eq.\,(\ref{eq:GHtt}).
The IR-divergent terms vanish for squared matrix elements in the combination
of virtual corrections containing a photon in the loop with real photons
emitted as soft bremsstrahlung off one of the $\tau$-leptons. Soft photons are
defined by the energy cut-off $E_{\rm{soft}}^{\rm{max}}$. 
As a prescription for the energy cut-off we use here a fraction of the mass of
the decaying particle, namely
$E_{\gamma}\leq E_{\rm{soft}}^{\rm{max}}=0.1m_{\cf}$. All
photons below this energy are considered as soft so that they are described by
the soft photon factor $\delta_{\text{SB}}$ multiplying the tree level
result,
\begin{align}
 \Gamma_{\text{\text{SB}}} = \delta_{\text{SB}}\,\Gamma^{\text{tree}}.
\end{align}
We use the result for $\delta_{\text{SB}}$ of Ref.\,\cite{Denner:1991kt} implemented in \texttt{FormCalc}\,\cite{Hahn:1998yk, Hahn:1999wr,
Hahn:2000jm, Hahn:2006qw, Hahn:2006zy}. More details on the separation of soft and hard, collinear and non-collinear QED corrections for this process can be found in Ref.\,\cite{Drees:2006um}.

\subsubsection{Self-energies involving mixing of neutral bosons}
\begin{figure}[ht!]
 \begin{center}
  \includegraphics[width=4.0cm]{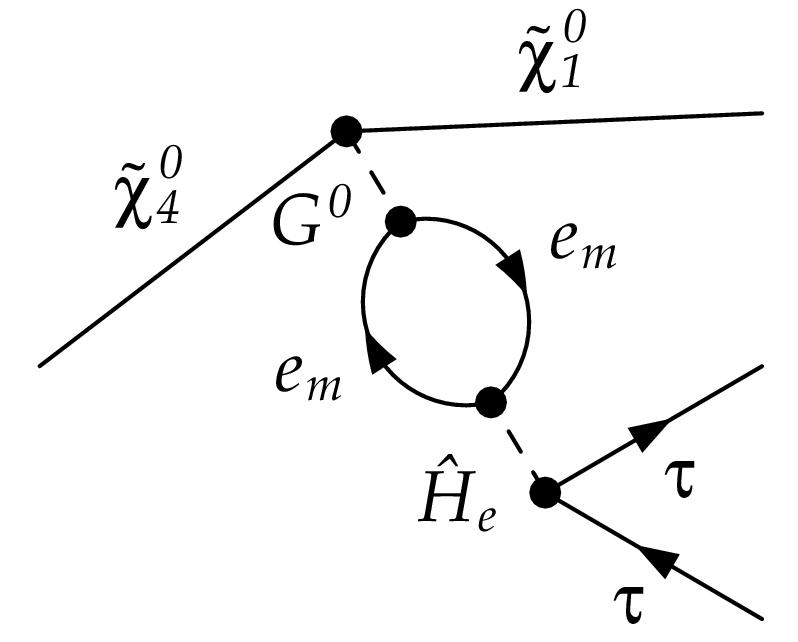}
  \includegraphics[width=4.0cm]{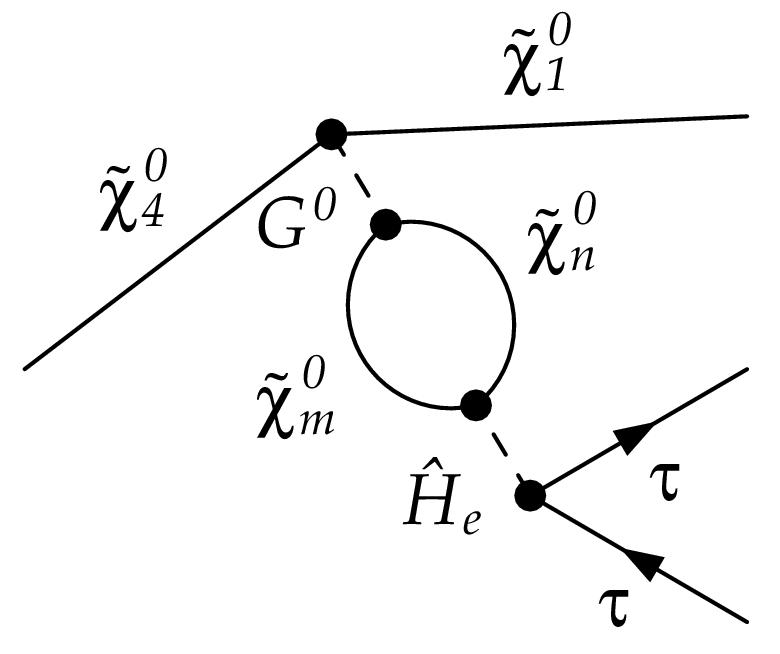}
  \includegraphics[width=4.0cm]{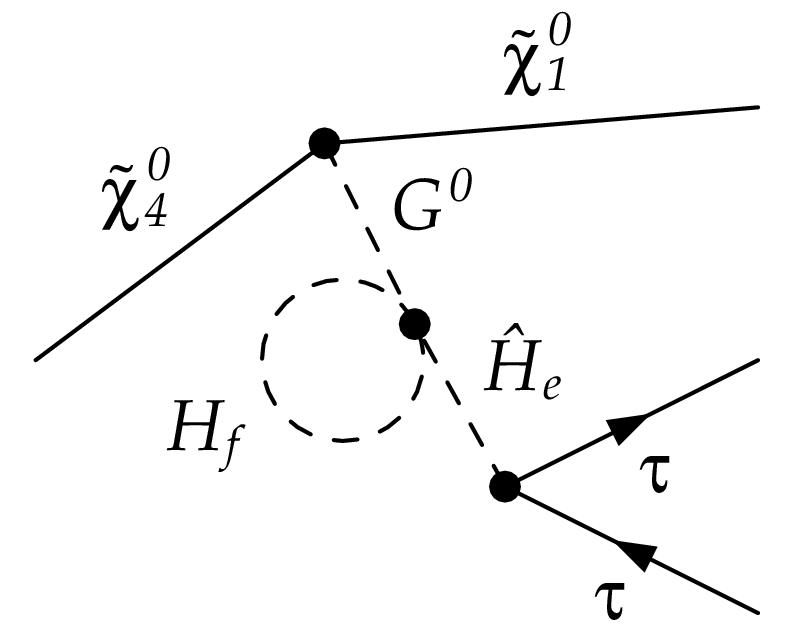}
  \includegraphics[width=4.0cm]{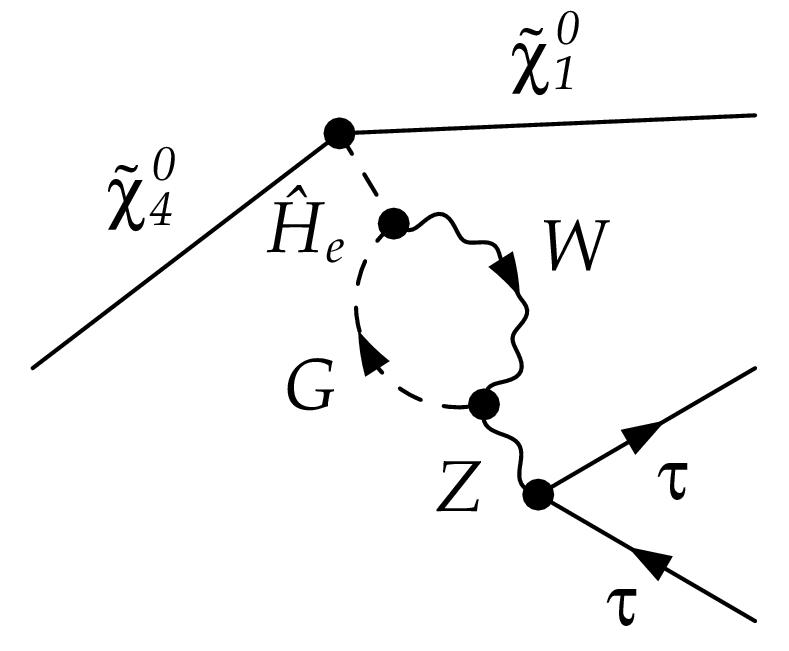}
\caption{Example self-energy diagrams contributing to the 3-body decay
$\cf\rightarrow\co\tau^{+}\tau^{-}$ with 1-loop corrections to the Higgs
propagator which mixes with the neutral Goldstone boson $G$ and the
$Z$-boson.
As in Fig.\,\ref{fig:Vert13body}, $\hat{H}_e$ denotes
a $\Zbf$-mixed neutral Higgs boson and $H_f$ an internal Higgs boson.}
\label{fig:SE3body}
 \end{center}
\end{figure}
The diagrams with self-energy corrections of the intermediate (``external'')
Higgs boson $\hat{H}_e$ are
classified in two categories. On the one hand, there are the mixing
contributions between the three neutral Higgs bosons (reduced to
$2\times2$ mixing in case of real MSSM parameters). They are approximated by
the $\hat{\mathbf{Z}}$-factors, which were checked to accurately reproduce the
full Higgs propagator mixing close to the complex pole (see
Sect.\,\ref{sect:Hmix} and Refs.\,\cite{Fowler:2010eba,HiggsMix:InPrep}). Consequently, no explicit propagator corrections with  Higgs self-energies are included. With the $\hat{\mathbf{Z}}$-factors, the strict one-loop order is extended to take more precise mixing effects in the Higgs 
sector into account. On the other hand, the $\hat{\textbf{Z}}$-factors do not
contain mixing with other neutral particles. Hence, the propagator corrections
of a Higgs with the neutral Goldstone boson $G$ and the $Z$-boson 
are calculated explicitly. Some example diagrams are shown in
Fig.\,\ref{fig:SE3body}. However, in case of $\CP$-conservation, the mixing
between $h/H$ and $G/Z$ vanishes.

\subsubsection{Box diagrams}
\begin{figure}[ht!]
 \begin{center}
  \includegraphics[width=3.2cm]{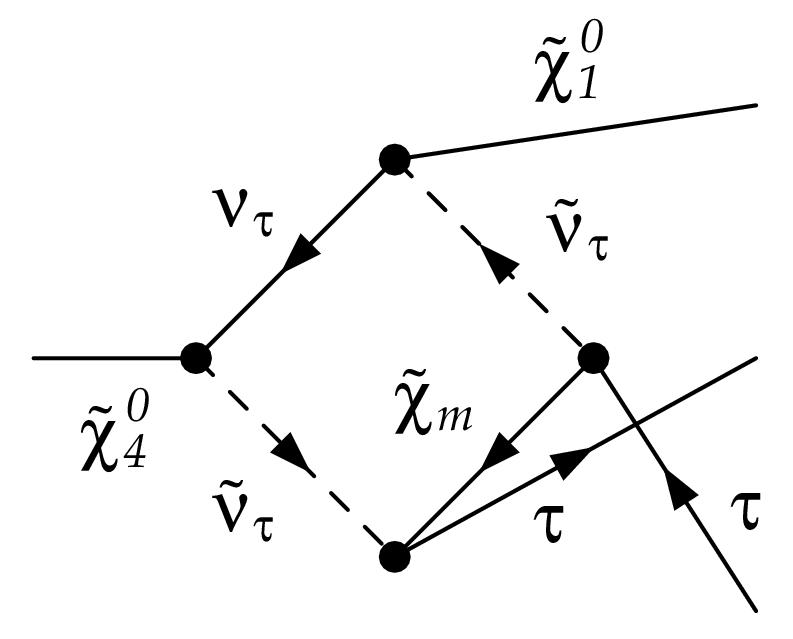}
  \includegraphics[width=3.2cm]{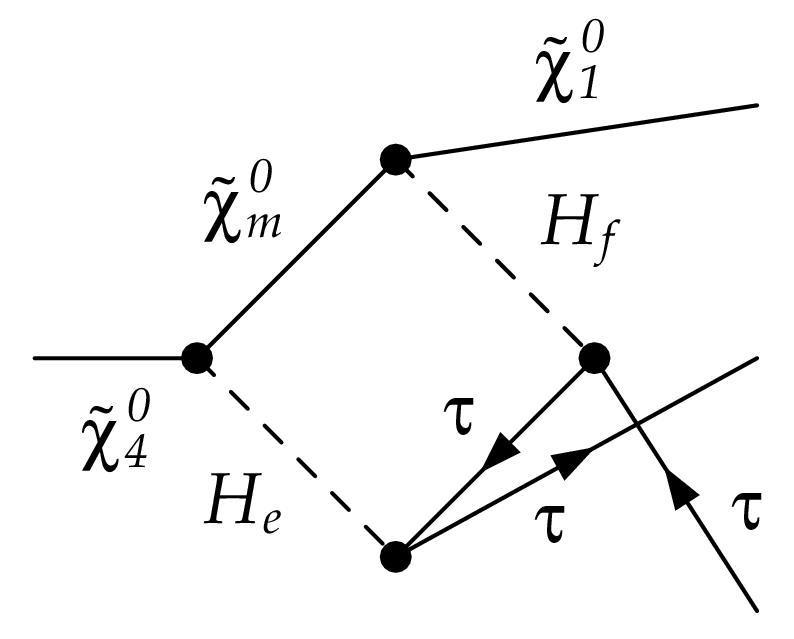}
  \includegraphics[width=3.2cm]{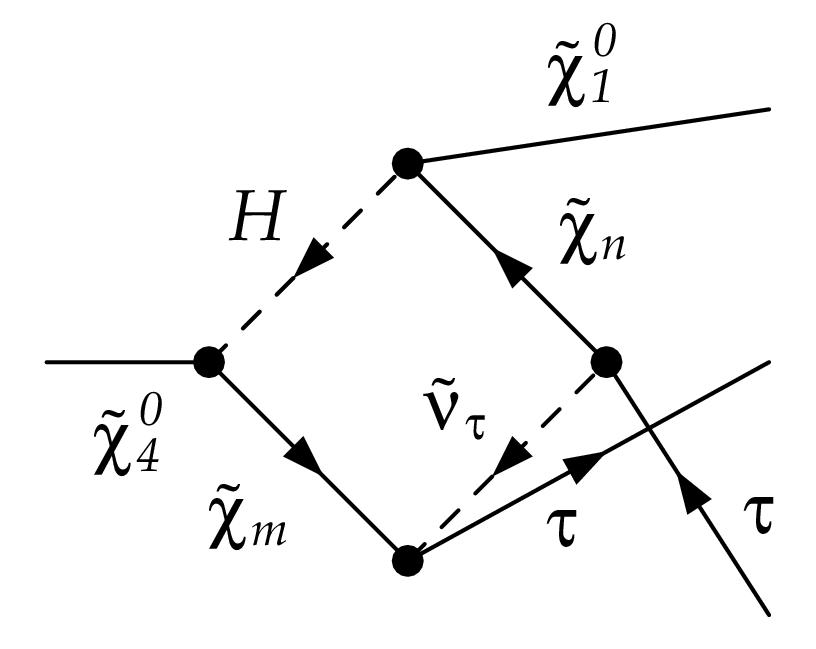}
  \includegraphics[width=3.2cm]{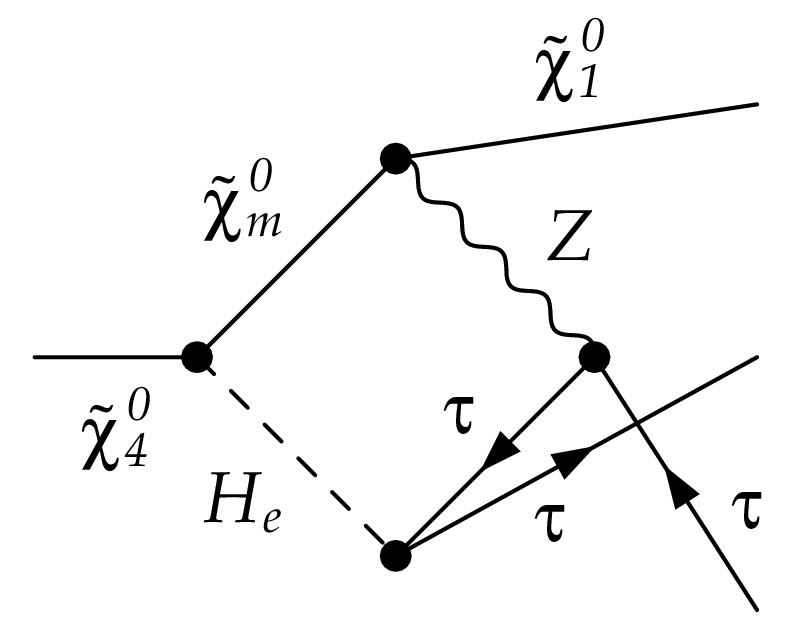}
  \includegraphics[width=3.2cm]{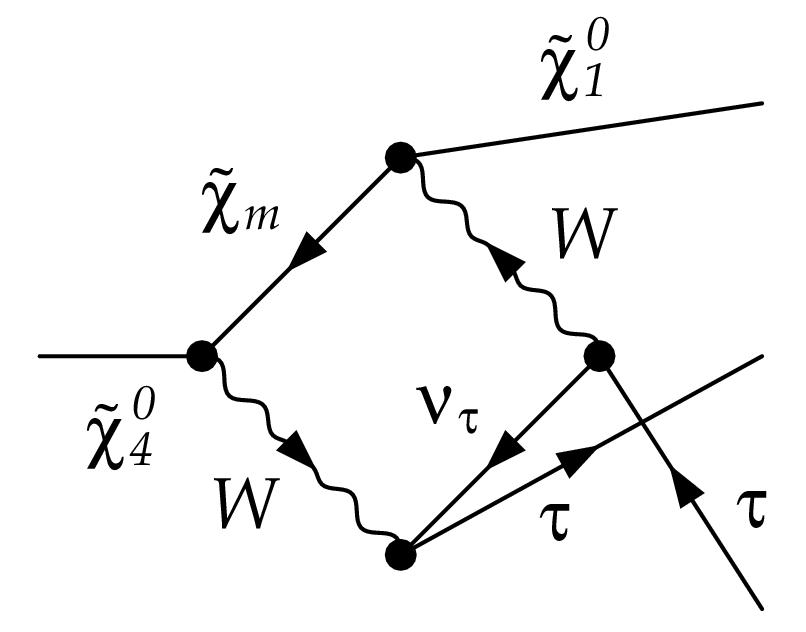}
\caption{Example box diagrams of the 3-body decay
$\cf\rightarrow\co\tau^{+}\tau^{-}$ (with and without Higgs bosons), 
where the particles are labelled as is
Fig.\,\ref{fig:Vert13body}. Only internal Higgs bosons $H_f$ appear in the
loop.}
\label{fig:Box}
 \end{center}
\end{figure}
\noindent
Finally, the $\cf$ cannot only decay into $\co\tau^{+}\tau^{-}$ via a resonant Higgs boson, but also through box diagrams. Fig.\,\ref{fig:Box} depicts some example diagrams with and without Higgs bosons. No counterterms are necessary because the boxes are UV-finite by themselves. The box diagrams are explicitly calculated including the full MSSM spectrum in the loops, but, as expected, those non-resonant contributions are found to be numerically suppressed. This is important for the comparison with the gNWA at the 1-loop level in Sect.\,\ref{sect:gNWAM1num} since the boxes cannot be factorised.

\subsection{Comparison of the tree level and 1-loop result}
\begin{figure}[ht!]
 \begin{center}
  \includegraphics[width=12cm]{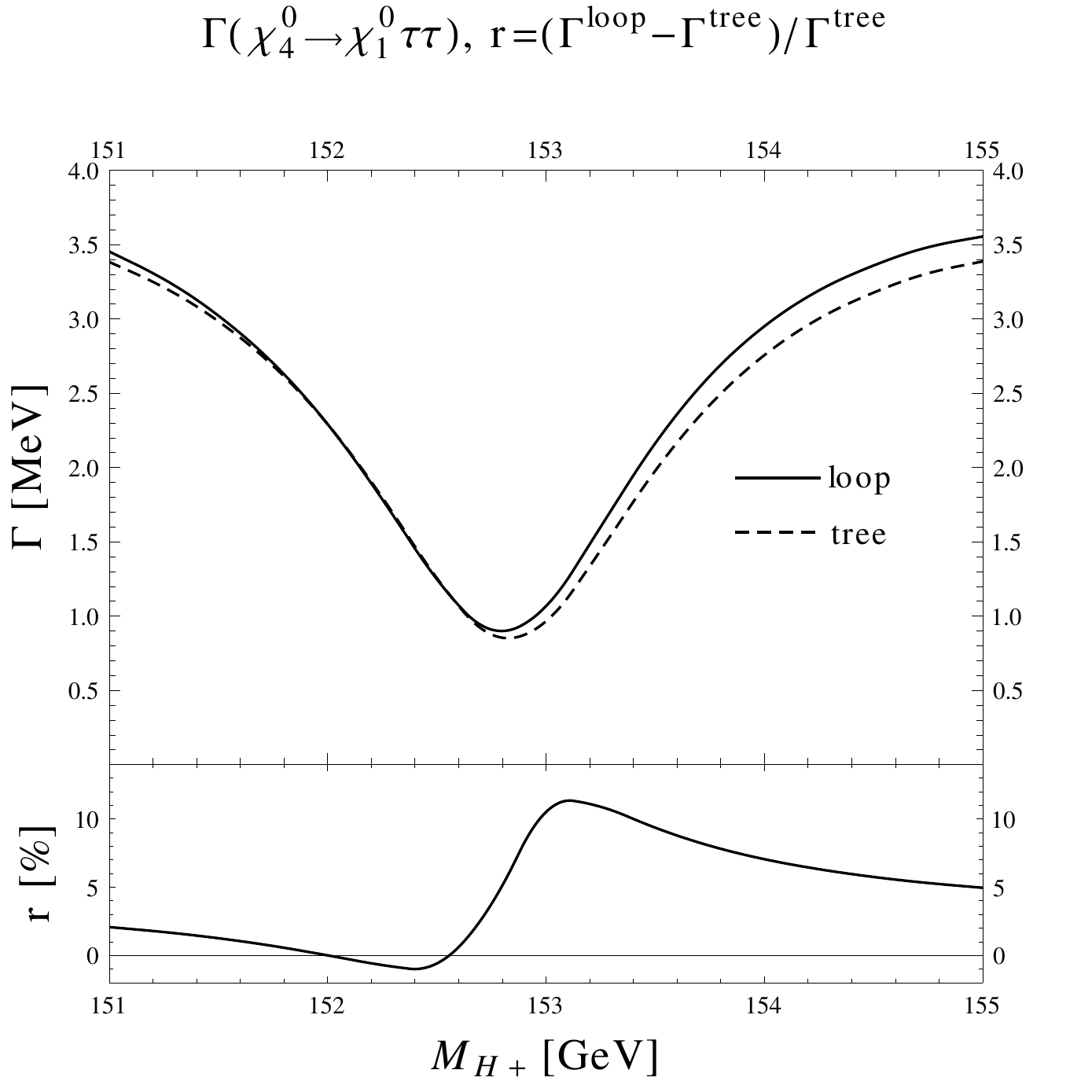}
\caption{The 1$\rightarrow$3 decay width $\Gamma(\cf \rightarrow \co \tau^{+}\tau^{-})$. \textbf{Upper panel:} Tree-level mediated by resonant $h, H$ including their interference (dashed) and full 1-loop result with vertex, soft photon and propagator corrections to the resonant $h,H$-exchange and, in addition, non-resonant box contributions (solid), both supplemented by higher-order Higgs masses, total widths and $\Zbf$-factors. \textbf{Lower panel:} Relative loop contribution $r=(\Gamma^{\text{loop}}-\Gamma^{\text{tree}})/\Gamma^{\text{tree}}$ in percent.}
\label{fig:rel}
 \end{center}
\end{figure}

Fig.\,\ref{fig:rel} shows the resulting decay width of $\cf$ into $\co$ and a
$\tau^{+}\tau^{-}$-pair as the full 3-body decay. As mentioned in
Sect.\,\ref{sect:ResultTree}, the $Z$-, $A$-, $G$- and slepton-exchange is not
included in this section, but the interference between all other contributions
to the
3-body decay is taken into account. The tree-level and 1-loop results are
based on the product of $\Zbf$-factors and Breit-Wigner propagators with
higher-order Higgs masses, total widths and $\Zbf$-factors. Despite being an
approximation of the complete Higgs propagator mixing, see
Eq.\,(\ref{eq:ZBW}), it is here referred to it as the full result that will
consistently serve as a reference for the validation of the gNWA at the 1-loop
level.

The full 1-loop decay width includes the vertex corrections at the production
and the decay vertex and box contributions as well as self-energy corrections
to the propagator and bremsstrahlung off the $\tau$-leptons in the final
state.
The NLO decay width (solid) is enhanced relative to the LO result (dashed) in most of the analysed parameter interval, up to $11\%$, as the plot of the ratio $r=(\Gamma^{\text{loop}}-\Gamma^{\text{tree}})/\Gamma^{\text{tree}}$ shows. However, around $\mhp\simeq 152$\,GeV, the 1-loop corrections vanish.

\section{Numerical validation of the gNWA at the loop level}\label{sect:ResultLoop}
In this example, the calculation of the full process at the 1-loop level is
still manageable, where \textit{full} here means the 3-body decays with
Breit-Wigner propagators and $\Zbf$-factors, though without the $Z$-, $A$- and
$G$-boson exchange. But we aim at validating the generalised narrow-width approximation at the 1-loop level so that it can be applied on kinematically more complicated processes for which the factorisation into production and decay is essential to enable the computation of higher order corrections.

Our strategy is to combine the NLO corrections for the production and decay subprocesses in such a way that the gNWA prediction can be consistently compared to the full 1-loop calculation. Only the box diagrams are left out in the gNWA compared to the 3-body decays.

\subsection{2-body decays}\label{sect:ResultLoop_2body}
 \begin{figure}[ht!]
 \begin{center}
  \subfigure[Higgs production]{\includegraphics[width=3.2cm]{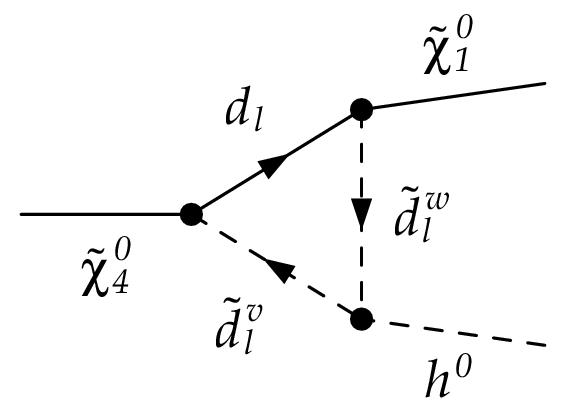}
  \includegraphics[width=3.2cm]{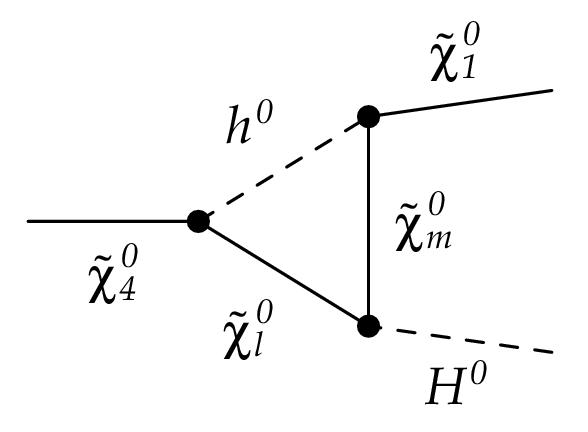}\label{fig:Hprod}}
  \subfigure[Higgs decay.]{\includegraphics[width=3.2cm]{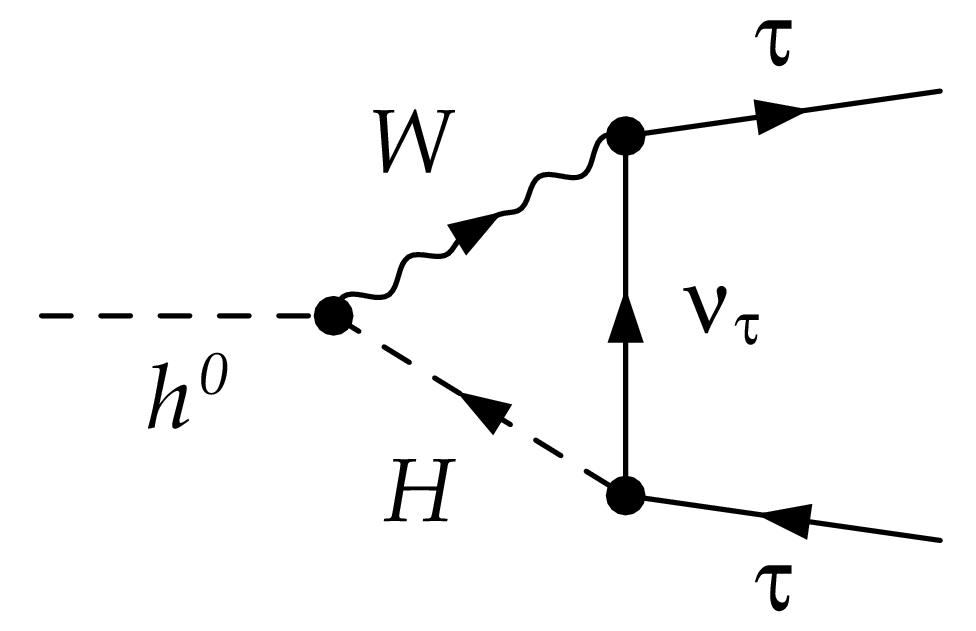}
  \includegraphics[width=3.2cm]{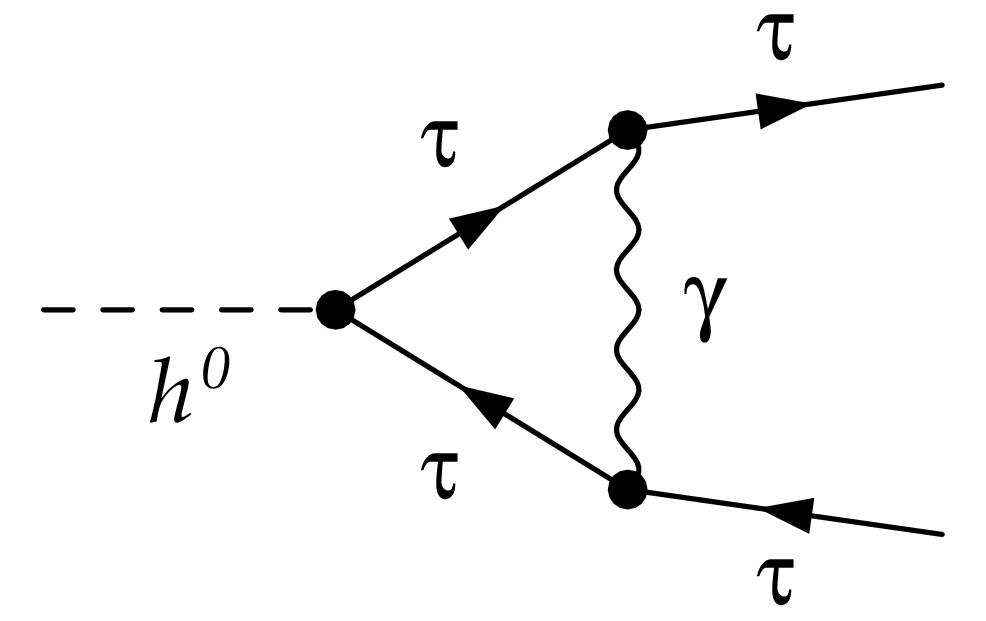}
  \includegraphics[width=3.2cm]{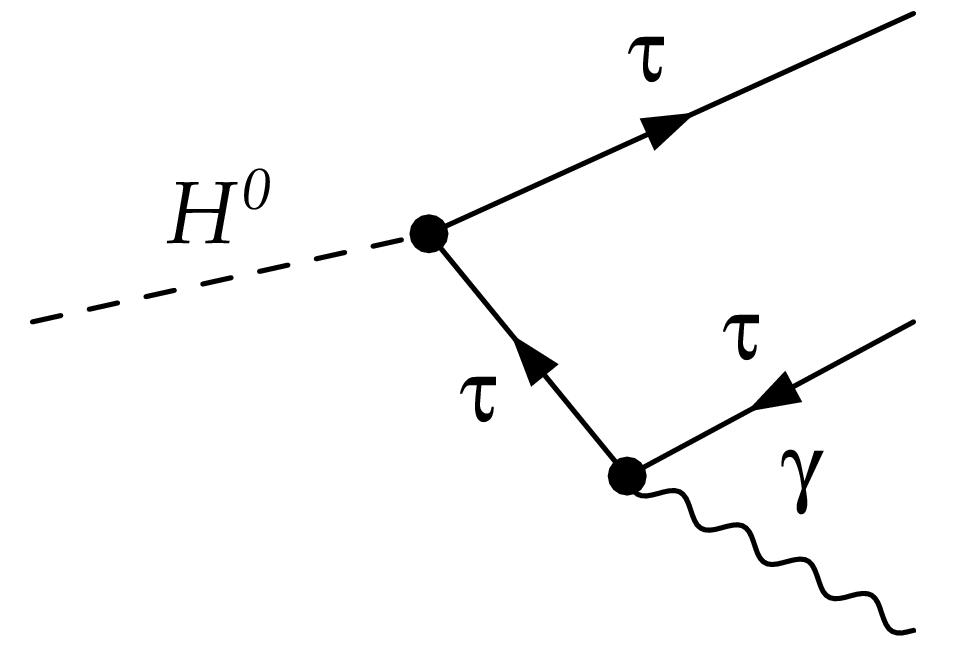}\label{fig:Hdec}}
\caption{Example diagrams of the 2-body decays for \textbf{(a)} Higgs production in $\cf\rightarrow \co h/H$ at NLO and \textbf{(b)} Higgs decay in $h/H\rightarrow \tau^{+}\tau^{-}$ at NLO with virtual and real corrections.}
\label{fig:Vert2body}
\end{center}
\end{figure}
\begin{figure}[ht!]
 \begin{center}
  \subfigure[Higgs production in $\cf\rightarrow \co h/H$.]{\includegraphics[height=4.6cm]{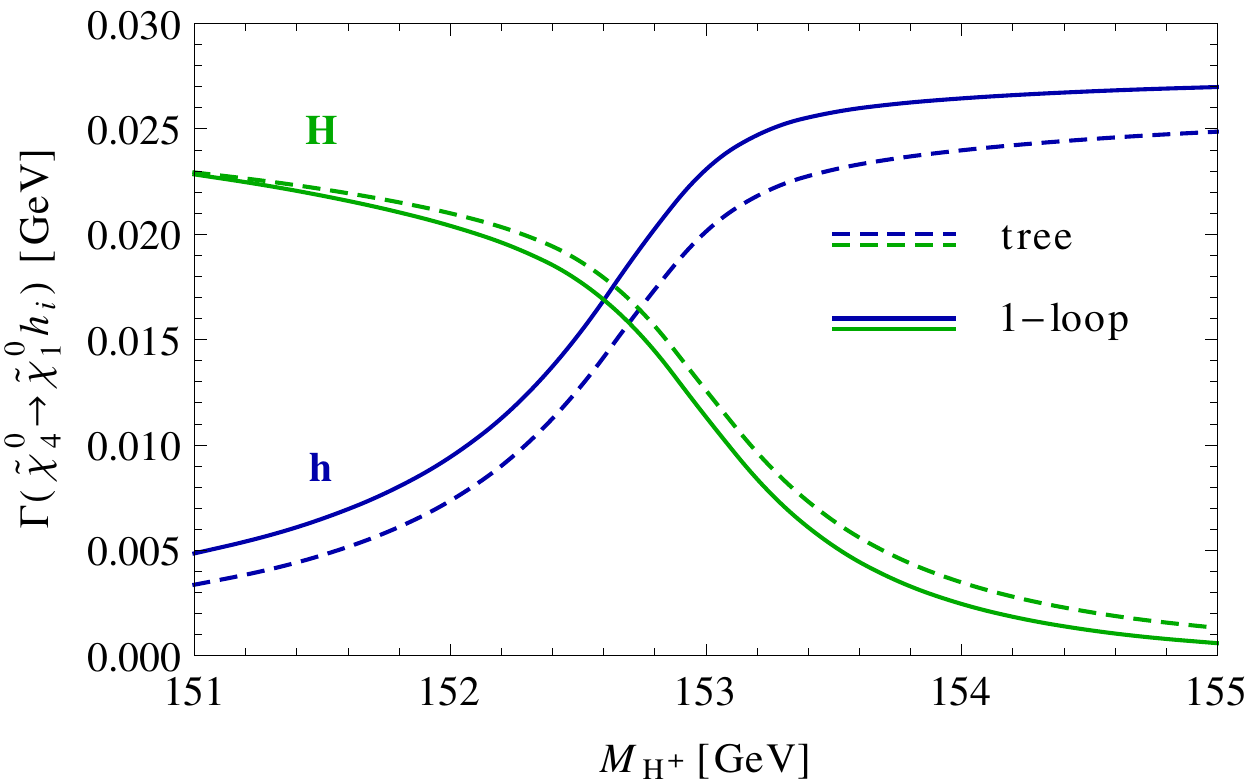}\label{fig:prod}}
  \subfigure[Relative loop contribution in $\cf\rightarrow \co h/H$.]{\includegraphics[height=4.5cm]{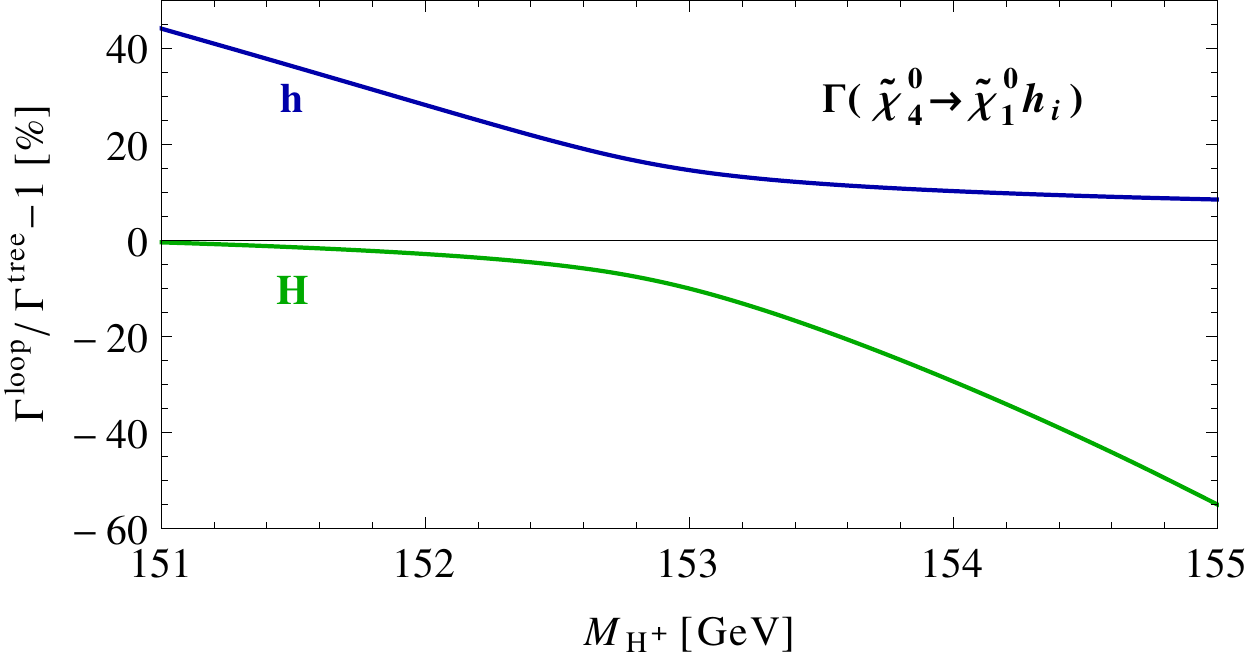}\label{fig:prodrel}}
  \subfigure[Higgs decay $h/H\rightarrow \tau^{+}\tau^{-}$.]{\includegraphics[height=4.6cm]{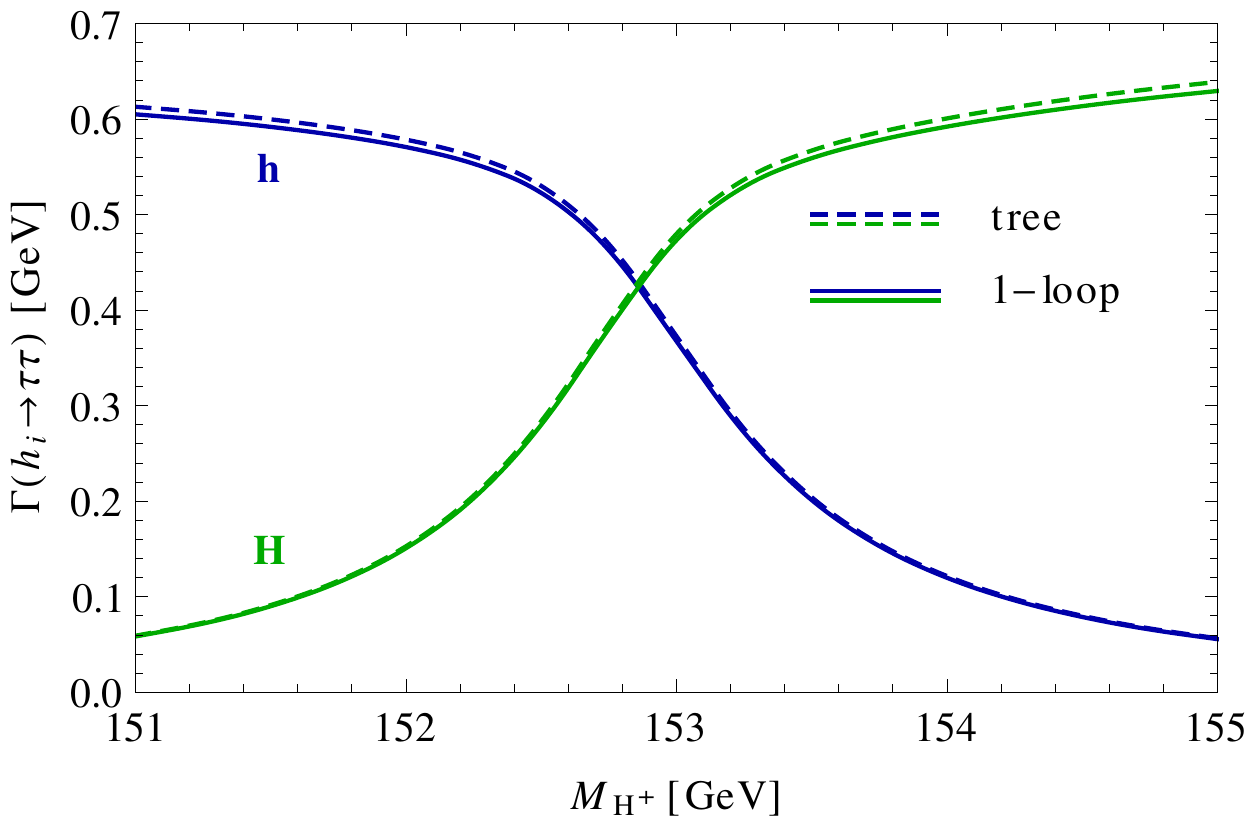}\label{fig:dec}}
  \subfigure[Relative loop contribution in $h/H\rightarrow \tau^{+}\tau^{-}$.]{\includegraphics[height=4.5cm]{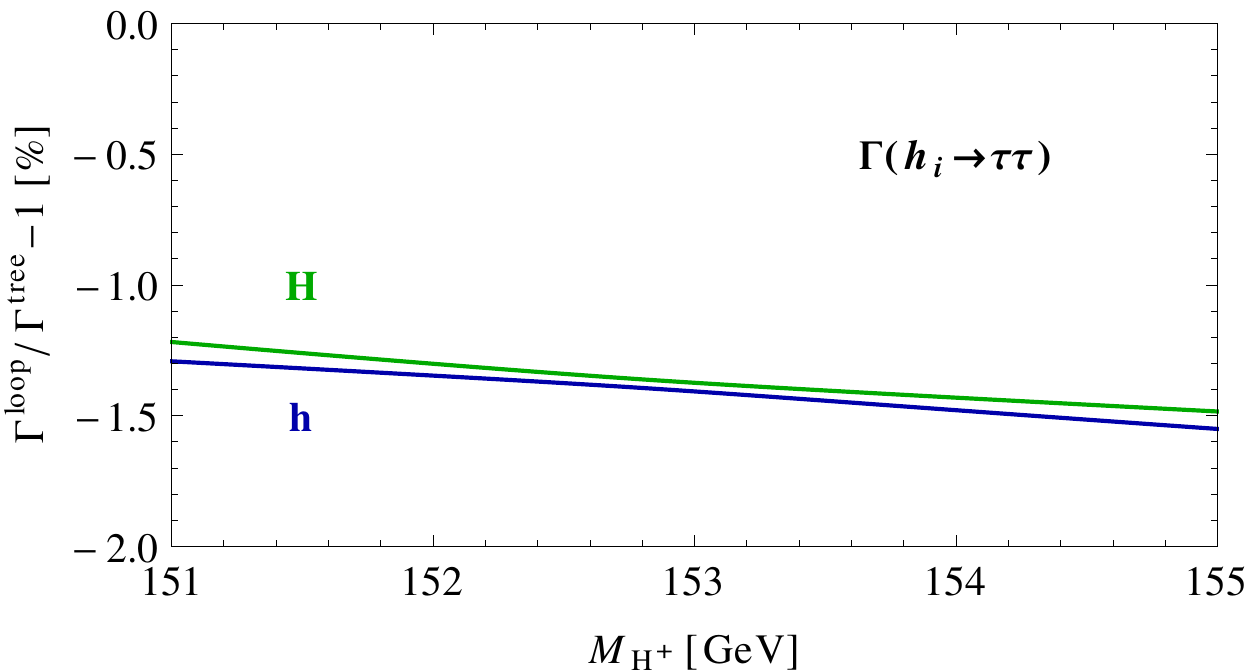}\label{fig:decrel}}
\caption{2-body decay widths of \textbf{(a)} $\cf\rightarrow \co h_i$ and
\textbf{(c)} $h_i\rightarrow\tau^{+}\tau^{-}$ with $h_i=h$ (blue) and $H$
(green) at the tree level (dashed) or at the 1-loop level (solid), and the
relative effect of the loop contributions \textbf{(b)}, \textbf{(d)}.}
\label{fig:2body}
 \end{center}
\end{figure}

The gNWA at NLO requires the 1-loop contributions to the 2-body decays as
subprocesses. For the production, we calculate the full 1-loop corrections to
$\Gamma(\cf\rightarrow \co h/H)$ in the NNN on-shell renormalisation scheme,
see Refs.\,\cite{Fowler:2009ay,Bharucha:2012nx,Bharucha:2012re}, with the same
choice of on-shell states as in the 3-body-decay described in
Sect.\,\ref{sect:VirtNeutralino}. Higgs mixing is taken into account by
$\hat{\textbf{Z}}$-factors, but mixing with $G$-/$Z$-bosons 
is generated explicitly, which, however, vanishes in this
$\mathcal{CP}$-conserving scenario. Some example diagrams for vertex
corrections are shown in Fig.\,\ref{fig:Hprod}. Fig.\,\ref{fig:prod} presents
the resulting 2-body decay widths for the production of $h$ (blue) and $H$
(green) at the tree level (dashed) and the 1-loop level (solid). While the
1-loop corrections increase $\Gamma(\cf\rightarrow\co h)$, they decrease the
production of $H$ from the decay of $\cf$. The substantial relative effect 
can be seen in Fig.\,\ref{fig:prodrel}.

For the decay, the full vertex corrections to $h_i\rightarrow \tau^{+}\tau^{-}$ are included. Furthermore, real soft photon emission off the $\tau$-leptons in the final state is included. In order to allow for a meaningful comparison between the gNWA and the full calculation, the energy cut-off is defined by the same value $E_{\rm{soft}}^{\rm{max}}=0.1m_{\cf}$ as in the 3-body decay. Example diagrams are displayed in Fig.\,\ref{fig:Hdec}, where the first diagram belongs to the IR-finite ones, but the second and third diagrams are IR-divergent. The emission of a real photon is not directly calculated as a 3-body decay, but still with the 2-body phase space in the soft-photon approximation. The numerical influence of the corrections of $\mathcal{O}(\alpha)$ on $\Gamma(h_i\rightarrow \tau^{+}\tau^{-})$ is shown in Fig.\,\ref{fig:dec}. The 1-loop and real corrections slightly decrease both decay rates (for $h_i=h,H$) by $1.2\%$ to $1.5\%$ as displayed in Fig.\,\ref{fig:decrel}.
\subsection{On-shell matrix elements and R-factor approximation}\label{sect:gNWAM1num}
The on-shell factorisation of the interference term has already been applied
at the leading order in Sect.\,\ref{sect:ResultTree}. In this section, we will
investigate its accuracy at the next-to-leading order. Since a wide range of
processes even with many external particles can be computed at
lowest order without applying the NWA, 
we use the full leading order result of the three-body decay 
(i.e., without NWA) 
and add the
1-loop contribution for which we use the gNWA. With this procedure, we apply
the on-shell approximation only when necessary without introducing an
avoidable uncertainty at the tree level\footnote{As a further step, one could
split the real photon contribution into IR-singular and finite terms and apply
the NWA only on the singular ones according to
Refs.\,\cite{Grunewald:2000ju,Denner:2000bj}.}.

\begin{figure}[ht!]
 \begin{center}
  \includegraphics[width=13cm]{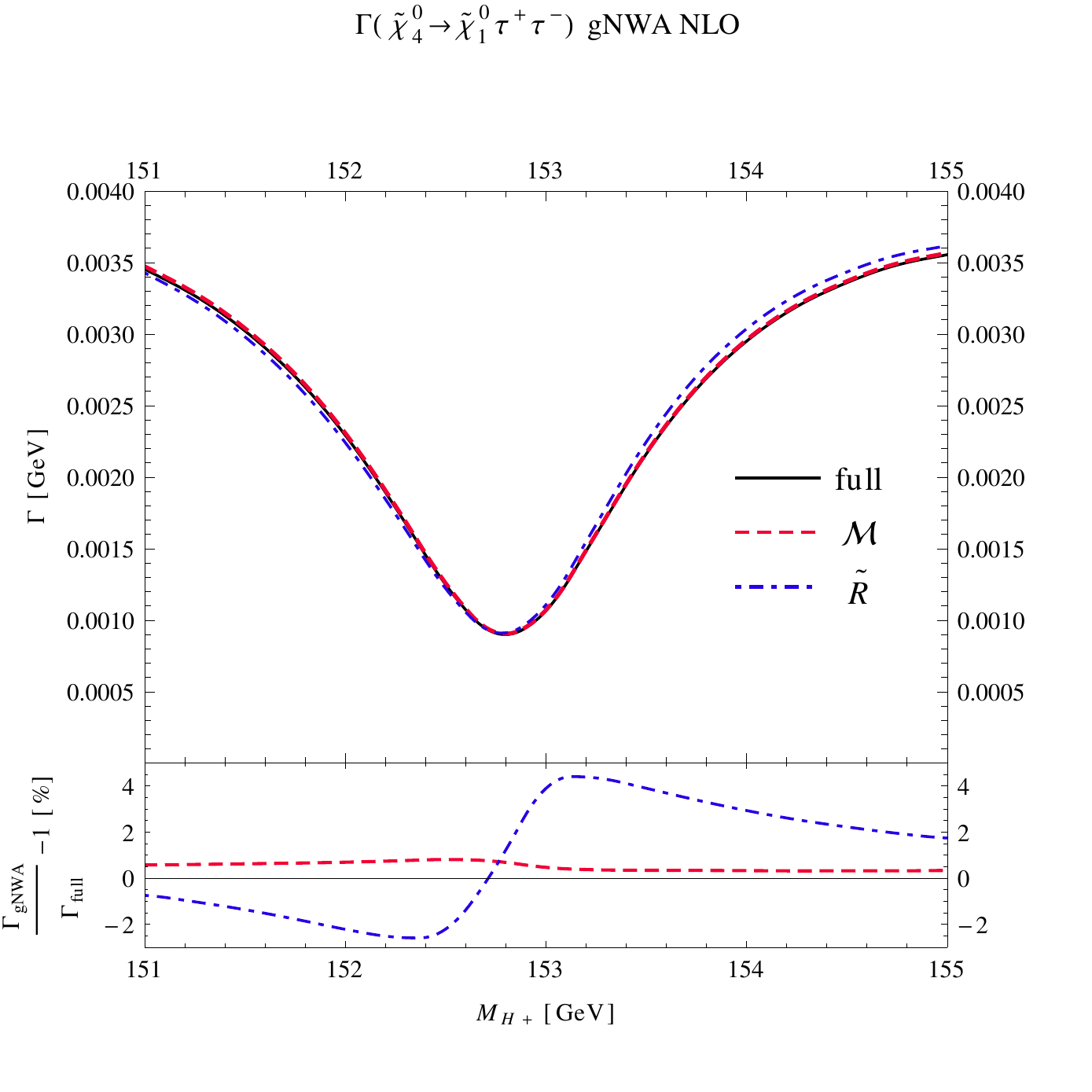}
\caption{\textbf{Upper panel:} The decay width
$\cf\rightarrow\co\tau^{+}\tau^{-}$ at the 1-loop level with resonant
$h,H$-exchange and, for the full 3-body decay (black, solid), with box
contributions. The gNWA with on-shell matrix elements is denoted by $\Mm$ (red,
dashed), and the gNWA with interference weight factors is denoted by $\tilde{R}$
(blue, dash-dotted). \textbf{Lower panel:} The relative deviation of the gNWA
(matrix element and R-factor approximation) from the full 1-loop result in
percent.}
\label{fig:gNWAMR1_abs_rel}
 \end{center}
\end{figure}

In Fig.\,\ref{fig:gNWAMR1_abs_rel}, we compare the numerical results of the
method of on-shell matrix elements using Eqs.\,(\ref{eq:M1strict}) and
(\ref{eq:Mloop}), denoted by $\Mm$, and of the interference weight factor
approximation from Eq.\,(\ref{eq:int1R0}), denoted by $\tilde{R}$, with the
full 1-loop result as calculated in Sect.\,\ref{sect:ResultLoop}. The upper
panel shows the prediction of the partial width $\Gamma(\cf\rightarrow \co
\tau^{+}\tau^{-})$. The lines of the gNWA based on matrix elements (red,
dashed) and the full 1-loop calculation (black, solid) lie nearly on top of
one another. Also the additional $\tilde{R}$-factor approximation (blue,
dash-dotted) yields a good qualitative agreement with the full result, but
less accurate than achieved by the on-shell matrix elements. The lower panel
visualises the relative deviation of the decay width predicted by the two
versions of the gNWA from the full result. As expected, the R-factor method
reproduces the full result best where the 
difference between $M_h$ and $M_H$ is smallest, i.e., in the centre of the
analysed parameter interval. But the assumption of equal masses 
becomes worse away from the centre of the analysed interval,
leading to a deviation from the
full 1-loop result of up to $4.5\%$. 
Thus, for those parameters the matrix
element method performs clearly better within an accuracy of better than
$1\%$.

\begin{figure}[ht!]
 \begin{center}
  \includegraphics[width=12cm]{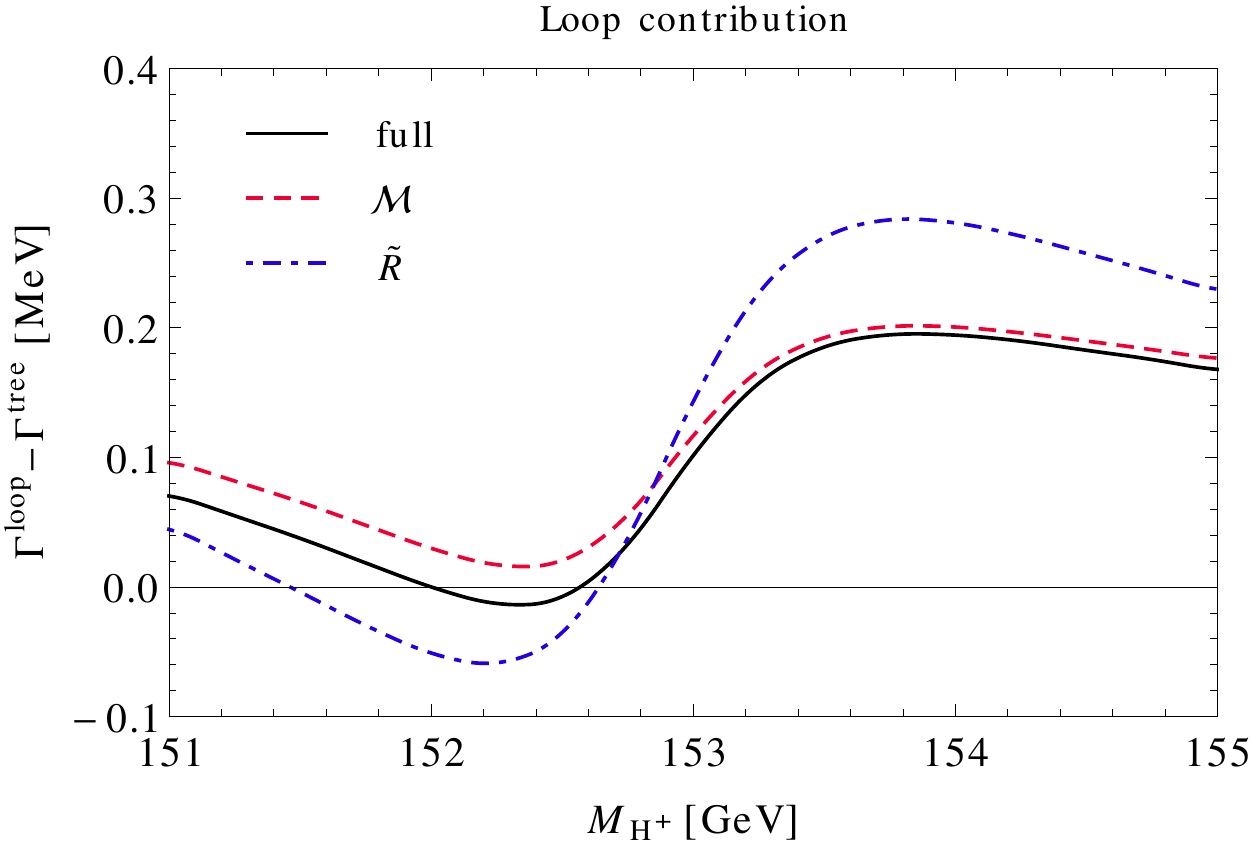}
\caption{Pure loop contributions in the full calculation (black, solid) and approximated by the gNWA using the matrix element method denoted by $\Mm$ (red, dashed) and using the R-factor approximation denoted by $\tilde{R}$ (blue, dash-dotted).}
\label{fig:TreeMinusLoop}
 \end{center}
\end{figure}

In order to further investigate how well the gNWA predicts the interference
term at the 1-loop level, we take a closer look in
Fig.\,\ref{fig:TreeMinusLoop} at the pure loop contribution
$\Gamma^{\text{loop}}-\Gamma^{\text{tree}}$ of the full three-body decay
(black, solid), the gNWA using on-shell matrix elements (red, dashed, denoted
by $\Mm$) and the $\tilde{R}$-factor approximation (blue, dash-dotted, denoted
by $\tilde{R}$). While at the tree level we found that both versions of the
gNWA work comparably well (see Fig.\,\ref{fig:sgNWA_tree}), the $\Mm$-method
provides a significantly better prediction of the interference term at the
1-loop level.
\begin{figure}[ht!]
 \begin{center}
  \includegraphics[width=12cm]{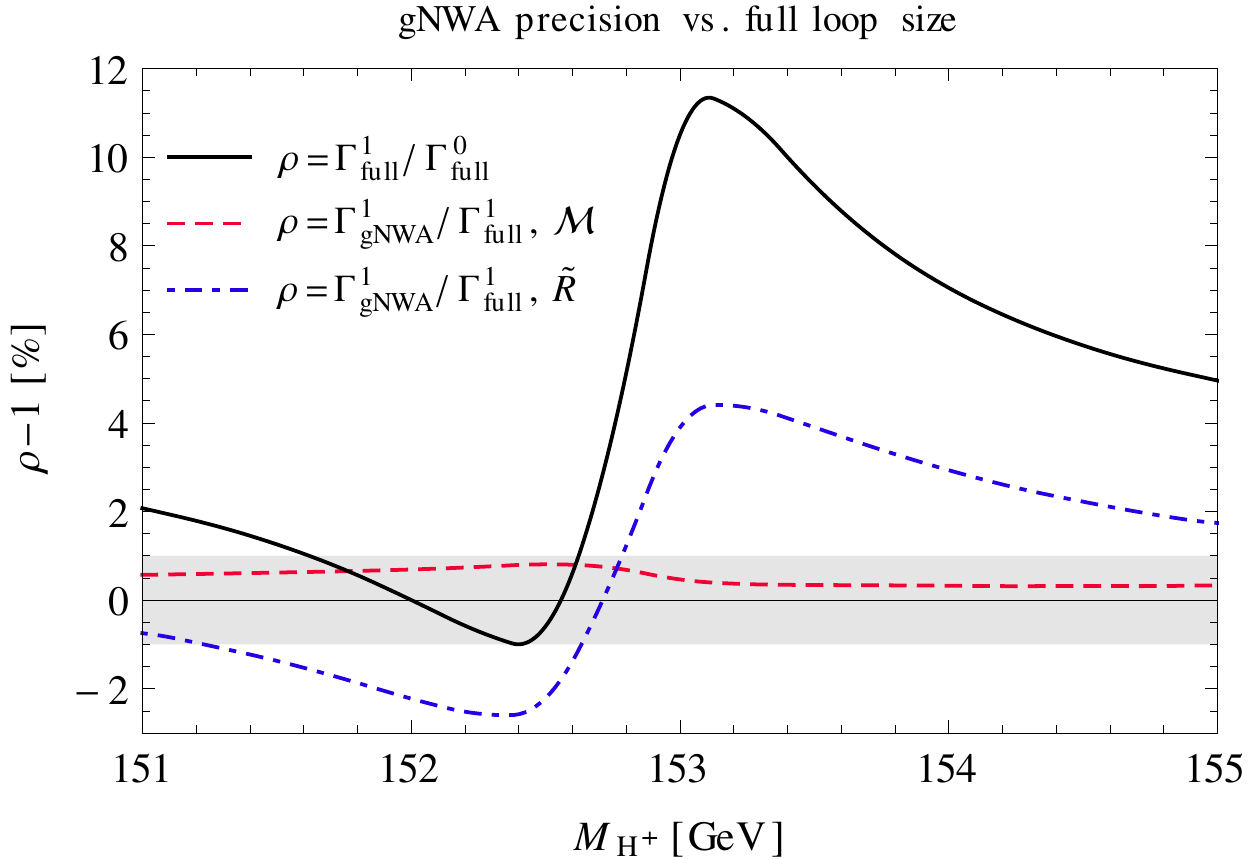}
\caption{Precision of the gNWA at the 1-loop level using the matrix element method denoted by $\Mm$ (red, dashed) and using the R-factor approximation denoted by $\tilde{R}$ (blue, dash-dotted) compared to the relative size of the loop contribution in the full calculation (black). The $\pm1\%$ region is indicated in grey.}
\label{fig:gNWAMR1_prec_loopsize}
 \end{center}
\end{figure}

When the gNWA is used to approximate one-loop effects, we need to compare the
accuracy of the approximation with the overall size of the loop correction. 
Fig.\,\ref{fig:gNWAMR1_prec_loopsize} provides a
comparison between the precision of the gNWA with respect to the full
calculation (for on-shell matrix elements denoted by $\Mm$ in red and the
R-factor approximation denoted by $\tilde{R}$ in blue) and the relative size
of the 1-loop correction to the 3-body decay width in black. While the loop
correction ranges from $-1\%$ to $11\%$ in this example case, 
the deviation of the matrix element method from the full result 
remains below $1\%$. The uncertainty of this approximation is therefore
significantly smaller than the typical size of the loop correction in this
case. The deviation of the R-factor approximation 
from the full result is found to be larger, within $-3\%$ to $4.5\%$ in this
case, but it is still about a factor of two smaller than the size of the loop
correction in the region where the latter is sizable.

The plot shows that the overall performance of the gNWA with the
$\mathcal{M}$-method is good except for the region around $\mhp\simeq
152$\,GeV--\,152.5\,GeV where the $\mathcal{M}$-method uncertainty exceeds the
relative size of the full loop correction slightly. But here the full loop
correction is in fact very small.
Keeping in mind that the full calculation is
subject to uncertainties itself (e.g.\ from missing higher-order corrections)
which might reach the level of $1\%$ 
(for illustration, the $\pm 1\%$ range is indicated in the plot), the
$\mathcal{M}$-method can be regarded as adequate to approximate loop
corrections to the interference term within the expected uncertainty of the
full result (as long as non-factorisable corrections remain numerically
suppressed).
On the other hand, the R-factor method gives rise to larger deviations
and should therefore be regarded as a 
simple estimate of 
the higher-order result including interference effects. 

\subsection{Separate treatment of photon contributions}\label{sect:treatphoton}

\begin{figure}[htb!]
\begin{center}
  \includegraphics[width=10cm]{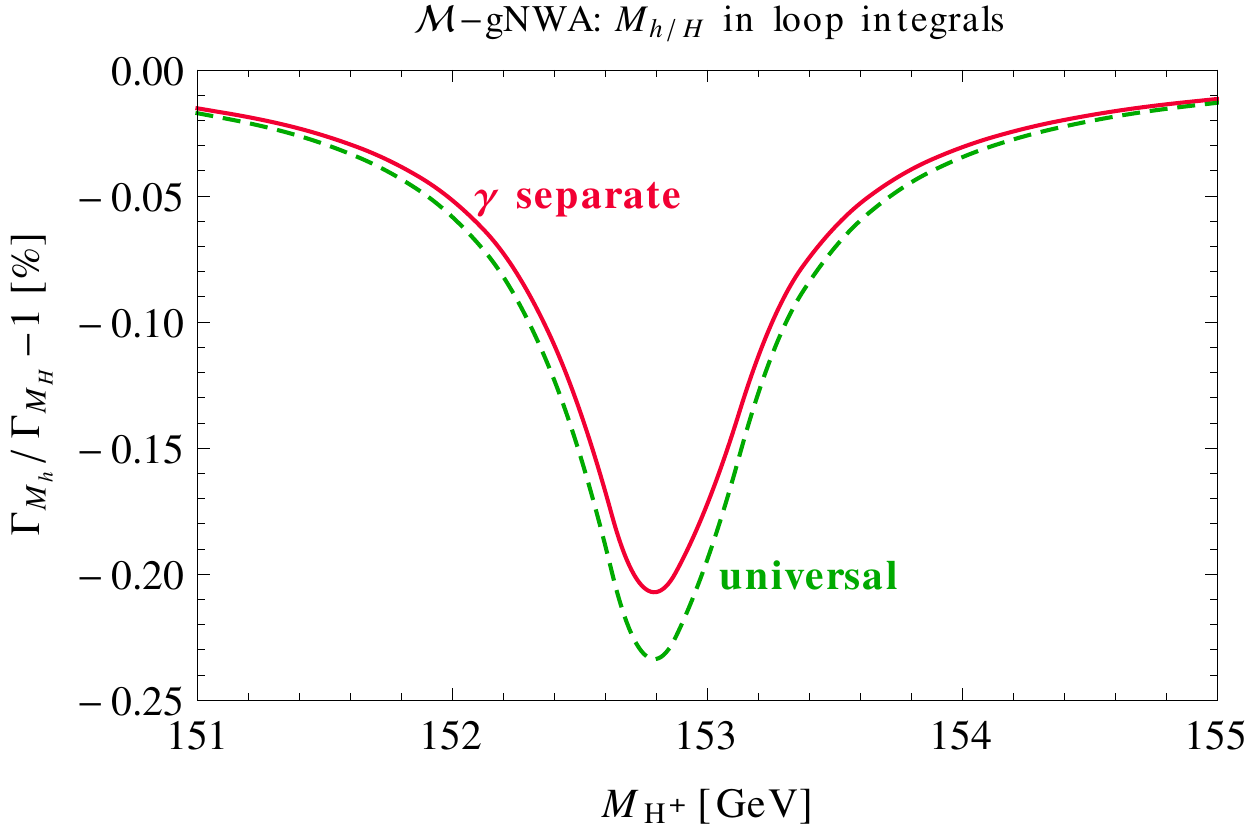}
\caption{Impact of the dependence of the gNWA on the choice of the mass
$\overline{M}$ (see text). 
The relative deviation between $\Gamma_{M_h}$ and
$\Gamma_{M_H}$, where $\Gamma_{M_i}\equiv
\Gamma_{\rm{gNWA}}^{\Mm}(\overline{M}^{2}=M_i^{2})$,
is shown for the universal
treatment of all one-loop matrix elements for the decay and for the case where
the photonic contribution is treated separately.}
\label{fig:SpecialhH}
\end{center}
 \end{figure}

As discussed in Sect.\,\ref{sect:SpecialTreatment}, the factor
$\delta_{\text{SB}}$, which multiplies the squared tree level matrix element
to account for the contribution of 
soft bremsstrahlung, and the IR-divergent loop integrals
must be evaluated at the same mass to enable the cancellation of
IR-singularities between real and virtual photon contributions. In order
to reduce the
ambiguity whether to choose the common mass $\overline{M}=M_h$ or $M_H$, the
IR-finite diagrams can be evaluated at their correct mass shell.
Fig.\,\ref{fig:SpecialhH} compares the dependence of the gNWA result on the
ambiguous mass choice, i.e., the relative deviation between
$\Gamma_{\rm{gNWA}}(\overline{M}=M_h)$ and
$\Gamma_{\rm{gNWA}}(\overline{M}=M_H)$, for the matrix element method. 
The dashed green line represents the
universal treatment where the loop integrals in all decay one-loop matrix
elements are evaluated at $\overline{M}^{2}$ whereas the solid red line shows
the separate calculation of the photonic contribution 
as described in Sect.\,\ref{sect:SpecialTreatment}.
The impact of the dependence of the gNWA 
on the choice of the mass $\overline{M}$ is found
to be rather small, giving rise to a maximum deviation of $0.23\%$ for the universal
treatment of all one-loop matrix elements for the decay. Restricting this
approximation just to the photonic contribution is seen to have an
insignificant effect in this example, reducing the deviation to $0.2\%$.

\subsection{gNWA prediction with most precise input values}\label{sect:best}

As a first step, we defined the gNWA at the 1-loop order for a consistent comparison 
between the gNWA and the full 1-loop calculation. As an exception, the Higgs masses, 
total widths and wave function normalisation
factors $\hat{\textbf{Z}}$ have been obtained from 
\texttt{FeynHiggs}\,\cite{Heinemeyer:1998np, Heinemeyer:1998yj,
Degrassi:2002fi, Heinemeyer:2007aq} 
at the 2-loop order and used both in the gNWA and the full calculation. 
In this section we want to exploit the factorisation and include all components 
at the highest available precision. 
This means for the gNWA with the on-shell matrix element method and the R-factor approximation 
that we use the calculated 1-loop production part and the \texttt{FeynHiggs} branching ratios in $\Gamma_P(\cf\rightarrow \co h_i)\cdot \text{BR}_D(h_i\rightarrow \tau^{+}\tau^{-})$. 
Furthermore, the product of on-shell matrix elements from Eq.\,(\ref{eq:Mloopsoft}) is expanded 
up to the product of 1-loop matrix elements in Eq.\,(\ref{eq:Mbest}). 
The higher-order extension of the R-factor approximation is defined in Eq.\,(\ref{eq:intbestR0}).

\begin{figure}[htb!]
 \begin{center}
 \subfigure[Including $A$, $Z$, $G$, $\tilde\tau$-exchange.]{\includegraphics[width=8.2cm]{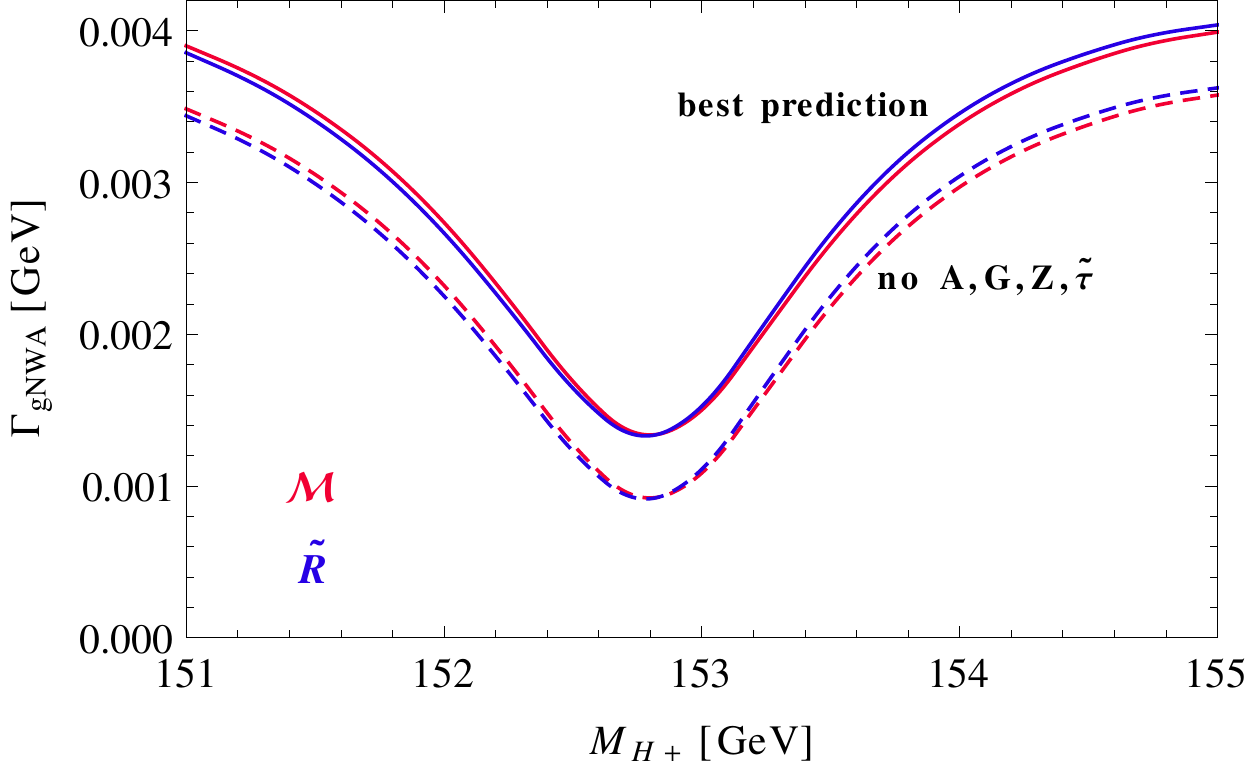}\label{fig:gNWAbestZ}}
 \subfigure[Relative effect on gNWA.]{\includegraphics[width=8.2cm]{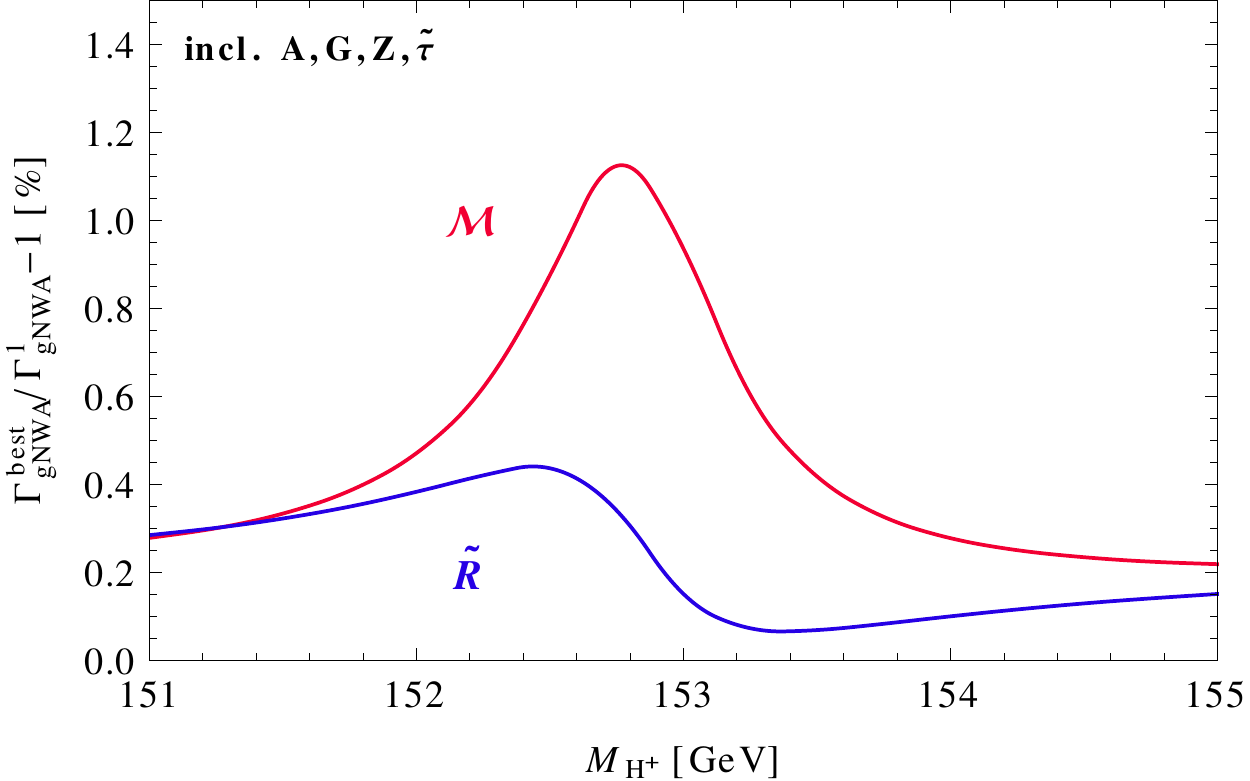}\label{fig:gNWAbestrel}}
 \caption{\textbf{(a)} The gNWA using the most accurate predictions for all
parts of the process, supplemented with a tree-level result with (solid)
and without (dashed) the additional $A,\,G,\,Z$ and 
$\tilde\tau$-exchange contributions, 
for the $\Mm$-method (red) and the $\tilde{R}$-approximation (blue).
\textbf{(b)} The relative effect of the most
precise branching ratios and the product of 1-loop terms on the prediction of the
gNWA with on-shell matrix elements (red, denoted by $\Mm$) and the R-factor
approximation (blue, denoted by $\tilde{R}$).}
\label{fig:gNWAbest}
 \end{center}
\end{figure}

So far we have neglected additional contributions that do not play a role 
in the discussion of the interference effects between contributions 
with $h$ and $H$ exchange in the decay of 
$\cf\rightarrow \co \tau^{+}\tau^{-}$ 
for the considered $\CP$-conserving scenario. In order to obtain a more 
phenomenological prediction of $\Gamma(\cf\rightarrow \co \tau^{+}\tau^{-})$
we now take into account also
the resonant exchange of the $\mathcal{CP}$-odd Higgs boson $A$, 
the neutral Goldstone boson $G$ and the $Z$-boson, as well as 
the non-resonant 3-body decay via a $\tilde\tau$.
We include the contributions from $A,\,G,\,Z$ and $\tilde\tau$-exchange
at the tree-level, while at the loop level we incorporate the most precise
gNWA result (where those additional contributions are neglected).
Fig.\,\ref{fig:gNWAbestZ} shows the prediction of the higher-order improved
gNWA, supplemented by the full tree-level contribution including $A,\,G,\,Z$ and 
$\tilde\tau$-exchange diagrams,  as solid lines using on-shell matrix elements 
(red) and the R-factor approximation (blue). The corresponding results where
the $A,\,G,\,Z$ and $\tilde\tau$-exchange contributions have been neglected 
are indicated by the
dashed lines.
The contributions from $A,\,G,\,Z$ and $\tilde\tau$ are found to yield a
non-negligible upward shift in this example. 

Fig.\,\ref{fig:gNWAbestrel} shows the impact of including the most precise branching ratios 
and the product of 1-loop matrix elements in the gNWA, denoted by $\Gamma^{\text{best}}_{\text{gNWA}}$. 
For the matrix element method (in red, denoted by $\Mm$), 
this amounts to up to $1.2\%$ relative to the 1-loop formulation used above
for the comparison with the result for the 3-body decay. 
For the R-factor approximation (in blue, denoted by $\tilde{R}$), 
the effect of up to $0.4\%$ is smaller 
because the effect on the interference term beyond the 1-loop order turns out
to be negative. 
With reference to the gNWA including only $h$ and $H$, the relative impact 
of the higher-order corrections is slightly higher 
($1.6\%$ for the matrix element method and $0.6\%$ for the R-factor approximation).

The numerical size of the contributions beyond the 1-loop order depends on the
process and scenario, but the gNWA allows for their inclusion also in the
interference term.

\section{Conclusions}\label{sect:conclusion}

In this paper, we have developed a generalisation of the standard 
narrow-width approximation (NWA) that extends the applicability of this
important tool to scenarios where interference effects between
nearly mass-degenerate particles are important. This can be the case in many
extensions of the SM where the spectrum of the new particles is such that the
mass difference between two or more particles is smaller than one of their
total decay widths.
In such a case, their resonances overlap so that the interference cannot be neglected 
if the two states mix. In order to still enable the convenient factorisation 
of a more complicated process
into production and decay of an intermediate particle, 
we have demonstrated how to factorise also the interference term. 
This is achieved by evaluating the production and decay matrix elements on the mass-shells 
of the resonant particles in analogy to the terms present in the standard NWA. 
If one additionally assumes equal masses of the intermediate particles, 
it is possible to further approximate the interference contribution 
by an interference weight factor, $R$, in terms of production cross sections, 
decay branching fractions, ratios of couplings and a universal, process independent 
integral over Breit-Wigner propagators.

We have developed this generalised narrow-width approximation (gNWA) both at
the tree-level and at one-loop order. Following the analytic derivations, we
have discussed the application to a simple example process 
in the context of the MSSM with real parameters. We have considered 
the three-body decay 
of the heaviest neutralino via a resonant neutral, $\mathcal{CP}$-even Higgs
boson, $h$ or $H$, 
into the lightest neutralino and a pair of $\tau$-leptons. 
This process is well-suited for a test of the gNWA 
since it is sufficiently simple so that the full process can be calculated at
the loop level and compared with the predictions of the gNWA. Within the gNWA
this process can be decomposed into basic kinematic building blocks, namely
two subsequent 2-body decays, and the 
interference contributions involve only scalar particles. 
The discussion of interference effects can therefore be disentangled from 
spin-correlation issues. 
Furthermore, the process involves charged external particles, so
that the issue of the cancellation of IR divergencies between virtual loop
corrections and bremsstrahlung contributions is relevant, while the fact that
only the final state particles are charged makes the treatment of the
IR-divergent contributions very transparent.

We have validated the gNWA at the Born level (supplemented by higher-order
Higgs masses, widths and mixing factors) 
and at the 1-loop level including 
corrections of $\mathcal{O}(\alpha)$ with respect to the lowest order. 
Within the considered parameter region, 
the chosen modified $M_h^{\rm{max}}$-scenario leads to a small difference between the loop-corrected masses of
$M_h$ and $M_H$ below their total widths. 
This configuration results in a large negative interference term 
so that in the standard NWA, where the interference contribution is not taken
into account, the 3-body decay width is overestimated by a factor of
up to five in this example. 
Hence, the standard NWA is clearly insufficient in this scenario. The
inclusion of the factorised interference term, however,
leads to an agreement with the unfactorised decay width within few percent. 
At the tree level, the method of on-shell matrix elements and the $R$-factor 
approximation lead to very similar results.

However, at the Born level the methods for calculating multi-leg processes
without further approximations are very advanced. Accordingly, a particular
interest in the NWA concerns its application to the loop level, where the
difficulty in computing processes involving a variety of different mass
scales grows very significantly with the number of external legs of the
process. In many cases the factorisation into different sub-processes 
provided by the NWA is essential to enable the computation of higher-order 
contributions. 
In cases where a full tree level calculation is feasible, the NWA can
therefore be applied just at the loop level in order to facilitate the
computation of the higher-order corrections, while the lowest order contributions
are evaluated without further approximations 
in order to avoid an unnecessary theoretical uncertainty.

For a validation of the gNWA beyond the LO we have performed the 1-loop 
calculation of $\Gamma(\cf\rightarrow \co \tau^{+}\tau^{-})$ 
including all vertex corrections, self-energies involving Higgs-Goldstone/$Z$ 
mixing, Higgs-Higgs mixing contributions via finite wave function normalisation 
factors, box diagrams, as well as soft photon radiation.
All higher order corrections except for the box diagrams factorise, 
which makes a separate calculation of the 1-loop production and decay part possible 
as long as the non-factorisable contributions remain sufficiently small.
We have shown that within the gNWA
the factorised interference term at the next-to-leading order 
is both UV- and IR-finite. In order to preserve the cancellations of IR-singularities 
between virtual and real photon contributions also in the on-shell matrix elements, 
all IR-divergent integrals in matrix elements and the soft-photon factor were
evaluated at the same mass value. This prescription could be further improved
by extracting the singular parts from the real photon contribution and
applying the NWA only to those terms which match
the singularities from the virtual photons.
Furthermore, we have extended the interference weight factor to the 1-loop level. 
In the numerical comparison to the 3-body decay width, the gNWA based on 1-loop on-shell matrix 
elements agrees with the full 1-loop result within an accuracy of better than
$1\%$, which is much below the typical size of the loop corrections in
this case.  
The gNWA with interference weight factors, on the other hand, deviates from
the full result by up to $4\%$, which is still about a factor of two smaller than the size of the loop correction in the region where the latter is sizable. 
Therefore the method of on-shell matrix elements appears to be a well-suited approach 
for predicting the interference term at 1-loop order within roughly the remaining
theoretical uncertainty of the full result, while the additional 
R-factor approximation may be of interest as a technically simpler rough
estimate of the higher-order result including interference effects.

In our discussion we have first focussed on the strict 
$\mathcal{O}(\alpha)$ contribution relative to the lowest order within the gNWA
(except for masses, total widths and wave function normalisation factors, for
which we have incorporated dominant 2-loop contributions throughout this work)
for the purpose of a consistent comparison with the 3-body decay width. 
In the most accurate 
final result the factorisation into subprocesses for production and
decay has the virtue that higher-order corrections can naturally be
implemented into each of the subprocesses, which formally corresponds to a
higher-order effect for the full process. This applies also to the
interference term, where we have discussed the incorporation of higher-order
contributions for the two considered versions of the gNWA.

While much of our discussion has been directed to the specific example process
that we have investigated, we have provided a generic formulation of the
gNWA and we have commented on various features that are relevant for more
complicated processes. The method presented here should therefore be
transferable to processes with more external legs, with a more complicated
structure of IR divergencies, and to cases where the interference arises
between particles of non-zero spin. 

Based on the methodical study presented here, a next step will be a more
detailed investigation of phenomenological applications of the gNWA. This will
be addressed in a forthcoming publication.

\section*{Acknowledgements}
We thank Alison Fowler for the initial step of this work in her PhD
thesis\,\cite{Fowler:2010eba} and for providing us with her FeynArts model
file. Furthermore, we are thankful to Aoife Bharucha for useful discussions on
the renormalisation of the neutralino sector and to Thomas Hahn and Sven
Heinemeyer for support with FormCalc and FeynHiggs. E.F.\ thanks Howard Haber
and the other SCIPP members at UC
Santa Cruz as well as the CERN theory group, especially Andreas Weiler, for hospitality and interesting discussions, and the Studienstiftung des deutschen Volkes, DAAD and the PIER Helmholtz
Graduate School for financial support.
The work of G.W.\ is supported in part by the Collaborative Research Centre
SFB~676 of the DFG, ``Particles, Strings and the Early Universe'', and by the
European Commission through the ``HiggsTools'' Initial Training Network
PITN-GA-2012-316704.

\bibliographystyle{utphys}
\bibliography{NeutralinoRen,mthesis_literature,gNWALoop,exposee}
\markboth{}{}
\end{document}